\documentclass[prb,aps,nofootinbib,amssymb,twocolumn,superscriptaddress,10pt]{revtex4-2}
\usepackage[T1]{fontenc}
\usepackage{graphicx}
\usepackage{xcolor}
\usepackage{dcolumn}
\usepackage{comment}
\usepackage{amsmath,amssymb,bbold,bm}
\usepackage{hyperref}
\usepackage{amsfonts}
\usepackage{float,latexsym}
\usepackage{multirow}
\usepackage{color}
\usepackage[normalem]{ulem}
\usepackage{cleveref}
\usepackage{physics}
\usepackage{tabularx}
\usepackage{array}
\usepackage[caption=false]{subfig}
\usepackage{siunitx}
\usepackage{pifont}
\usepackage{placeins}
\usepackage{mathtools}
\usepackage{longtable}
\usepackage{multirow}
\usepackage{nicematrix}
\usepackage{tikz}
\usepackage{xstring}
\usepackage{makecell}
\usepackage{xr}
\usepackage{ifthen}
\usepackage{svg} 
\usetikzlibrary{decorations.markings}
\usetikzlibrary{positioning}
\usetikzlibrary{arrows.meta}
\usepackage{tikz-feynman}
\usetikzlibrary{calc}
\tikzfeynmanset{warn luatex=false}
\let\vec\mathbf

\makeatletter
\def\maketag@@@#1{\hbox{\m@th\normalfont\normalsize#1}}
\makeatother

\crefname{appendix}{Appendix}{Appendices}
\crefname{equation}{Eq.}{Eqs.}
\crefname{figure}{Fig.}{Figs.}
\crefname{table}{Table}{Tables}
\crefname{section}{Section}{Sections}
\crefname{enumi}{Point}{Points}
\creflabelformat{appendix}{[#2#1#3]}
\crefmultiformat{figure}{Figs.~#2#1\xdef\mycreffirstarg{#1}#3}%
 { and~#2#1#3}{, #2#1#3}{ and~#2#1#3}

\makeatletter     
\renewcommand\onecolumngrid{
\do@columngrid{one}{\@ne}%
\def\set@footnotewidth{\onecolumngrid}
\def\footnoterule{\kern-6pt\hrule width 1.5in\kern6pt}%
}

\makeatletter
\AtBeginDocument{\let\LS@rot\@undefined}
\makeatother 

\usepackage{svg} 

\crefname{appendix}{Appendix}{Appendices}
\crefname{equation}{Eq.}{Eqs.}
\crefname{figure}{Fig.}{Figs.}
\crefname{table}{Table}{Tables}
\crefname{section}{Section}{Sections}
\creflabelformat{appendix}{[#2#1#3]}

\makeatletter
\AtBeginDocument{\let\LS@rot\@undefined}
\makeatother

\makeatletter
\renewcommand\onecolumngrid{\do@columngrid{one}{\@ne}\def\set@footnotewidth{\onecolumngrid}\def\footnoterule{\kern-6pt\hrule width 1.5in\kern6pt}}

\allowdisplaybreaks

\begin{document}
\title{\texorpdfstring{Hidden antiferromagnetism, persistent valley fluctuations, and $U(6)$ crossovers in triangular-lattice M-point moir\'e materials via determinantal quantum Monte Carlo}{Hidden antiferromagnetism, persistent valley fluctuations, and U(6) crossovers in triangular-lattice M-point moir\'e materials via determinantal quantum Monte Carlo}}
\author{Konstantinos Vasiliou}
	\affiliation{Rudolf Peierls Centre for Theoretical Physics, University of Oxford, Oxford OX1 3PU, United Kingdom}
    \author{Dumitru C\u{a}lug\u{a}ru}
\affiliation{Rudolf Peierls Centre for Theoretical Physics, University of Oxford, Oxford OX1 3PU, United Kingdom}
    \author{Johannes S. Hofmann}
	\affiliation{Max-Planck-Institut f\"ur Physik Komplexer Systeme, N\"othnitzer Strasse 38, 01187 Dresden, Germany}
	\author{S.A. Parameswaran}
	\affiliation{Rudolf Peierls Centre for Theoretical Physics, University of Oxford, Oxford OX1 3PU, United Kingdom}

\let\oldaddcontentsline\addcontentsline
\begin{abstract}
A new moiré material platform was recently proposed based on twisting two-dimensional atomic monolayers whose low-energy states lie at the three M-points of the Brillouin Zone. Continuum and {\it ab initio} modeling suggest that electrons in the conduction bands of these materials realize three-valley Hubbard models with valley-selective, quasi-one-dimensional hopping. Remarkably, the onsite Hubbard repulsion is almost $U(6)$-symmetric without fine-tuning. Here, we show that this class of systems naturally admits sign-free determinantal Quantum Monte Carlo simulations at a filling of three electrons per moiré unit cell. We use these to explore the phase diagram for interactions of various strengths and $U(6)$-breaking anisotropies. We show that for near-isotropic interactions as relevant to, e.g., AA-stacked twisted SnSe$_2$, the system exhibits an extended intermediate-coupling regime in which local-moment formation and itinerancy compete, and the crossover to a putative low-temperature ordered state can be understood in terms of fluctuating $U(6)$ local moments. We argue that many of these features persist beyond the idealized sign-problem-free limit.
\end{abstract}
\maketitle
\noindent
{\textit{Introduction.}}--- The advent of moir\'e materials has stimulated new directions in the study of strongly-correlated electron systems, such as explorations of  the interplay of band topology with interactions. It has also
revitalized classic questions linked to the competition between itineracy and localization  familiar from Hubbard or heavy-fermion settings. Three aspects of the moir\'e platform are especially relevant in this regard~\cite{AND21,KEN21,NUC24}. First, their inherent tunability and two-dimensional (2D) nature allows convenient and controlled access to a wide range of parameter regimes and experimental probes. Second, the geometry of the emergent moir\'e superlattices can often frustrate ordering,  enriching phase structure. Third, the separation of the moir\'e and atomic lattice scales  facilitates the powerful experimental tool of electrostatic doping, while also  (approximately) enhancing symmetries: e.g. valley symmetry can be viewed as continuous on the moir\'e scale despite its microscopic point-group origins. For these reasons, various phenomena familiar from more traditional strongly-correlated systems often emerge with unconventional features in the moir\'e setting.

Hitherto, nearly all moir\'e materials have been assembled from  graphene and transition metal dichalcogenide monolayers with low-energy degrees of freedom near the $\Gamma$ or $K$ points in the Brillouin Zone (BZ)~\cite{Cao_2018_CI,Cao_2018_SC,ANG21,CLA22a,PI26}. Recently, a new platform based on twisting monolayers with low energy degrees of freedom near the M point has been proposed~\cite{Calugaru_2025,Lei_2025}. This results in a three-valley  moir\'e problem where the bandstructure in each valley involves states from a distinct monolayer M-point, and the valleys are related by threefold ($C_{3z}$) rotations.  Absent external fields, the bands in each valley are time reversal invariant and topologically trivial and hence admit a low-energy Wannier orbital representation. Restricting this to a single band in each valley yields a three-orbital tight-binding model on the triangular lattice (we retain the term `valley' to reflect the origin of the orbitals), with anisotropic, valley-selective hoppings. While this structure is reminiscent of multi-orbital systems of Kugel-Khomskii type~\cite{Kugel_1982,Li_1998,Feiner_1997,Tokura_2000,Khaliullin_2005,Nussinov_2015} and might thus have been anticipated on general grounds, the moir\'e setting injects two unusual features into the problem. (1) In many cases there is an emergent momentum-space nonsymmorphic symmetry ($\tilde{M}_z$) that forbids certain hoppings, enhancing the anisotropy to the extent that the kinetic energy becomes quasi-1D within a valley; since interactions couple the valleys, the system is approximately ``mixed dimensional''. (2) The suppression of intervalley Hund's couplings at small twist angles and the spatial proximity of the three on-site orbitals mean that interactions (which remain fully 2D) are almost $U(6)$ symmetric. While both these features are only approximate and e.g., are affected by relaxation,  modeling suggests that they remain substantially valid in realistic systems, such as the  leading M-point material candidate, twisted AA-stacked SnSe$_2$ (AA t-SnSe$_2$). Preliminary analyses~\cite{Li_2025,Bao_2025,debeule_2025,ingham_2025} of this system using mean-field techniques or in  idealized regimes
suggest striking phenomenology, but --- as in many cases --- a controlled window into the intermediate coupling regime is an open problem whose resolution is of great relevance to experiments planned in the near term.

Here and in a companion paper~\cite{Calugaru_2026}, we show that the M-point moiré problem constitutes a qualitatively new setting in which such access becomes possible. Specifically, we show that this  class of systems naturally admits two distinct, unbiased sign-free Quantum Monte Carlo (QMC) formulations along complementary axes in parameter space, which separately leverage the mixed dimensionality and the symmetry structure. In the companion work~\cite{Calugaru_2026}, we establish that in the quasi-1D ``chain'' limit --- where electrons in a given valley propagate along one of three principal directions --- Stochastic Series Expansion (SSE) QMC~\cite{Sandvik_1992,SEN02,SAN19,XU15} applies at arbitrary electron filling $\nu \in [0,6]$ (including both spins and all 3 valleys). In the present work, we focus on half-filling ($\nu=3$), and show that upon introducing second-neighbor inter-chain hopping $t_\perp$ (while preserving $\tilde{M}_z$ symmetry), a hidden bipartite structure emerges that enables determinantal QMC (DQMC)~\cite{Blankenbecler_1981,Scalapino_1981,White_1989,Assaad_2008}. This structure supports a form of ``hidden'' N\'eel antiferromagnetic order within a single valley at strong coupling, in contrast to the geometric frustration expected on the triangular lattice. However, as the interaction anisotropy parametrized by $\alpha$ approaches the $U(6)$-symmetric point ($\alpha \to 1$), enhanced valley degeneracy leads to strong fluctuations that suppress the onset of a correlated insulating state, generating an extended intermediate-coupling regime in which local-moment formation and itinerancy compete down to low temperatures. We find that substantial valley fluctuations persist even beyond the onset of magnetic order and show that, for near-isotropic interactions, the physics can be understood in terms of a $U(6)$ local-moment crossover at intermediate temperatures. Our results identify the M-point problem as a new class of moir\'e systems whose intrinsic non-perturbative accessibility provides a beachhead for systematic exploration of strongly correlated multi-orbital physics.
\begin{figure}[t]
    \centering 
    \includegraphics[width=\linewidth]{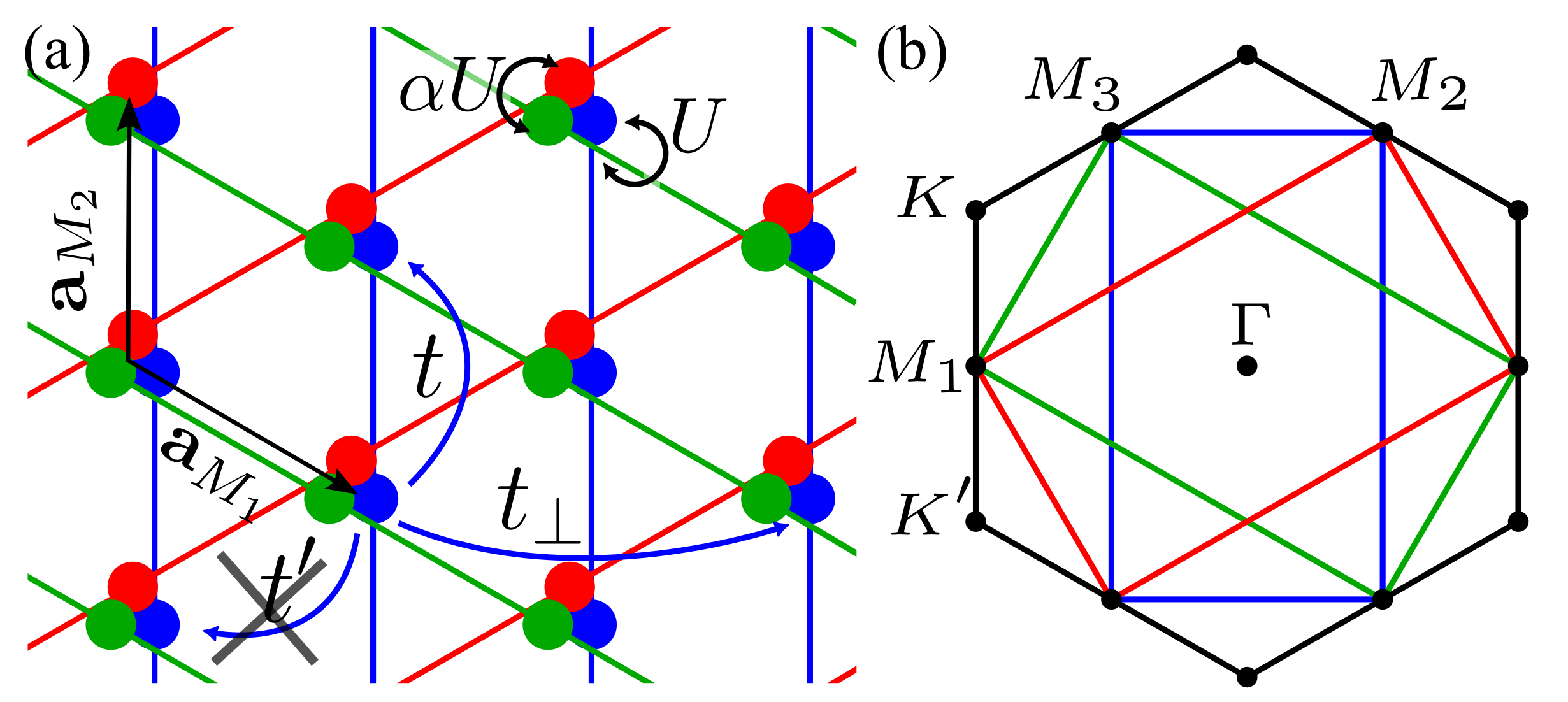}
    \caption{(a) Each dot represents  a spinful Wannier orbital, with colors denoting different valleys $\eta$. We set symmetry-forbidden or subdominant terms beyond $t,t_\perp$ to zero. (b) The reciprocal Wigner-Seitz cell of the triangular lattice (black) defines the moir\'e Brillouin zone (mBZ) but each valley experiences an effective rectangular BZ (colors).}
    \label{fig:main:lattice diagram}
\end{figure}

\noindent
{\textit{Model, Methods, and Measures.}}--- We focus on the $C_{3z}$-symmetric triangular lattice model obtained by Wannierizing the lowest-energy conduction band of the continuum moir\'e model for the M-point materials. Each moir\'e unit cell hosts one orbital for each valley $\eta\in \{0,1,2\}$ (colored dots in Fig.~\ref{fig:main:lattice diagram}a) that can be occupied by electrons of either spin $s\in\{\uparrow,\downarrow\}$. 
A special feature is that electrons only hop within the same valley, and do so anisotropically: in valley $\eta$, the dominant hopping is the nearest-neighbor $t$ along $C_{3z}^{\eta}\mathbf{a}_{M_2}$, with a weaker second-neighbor  $t_\perp$ perpendicular to $C_{3z}^{\eta}\mathbf{a}_{M_2}$. Here, $\mathbf{a}_{M_2}=\hat{\mathbf{y}}$ denotes a moir\'e lattice vector in units of the moir\'e scale. The nearest-neighbor inter-chain hopping ($t'$ in Fig.~\ref{fig:main:lattice diagram}a) is forced to vanish by the $\tilde{M}_z$ symmetry~\cite{Calugaru_2025,Li_2025}.  Thus, the kinetic term takes the form
\begin{equation}
        H_t = \sum_{\mathbf{R},\Delta \mathbf{R}} \sum_{\eta,s}  t^{\eta}_{\Delta \mathbf{R}}\hat{d}^\dagger_{\mathbf{R},\eta,s} \hat{d}^{\phantom{\dagger}}_{\mathbf{R}+\Delta\mathbf{R},\eta,s},
    \label{Eq:Hamiltonian:Kin}
\end{equation}
where $\hat{d}^\dagger_{\mathbf{R},\eta,s}$ creates an electron of valley  $\eta$ and spin $s$ at $\mathbf{R}$, and $t^{\eta}_{\Delta \mathbf{R}}$ is just the $t,t_\perp$ appropriate to valley $\eta$. A convenient scale for $H_t$ is its {half-width}, $W=2(t+t_\perp)$. Since $t'=0$, $H_t$  disconnects into two decoupled rectangular lattices in each valley.

The electrons interact via density-density interactions originating from the microscopic Coulomb repulsion, truncated at the nearest-neighbor (NN) level. Because the Wannier centers of the three valleys are not exactly coincident, the interactions have a small degree of valley anisotropy. We focus here on the anisotropy of the on-site term, parametrized by $\alpha$ such that interactions are $U(6)$ symmetric for $\alpha=1$, but the symmetry reduces to $U(2)^{\otimes 3}$ for $\alpha<1$ ($\alpha>1$ is not physically realizable in this setting).
Including a nearest-neighbor interaction $V$ (that we keep isotropic throughout for simplicity, as it does not qualitatively affect our results), we have
\begin{equation}
\begin{split}
H_U = \sum_{\mathbf{R}} \Big[ \frac{U}{2} \Big(
    \sum_{\eta}&\hat{n}_{\mathbf{R},\eta}^2
    + 2\alpha\sum_{\eta > \eta'} \hat{n}_{\mathbf{R},\eta}\hat{n}_{\mathbf{R},\eta'} \Big)\\
    &+ \frac{V}{2} \sum_{|\Delta\mathbf{R}|=1}
    \hat{n}_\mathbf{R}\hat{n}_{\mathbf{R}+\Delta\mathbf{R}}
\Big] \,,
\end{split}
\label{Eq:Hamiltonian:Int}
\end{equation}
with $\hat{n}_{\mathbf{R},\eta}=\sum_s \hat{d}^\dagger_{\mathbf{R},\eta,s}\hat{d}^{\phantom{\dagger}}_{\mathbf{R},\eta,s}$ and $\hat{n}_\mathbf{R} = \sum_\eta \hat{n}_{\mathbf{R},\eta}$. Owing to valley-spin degeneracy, $H=H_t+H_U$ allows filling $\nu\in[0,6]$. Apart from the discrete $\tilde{M}_z$,$C_{3z}$,$C_{2x}$ symmetries, the system also enjoys continuous spin-rotation and charge conservation within each valley, which combine to U$(2)^{\otimes 3}$. Note that  $H_t$ breaks $U(6)$ to U$(2)^{\otimes 3}$ even if $\alpha=1$. 

We work exclusively at half-filling, $\nu=3$, where the  system exhibits an anti-unitary particle-hole symmetry (PHS) $\mathcal{P}$ which makes it amenable to sign-problem-free DQMC~\cite{Wu_2005,TRO05,LI19b,Hofmann_2022} for $0<V/U<(2\alpha+1)/9$. Heuristically, this is because even though the sites lie on a triangular lattice, $H_t$ is bipartite in each valley, leading to a PHS similar to that on the square lattice. Rewriting Eq.~\ref{Eq:Hamiltonian:Int} in a form amenable to a Hubbard-Stratonovich transformation, which is necessary for DQMC, leads to the restriction on $V/U$ as in the extended Hubbard model ~\cite{Yao_2022,Sousa_2024,Golor_2015,SupMat}. 

We implement DQMC using the ALF package~\cite{ALF_article,ALF_codebase} on systems of $L\times L$ unit cells for $L\leq 12$, down to temperatures $T = 0.01t$. We organize our discussion around four observables: 
(1)~the momentum-resolved spin-spin correlations $\mathcal{S}^{(\eta)}_\mathbf{q}=\langle \hat{\mathbf{S}}^{(\eta)}_{\mathbf{q}}\cdot\hat{\mathbf{S}}^{(\eta)}_{\mathbf{-q}} \rangle$ with the spin operator defined as $\hat{\mathbf{S}}^{(\eta)}_{\mathbf{R}}=\left(1/2\right)\times\sum_{ss'}\hat{d}^\dagger_{\mathbf{R},\eta,s}\bm{\sigma}_{ss'}\hat{d}_{\mathbf{R},\eta,s'}$, which serves as a witness of magnetic ordering; 
(2)~the imaginary-time Green's function  $\mathcal{G}^{(\eta)}(\mathbf{k},\tau)=\langle\mathcal{T}_{\tau} \hat{d}_{\mathbf{k},\eta,s}(\tau)\hat{d}^\dagger_{\mathbf{k},\eta,s}(0) \rangle$ which provides a window into the charge spectrum;
(3)~the valley fluctuations $\mathcal{F}(\mathbf{k},\tau) = \left(1/6\right)\times \sum_{\eta\neq\eta'}\langle \mathcal{T}_\tau \hat{d}^\dagger_{\mathbf{k},\eta,s}(\tau) \hat{d}_{\mathbf{k},\eta',s'}(\tau) \hat{d}^\dagger_{-\mathbf{k},\eta',s'}(0) \hat{d}_{-\mathbf{k},\eta,s}(0)      \rangle$ with arbitrary $s,s'$ spins; 
and (4)~the fully-resolved, local, equal-time density-density correlator $\langle \hat{n}_{\vec{R},\eta,s} \hat{n}_{\vec{R},\eta',s'}\rangle$. Note  that $\mathcal{G}^{(\eta)}$ and $\mathcal{S}^{(\eta)}$ are valley-diagonal due to the U$(2)^{\otimes 3}$ symmetry. We extract (1-4) from DQMC data using standard techniques~\cite{SupMat}.

\noindent
{\textit{Magnetic properties.}}--- As flagged earlier, electrons in each valley hop on one of two decoupled bipartite rectangular sublattices. It is hence useful to view each valley as defining a hidden rectangular Brillouin zone (BZ) within the larger moir\'e BZ (mBZ) of the triangular lattice (Fig.~\ref{fig:main:lattice diagram}b). Since  inter-valley interactions are purely density-density, it follows then that collinear antiferromagnetism is possible within these sublattices, evading the frustration of the superficially triangular geometry. This is immediately manifest in our DQMC results:
we find a sharp peak in the momentum-space structure of $\mathcal{S}^{(\eta)}_{\mathbf{q}}$ linked to these `hidden' AFM correlations 
at the rectangular BZ corners  $\mathbf{Q}^{(\eta)} = C_{3z}^{\eta}\left(\pi/\sqrt{3},\pi\right)$, which coincide with the $M$-points of the mBZ  (Fig~\ref{fig:main:lattice diagram}b).
\par
\begin{figure}[t!]
    \centering
    \includegraphics[width=\linewidth]{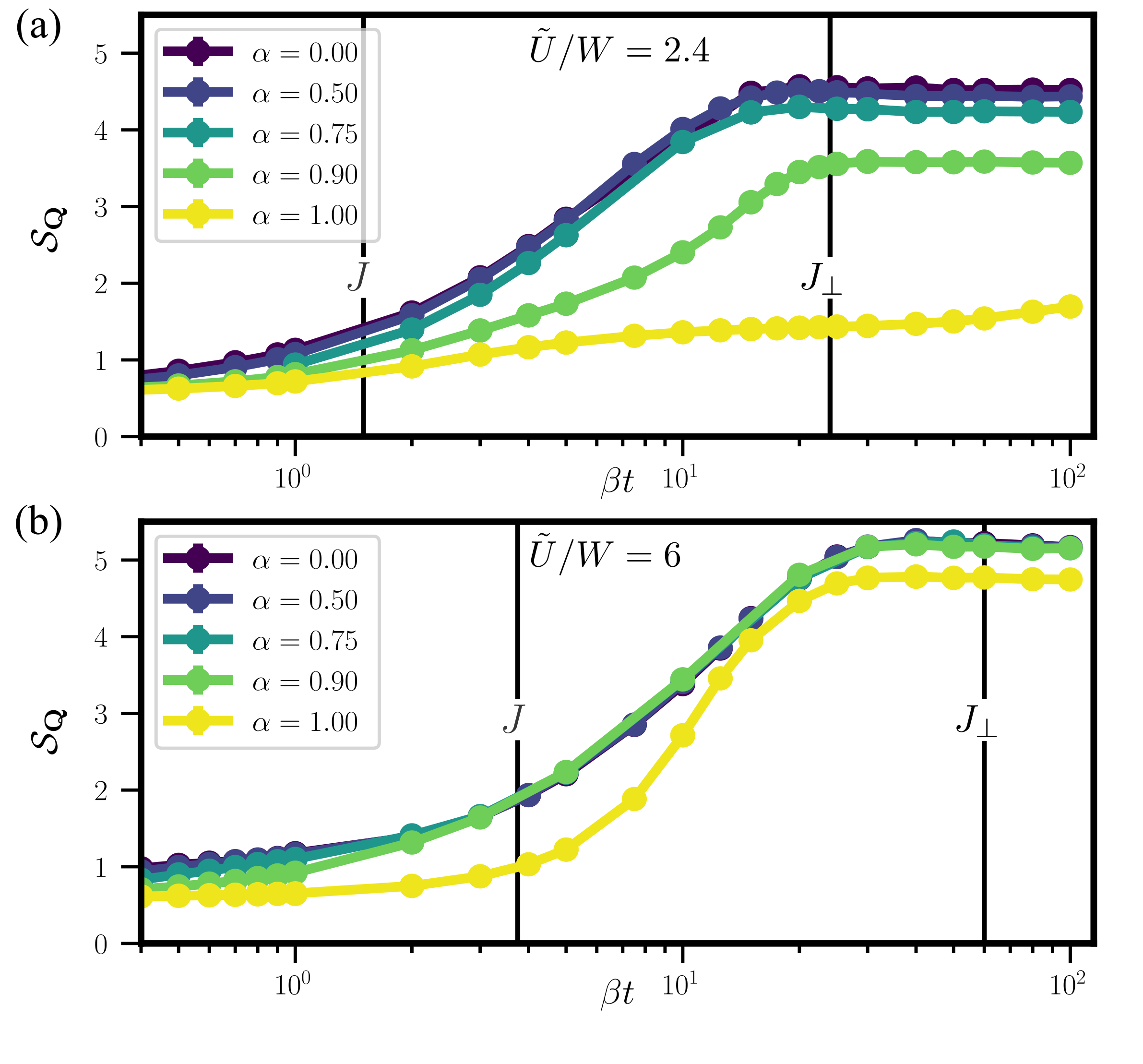}
    \captionsetup{skip=2pt}
    \caption{The temperature dependence of the spin correlations at the ordering vector $\mathbf{Q}^{(\eta)}$ for $V=0,t_\perp=0.25$ and $L=12$, for $\tilde{U}/W=2.4$ and $\tilde{U}/W =6$ ($U/t=6,15$ respectively). The vertical lines denote $J=4t^2/U,J_\perp=4t_\perp^2/U$. For $\alpha>0.90$ in the top panel, the system remains in the `intermediate-coupling' regime, whereas all other $(U,\alpha)$ pairs can be characterized as `strongly-coupled'.}
\label{fig:main:AFM correlations vs beta}
\end{figure}
A more subtle question is whether these correlations reflect an incipient long-range $T=0$ AFM order, permitted by a nonzero $t_\perp$. (Since $d=2$ is the lower critical dimension for  finite-$T$ continuous symmetry breaking, the correlation length diverges exponentially as $T\to0$ and so $T=0$ order appears as a finite-$T$  ``renormalized classical''  regime.)
Note that owing to the hidden bipartite structure in the hopping, the non-interacting Fermi surface of valley$-\eta$ electrons is perfectly nested  at $\mathbf{Q}^{(\eta)}$, so that the system is unstable to the formation of Slater-like AFM order for arbitrarily weak $U/t\to 0^+$~\cite{SupMat}; however, for $U\ll t$ the corresponding temperature scale at which finite samples appear ordered is exponentially small in $U/t$ and hence is numerically inaccessible. We focus instead on the crossover physics for large $U/t$ and work down to low but nonzero $T \approx 0.01t$, with the caveat that the ultimate nature of the $T=0$ state in some parameter regimes may differ from that of the low-$T$ crossover.

\begin{figure*}
    \centering
    \includegraphics[width=\textwidth]{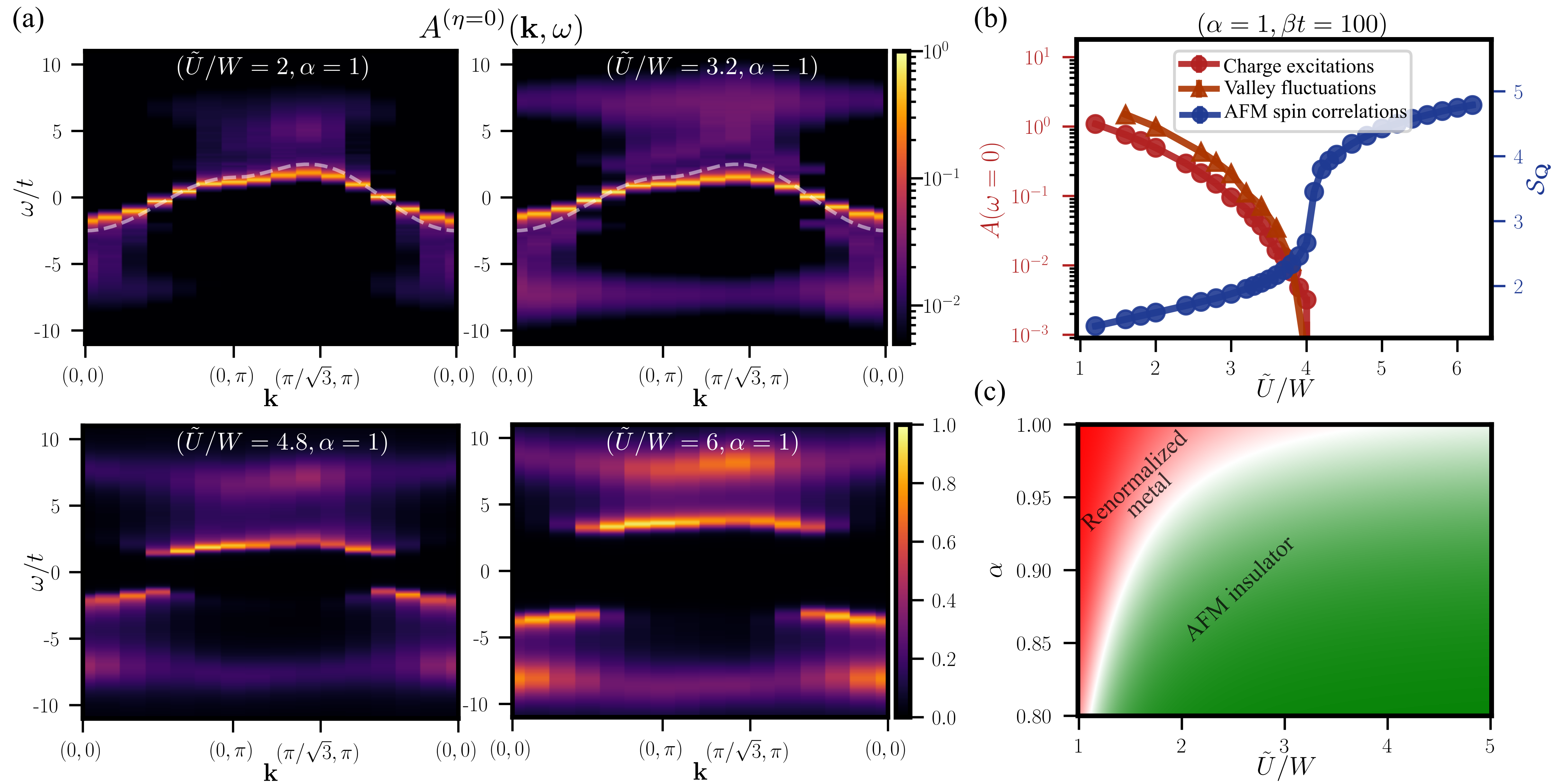}
    \caption{(a)~Evolution of the analytically continued spectral function $A^{(\eta=0)}(\mathbf{k},\omega)$  in valley $0$ and at  $T=0.05t$ from $U/t=5$ ($\tilde{U}/W=2$) to $U/t=15$ ($\tilde{U}/W=6$). (b)~AFM spin correlations (blue) and the {local} (BZ-integrated) $\omega=0$ charge and valley spectral weights (red)  extracted from imaginary-time data at $T = 0.01t$.
    Note that AFM correlations rise concurrently with a spectral gap opening at $U\approx 10t$. (c)~In the qualitative low-$T$ phase diagram, an insulator with long-range AFM order crosses over to a strongly-renormalized finite-$T$ metal with only short-range AFM correlations, with the latter two features best captured via a $U(6)$ local-moment picture (cf. Fig.~\ref{fig:main:AFM correlations vs beta}). Throughout, we set $t_\perp=0.25t$, $V=0$ and consider a $12 \times 12$ system.}
    \label{fig:main:Phase diagram}
\end{figure*}
We work at small $V/U$ to exclude proximate competing charge-density wave (CDW) orders, and in this regime we find that a purely on-site interaction  $\tilde{U}=U-V$ leads to qualitatively similar physics~\cite{Schuler_2013}. Fig \ref{fig:main:AFM correlations vs beta} shows how AFM correlations grow as the system is cooled. For $\tilde{U} \gtrsim W$, the local moments are mostly well-formed once $T\lesssim \tilde{U}$, and  begin to align coherently for $T\lesssim J=4t^2/\tilde{U}$, the super-exchange scale. This is evident as a rise in correlations at $T\sim J$. Less expected is the dramatic suppression of the correlations near the isotropic point  $\alpha=1$. A closer analysis of the $\alpha=1$ correlations reveal that they are  sensitive to a second, system-size dependent energy scale $T\ll J$ (an $L\rightarrow \infty$ extrapolation is performed in the SI~\cite{SupMat}). This leads to a very broad intermediate-coupling regime $W \lesssim \tilde{U} \lesssim 4W$ (Fig \ref{fig:main:AFM correlations vs beta}a). In this regime, there are apparently low-energy processes that disrupt the formation of long-range AFM order. Once the system enters what might be termed the `true' strong-coupling regime, these low-energy processes are suppressed, the sensitivity to $\alpha$ diminishes, and true long-range AFM order  sets in (Fig \ref{fig:main:AFM correlations vs beta}b). We next turn to understanding this intermediate coupling regime and its striking $\alpha$-sensitivity.

\noindent
{\textit{Charge and valley fluctuations.}}--- To clarify the nature of the low-energy processes at intermediate coupling, we probe the charged excitations present in the system via the full real-frequency spectral function $A^{(\eta)}(\mathbf{k},\omega)=-\text{Im}G^{(\eta)}(\mathbf{k},\omega)$, analytically continued from the imaginary-time dynamical Green's function $\mathcal{G}^{(\eta)}(\mathbf{k},\tau)$ using classical maximum-entropy methods~\cite{Bryan_1990,Levy_2017,ALF_codebase}. We focus on $A^{(\eta)}(\mathbf{k},\omega)$ since it is the most intuitive, but our results are consistent with measures that rely only on imaginary-time or thermodynamic data, such as $\mathcal{G}^\eta$ or the compressibility $\kappa=\partial \langle \hat{N} \rangle/\partial\mu$~\cite{SupMat}.

The weak-coupling spectral weight closely tracks the non-interacting bands, remaining relatively insensitive to the interaction anisotropy $\alpha$, with any nesting-driven quasiparticle gap unresolvable even at $T=0.01t$ below $\tilde{U}\lesssim W$ ~\cite{SupMat}. For $\tilde{U}\gtrsim W$ and sufficiently large anisotropy (i.e. $(1-\alpha)U\gtrsim J$), quasiparticles clearly show a spectral gap. Upon further increasing $\tilde{U}/W$,  spectral weight shifts into the Hubbard bands, indicating strong coupling~\cite{SupMat} as in the usual Slater-Mott crossover~\cite{Borejsza_2004,hajieleftheriou_2024}. 

This picture changes for $\alpha\approx 1$: the quasiparticles remain gapless up to $\tilde{U}\lesssim 4W$ even as the Mott bands start to accumulate spectral weight, showing the coexistence of localized and itinerant degrees of freedom (Fig.\ref{fig:main:Phase diagram}a). For $\tilde{U}\gtrsim 4W$,  quasiparticles are gapped at $T=0.01t$ but retain spectral weight. Further increasing $\tilde{U}$ shifts  weight  into the Hubbard bands, until the full depletion of the quasiparticle weight signals strong coupling.

More insight can be gleaned by considering intervalley correlations. It is illuminating to perturb around the `atomic' $t=t_\perp=0$ limit with three electrons per site. For $\alpha=1$, the atomic problem is $U(6)$ symmetric, so that (in an obvious notation) the `valley-singlet' $(111)$ state is degenerate with the `valley-imbalanced' $(210)$ configurations. This degeneracy is lifted in the atomic limit at  $O((1-\alpha)U)$, and away from it by valley-selective hopping at $O(t^2/U)$. Such soft valley fluctuations are known  to enhance the critical $U_c$ for a metal-insulator transition in a fully $U(6)$ Hubbard model~\cite{Florens_2004}. Although our model is never truly $U(6)$-symmetric even for $\alpha=1$ (due to the hopping), for intermediate-to-strong coupling where $t^2/U$ terms are suppressed and the dominant $U(6)$-breaking is onsite,  the local-moment physics can be approximately $U(6)$-symmetric over a range of $T$ that grows as $\alpha\to 1^-$. (Note that this also explains why the weak-coupling physics is much less sensitive to $\alpha$ since there the leading $U(6)$ breaking is from the hopping)

The role of charge-neutral valley fluctuations is particularly significant at  intermediate coupling. In this regime,  which is quite extended in $\tilde{U}/W$ when $\alpha \approx 1$, intervalley correlations are strong and coherent over short distances. The fact that this behavior coincides with the presence of itinerant charged modes in $A(\mathbf{k},\omega)$, suggests that the two phenomena are linked: we conjecture that {in the intermediate-coupling regime, the  low-energy charged quasiparticles are electrons dressed by neutral valley fluctuations, as would be the case in a U$(6)$ Hubbard model}. At fixed low $T$ and for sufficiently large $\tilde{U}/W$, the itinerant quasiparticles eventually gap out, accompanied by a rise in AFM correlations $\mathcal{S}_{\mathbf{Q}}$~(Fig.~\ref{fig:main:Phase diagram}b). We probe the low-frequency spectral weight of the Green's function, as well as the valley fluctuations directly in imaginary time (i.e. without analytical continuation) by using the (BZ-averaged) proxies for the $\omega=0$ spectral weight $A_0 \approx\beta\sum_{\mathbf{k}}\mathcal{G}(\mathbf{k},\beta/2)$ and $A^{{v}}_0 \approx \left(\beta^2/\pi \right)\sum_{\mathbf{k}} \mathcal{F}(\mathbf{k},\beta/2)$~\cite{Trivedi_1996,Wang_2020}.
The evolution of $A_0, A^v_0$ with $\tilde{U}/W$ (Fig.~\ref{fig:main:Phase diagram}b) shows the intertwined nature of the itinerant quasiparticles and valley fluctuations and their competition with magnetic order: gapless itinerant quasiparticles impede the coherence of long-range AFM correlations, while gapped quasiparticles facilitate it, with the competition between these tendencies controlled by the scale of the valley fluctuations.

\begin{figure}[t!]
    \centering
    \includegraphics[width=\linewidth]{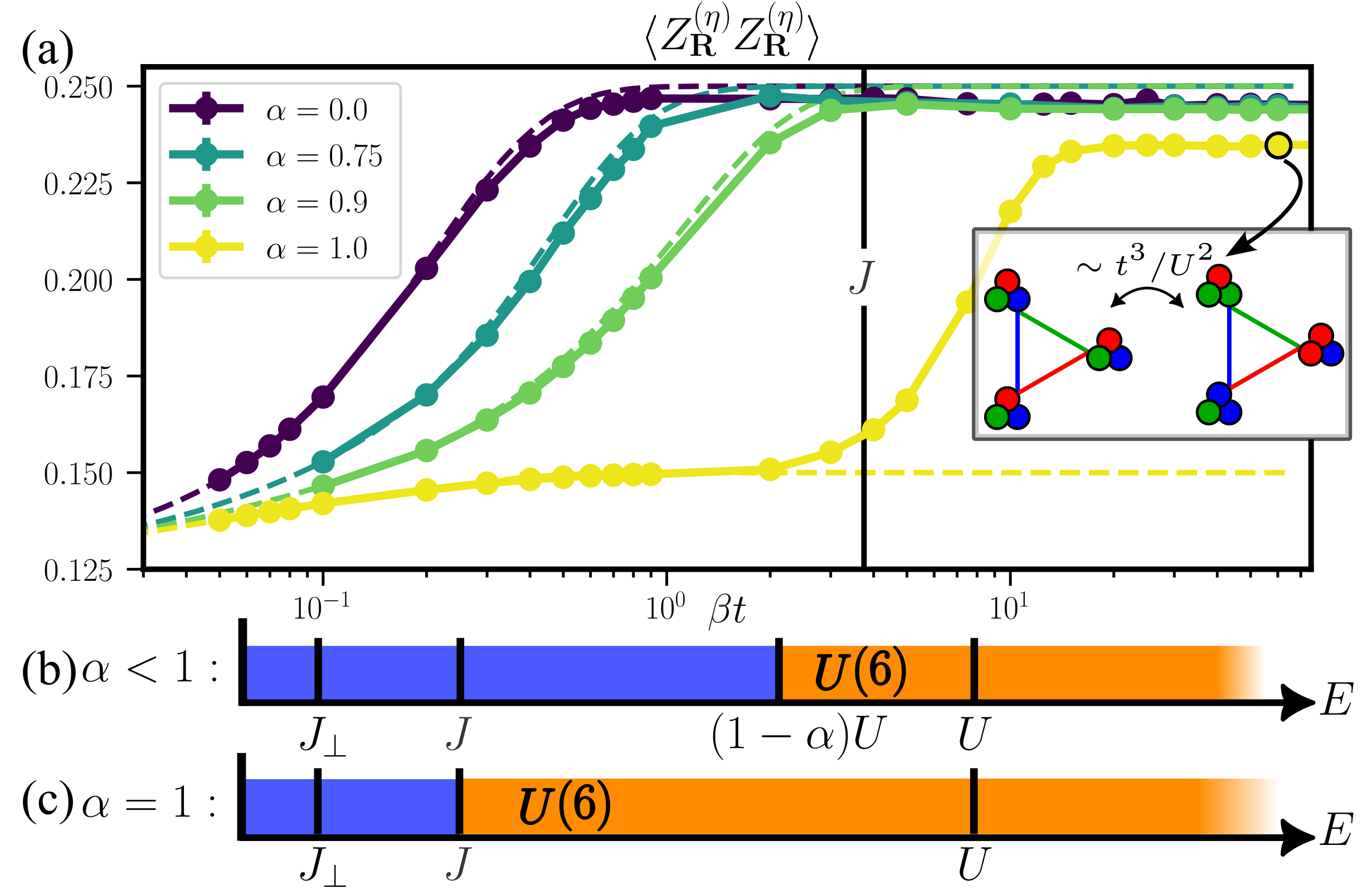}
    \caption{(a) The square of the local moment as a function of inverse temperature in the strong-coupling regime, at $\tilde{U}/W=6$. The dashed lines correspond to the atomic problem and the inset displays higher-order ring-exchange processes reducing the ground state total moment for the $U(6)$-symmetric interactions. (b)\&(c) Show the energy hierarchy for $U(6)$-breaking and $U(6)$-symmetric interactions. }
    \label{fig:main: local charge correlations}
\end{figure}

At strong coupling, where the spectral weight is primarily in the Mott bands, the neutral valley excitations remain active, although they lose real-space coherence and are hence local. They continue to influence  local observables, e.g. the size of the local moment (Fig \ref{fig:main: local charge correlations}a). For $J \lesssim T$ and $\alpha =1$, the U$(6)$-breaking $O(t^2/U)$ terms are incoherent and local observables approach their U$(6)$ values, while even as $T\to0^+$, low-weight valley-ring-exchange processes deplete the local moments~\cite{SupMat}. Many of these features are captured by a parton mean-field theory~\cite{SupMat} of coupled charge and neutral valley rotors~\cite{Florens_2002,Florens_2004}. As $\alpha\to 1^-$, the latter go soft, taking the charge mode with them and pushing the Mott phase to large $\tilde{U}/W$.

These considerations lead us to deduce the finite-$T$ crossover diagram of Fig.~\ref{fig:main:Phase diagram}c. At fixed $\tilde{U}/W$, for sufficiently large anisotropy, the valleys largely decouple. The system is Mott insulating with a large gap $\sim U$ and AFM correlations with $\xi \gg L$. On the other hand, if the anisotropy is small, the valley fluctuations reorganize the single-particle spectrum leading to the presence of (nearly) gapless strongly-renormalized electrons which disrupt the coherence of AFM correlations. This regime is a correlated finite-$T$ metal. We expect that qualitatively similar features pertain also to realistic systems as long as there is a broad intermediate-coupling window where the $U(6)$-breaking scale is set by $J\sim t^2/U$ rather than $(1-\alpha) U$, which requires $U/t\lesssim 1/\sqrt{1-\alpha}$. We have verified this for   $\alpha\approx0.98$ relevant to $\text{AA t-SnSe}_2$.

\noindent
{\textit{Concluding Remarks.}}--- In this work, we have identified a special yet physically relevant limit of the interacting M-point moir\'e problem, wherein the unusual hopping pattern of the idealized moir\'e bandstructure leads to a hidden bipartite structure such that at half-filling DQMC is sign-problem-free. This allowed us to explore the half-filled phase diagram for a wide range of interactions and down to low temperatures. Our key finding is that the near-isotropy of the on-site Hubbard repulsion leads to a broad thermal crossover regime at intermediate coupling that can be best understood in terms of fluctuating $U(6)$ moments. At small nonzero temperatures, the low-energy charged quasiparticles can be viewed as dressed by substantial neutral valley fluctuations that persist (as low-energy virtual processes) all the way to the strong-coupling Hubbard limit. We validated this intuitive picture by comparing our large-scale DQMC studies with both a perturbative approach and a parton mean-field description. We expect that our work will have direct relevance to future experimental studies of half-filled AA t-SnSe$_2$, where the valley-fluctuation-driven intermediate-coupling physics is likely prevalent over a large region in parameter space. The underlying symmetry structure that facilitated sign-free DQMC  at half-filling is likely applicable to a broad class of M-point moir\'e materials, which are thus intriguing settings for further study of strong correlation physics guided by precision numerics.

In closing, we comment on how deviations from the idealized sign-problem-free limit might modify our conclusions. As argued above, the weak-to-intermediate coupling physics of the sign-problem-free model is driven primarily by charge and valley fluctuations. These features are expected to be relatively insensitive to the inclusion of small terms that add realism while reintroducing the sign problem.  In contrast, the strong-coupling spin physics at $T=0$ is much more sensitive to such perturbations. The three most important terms we neglected are: (1)~the nearest-neighbor intravalley interchain hopping $t'$, which is forced to vanish by $\tilde{M}_z$ but can have a small non-zero value on including relaxation; (2)~a second-neighbor intravalley hopping $t''$ along the chain direction; and (3)~the on-site Hund's coupling $J_H$ between spins in different valleys that originates from both the inter-valley Coulomb interaction and electron-phonon couplings, which respectively give FM and AFM contributions to $J_H$~\cite{Calugaru_2026,SupMat}. 

To analyze the relative importance of these terms  it is easiest to work at strong coupling, where  $t',t''$ respectively give rise to nearest-neighbor interchain and second-neighbor intrachain AFM exchange terms $J'\sim t'^2/U$, $J''\sim t''^2/U$; we expect these to be much smaller than the bipartite superexchange scale $J\sim t^2/U$. Although the FM contribution to $J_H$ from intervalley Coulomb interactions is  suppressed relative to $U$ by the ratio of the microscopic and moir\'e lattice scales, $O(1)$ factors  conspire so that $|J_H|_{iv}\sim J$~\cite{SupMat,Calugaru_2026}. However, the AFM coupling from electron-phonon interactions can be comparable~\cite{SupMat,Calugaru_2026}, so it is reasonable to take $J_H$ as variable in sign and small in magnitude and hence comparable to $J_\perp, J',J''$.
 
Each of these terms destroys the bipartite structure, so that the strong-coupling physics is that of a {\it frustrated} antiferromagnet. As such, we expect that they will all further suppress the ordering temperature (if any) thereby widening the range of $T$ over which the crossover physics  is apparent. Going beyond this and definitively establishing the nature of the $T=0$  state   requires fundamentally different techniques, owing to the sign problem. However, the special case $J_H=J''=0$, $J' \lesssim J_\perp\ll J$ allows for some analytical control~\cite{SupMat}: In this limit, reinstating $J'$ leads to a model of weakly coupled N\'eel orders on the two independent bipartite rectangular sublattices in each valley. A variant of the `order by disorder' calculation of Refs.~\cite{Villain_80,Henley_89,Jolicoeur_1990} then  leads to a striped `spin-nematic' order within each valley, whose orientation in each valley is free to rotate independently. Taking $J_H, J''\neq0$ will frustrate this order once again. Further investigation of the $T=0$ phases, especially given the possibility of quantum spin liquid behaviour, is an intriguing  future direction.

\noindent
{\textit{Acknowledgments.}}--- We thank Jixun K. Ding, Ziwei Wang and Haoyu Hu, for insightful discussions, and Haoyu Hu, B. Andrei Bernevig, and Werner Krauth for collaboration on a companion paper. K.V acknowledges funding from
Leverhulme Trust International Professorship Grant No. LIP-202-014. We acknowledge support from a UKRI Frontier Research Consolidator Grant
 (under the Horizon Europe Guarantee, Grant No. EP/Z002419/1, S.A.P. and D.C.). 
J.~H. acknowledges financial support by the Deutsche Forschungsgemeinschaft (DFG, German Research Foundation) through the W\"urzburg-Dresden Cluster of Excellence ctd.qmat --- Complexity, Topology and Dynamics in Quantum Matter (EXC 2147, project-id 390858490)
and via the project A07 of the Collaborative Research Center SFB 1143 (Project No.~247310070). S.A.P. and K.V. acknowledge the hospitality and support of the Max Planck Institute for Complex Systems (via a Gutzwiller Award to S.A.P.) a  Spring/Summer 2024, during which they held early discussions with J.S.H. about the applicability of DQMC in the moir\'e setting. 
\renewcommand{\addcontentsline}[3]{}

\let\addcontentsline\oldaddcontentsline

\renewcommand{\thetable}{S\arabic{table}}
\renewcommand{\thefigure}{S\arabic{figure}}
\renewcommand{\theequation}{S\arabic{section}.\arabic{equation}}
\onecolumngrid
\pagebreak
\thispagestyle{empty}
\newpage
\begin{center}
	\textbf{\large Supplementary Information for ''\texorpdfstring{Hidden antiferromagnetism, persistent valley fluctuations, and $U(6)$ crossovers in triangular-lattice M-point moir\'e materials via determinantal quantum Monte Carlo}{Hidden antiferromagnetism, persistent valley fluctuations, and U(6) crossovers in triangular-lattice M-point moir\'e materials via determinantal quantum Monte Carlo}{}``}\\[.2cm]
\end{center}

\appendix
\renewcommand{\thesection}{\Roman{section}}
\tableofcontents
\let\oldaddcontentsline\addcontentsline
\newpage
\section*{Outline }
Here, we provide an outline of what is included in the Supplementary Information. In Section~\ref{section:supp:model}, we review the model studied in this work. The starting point here is the moir\'e-scale lattice model for the low-energy Wannier-orbitals of  twisted AA-stacked $\text{SnSe}_2$. A derivation of the lattice model starting from the continuum model on the moir\'e scale is not presented here, as it is covered in depth in the companion paper~\cite{Calugaru_2026}, as well as in Refs.~\cite{Calugaru_2025,Li_2025}. 

In Section~\ref{section:supp:dqmc methods} we provide an in-depth review of the Determinantal Quantum Monte Carlo (DQMC) method utilized in this work. Specifically, we describe in detail the interaction decomposition into squares of fermion bilinear operators, and we prove the absence of the sign-problem at half-filling. We also give details on the observables we measured.

Proceeding, in Section~\ref{section:supp:benchmarks} we benchmark and crosscheck the DQMC results in many ways: Using small-scale Exact Diagonalization (ED), comparing with Stochastic Series Expansion (SSE) results of ~\cite{Calugaru_2026}, and with internal DQMC checks. We also demonstrate the quality of the collected DQMC data presented in the rest of this work, by analyzing the sampling distributions.

In Section~\ref{section:supp:additional data}, we provide a wealth of additional data for different observables and parameter regimes, which support and strengthen the conclusions in the main text.

As is typical, for sufficiently weak or sufficiently strong interactions, the model is  amenable to perturbative treatments. This is why the main text focused on the intermediate-coupling regime where in the absence of DQMC one would be severly limited in their approach to understand the system. With that said, in Section~\ref{section:supp:weak_coupling} we study the model by perturbing away from the non-interacting case, using Random Phase Approximation (RPA), and in Section~\ref{section:supp:perturbation theory}, we study the effective spin model of the system, perturbing away from the atomic problem. These perturbative treatments help connect the weak-, intermediate- and strong-coupling pictures of the model.

Lastly, in Sections~\ref{section:supp:parton mean field} and~\ref{section:supp:ObD} we provide some complementary analytical treatments of the model: In Section~\ref{section:supp:parton mean field}, we solve the model using a parton approach, treating the resulting system at the Mean Field level. This simple approach qualitatively reproduces certain behaviors seen via the exact DQMC numerics, namely how the extend of the metallic regime depends very sensitively on the interaction anisotropy. It also allows for a sharpening of the interpretation of the low-energy degrees of freedom in the metallic regime as being valley fluctuations. Lastly, in Section~\ref{section:supp:ObD} we flesh out the physics of the last part of the main text relating to the ultimate fate of the magnetic order in the strong-coupling regime allowing for some ---but not all--- additional perturbations ignored in the more idealized model considered in the main text.
\section{Model}
\label{section:supp:model}
In this appendix we provide a short review of the lattice model studied in this work, using notation consistent with that of Refs.~\cite{Calugaru_2025,Li_2025,Calugaru_2026}. We also describe the symmetries of the model and provide the reader with a table of parameters for realistic samples of $\text{AA t-SnSe}_2$.
\subsection{Model and Notation}
Starting from the monolayer description of the material, upon twisting each layer $\ell=\pm$ by an angle $\ell \theta/2$, leading to a relative twist $\theta$, the M-points of their respective Brillouin zones are separated by $k_\theta= |\mathbf{K}_M^{\ell=-}-\mathbf{K}_M^{\ell=+}|=2|\mathbf{K}_M|\sin{(\theta/2)}$ which procides the scale of the reciprocal moire lattice. Upon constructing the continuum model and solving the single-particle Hamiltonian for the conduction bands of AA-t$\text{SnSe}_2$, we can then construct real-space Wannier orbitals for the first conduction band per valley. In second quantization these are associated with the operators $\hat{d}^\dagger_{\mathbf{R},\eta,s}$ with $s\in\{\uparrow,\downarrow\}$ being the spin, $\eta\in\{0,1,2\}$ being the valley, and $\mathbf{R}$ labeling the real-space moire lattice unit cell, generated by
\begin{equation}
    \mathbf{a}_{M_1} = a_M\left(\frac{\sqrt{3}}{2},-\frac{1}{2}\right), \hspace{2cm} \mathbf{a}_{M_2} = a_M\left(0,1\right)\,.
\end{equation}
The moir\'e lattice scale is $a_M=2\pi/k_\theta$, and is taken to be $1$ in this work. Wannier orbitals of valley $\eta$ correspond to low-energy electron excitations about different$\text{M}_\eta$ points of the microscopic model.

The starting point for this work, is the (moir\'e) lattice Hamiltonian
\begin{equation}
\begin{split}
\hat{H} =& \hat{H}_t +\hat{H}_U \\
    =& \sum_{\mathbf{R},\Delta \mathbf{R}} \sum_{\eta,s}  t^{\eta}_{\Delta \mathbf{R}}\hat{d}^\dagger_{\mathbf{R},\eta,s} \hat{d}^{\phantom{\dagger}}_{\mathbf{R}+\Delta\mathbf{R},\eta,s}-\mu \sum_{\vec{R}}\hat{n}_{\vec{R}}\\
    &+\sum_{\mathbf{R}} \Big[ \frac{U}{2} \Big(
    \sum_{\eta}\left(\hat{n}_{\mathbf{R},\eta}-1\right)^2
    + 2\alpha\sum_{\eta > \eta'} \left(\hat{n}_{\mathbf{R},\eta}-1\right)\left(\hat{n}_{\mathbf{R},\eta'}-1\right) \Big)
    + \frac{V}{2} \sum_{|\Delta\mathbf{R}|=1}
    \left(\hat{n}_\mathbf{R}-3\right)\left(\hat{n}_{\mathbf{R}+\Delta\mathbf{R}}-3\right)
\Big]\\
\end{split}
    \label{eq:supp:Hamiltonian}
\end{equation}
with $\hat{n}_{\vec{R},\eta}=\sum_{s}\hat{d}^\dagger_{\vec{R},\eta,s}\hat{d}_{\vec{R},\eta,s}$ , $\hat{n}_{\vec{R}}=\sum_\eta \hat{n}_{\vec{R},\eta}$. The hopping amplitudes considered are the following:
\begin{equation}
    t^{\eta}_{\pm C^{\eta}_{3z}\mathbf{a}_{M_2}}=t, \qquad t^{\eta}_{\pm C^{\eta}_{3z}\left(2\mathbf{a}_{M_1}+\mathbf{a}_{M_2}\right)} = t_\perp, \qquad t^{\eta}_{\pm C^{\eta}_{3z}\mathbf{a}_{M_1}}=t^{\eta}_{\pm C^{\eta}_{3z}\left(\mathbf{a}_{M_1}+\mathbf{a}_{M_2}\right)}=t', \qquad t^{\eta}_{\pm C^{\eta}_{3z}2\mathbf{a}_{M_2}}=t''\,. 
    \label{eq:supp:hoppings}
\end{equation}
Crucially, due to the emergent \textit{non-symmorphic} $\tilde{M}_z$ symmetry~\cite{Calugaru_2025}, the inter-chain nearest-neighbor hoppings associated with $t'$ are taken to be zero. These are the $t'$ processes shown in Fig.~\ref{fig:main:lattice diagram}. Strictly speaking, $\tilde{M}_z$ is a symmetry only at zero twist angle, but as seen in Table~\ref{supp:model:tab:parameters_master} it remains as an approximate symmetry up to quite large twist angles.
Further, $t''$ is subleading due to the separation between the Wannier orbitals. While we briefly comment on the inclusion of $t',t''$ in Section~\ref{section:supp:ObD}, we otherwise consider the model with
\begin{equation}
    t' = t'' = 0\,.
\end{equation}
It should also be noted that $\hat{H}_U$ in Eq.~\ref{eq:supp:Hamiltonian} differs from Eq.~\ref{Eq:Hamiltonian:Int} in the main text by a chemical potential term, making the particle-hole symmetry of  $\hat{H}_U$ explicit.

For the model to be sign-problem-free within DQMC, the system must possess an anti-unitary particle-hole symmetry which fixes the chemical potential to $\mu = 0$. There are a number of additional processes of the Wannier model for the low-energy bands of the M-point materials, which are ignored in most of the analysis because (1) they are subleading, as they are suppressed by the moir\'e scale or the symmetries of the model; and (2) because they would introduce a sign-problem to the system. We argue in the main text that we expect many of the generic perturbations of the model away from its idealized sign-problem-free form to not qualitatively alter the intermediate-coupling picture of the model we develop here. Nonetheless, we briefly comment on  the amplitude and sign of the valley-Hund's coupling here, and also on the $\tilde{M}_z$-breaking $t'$ hopping in Sec.~\ref{section:supp:ObD}.
\subsection{Symmetries}
\noindent
Besides the \textit{non-symmorphic} $\tilde{M}_z$ symmetry~\cite{Calugaru_2025}, the system also possesses the following symmetries:
\subsubsection{\texorpdfstring{$C_{3z}$ and $C_{2x}$ symmetries}{C3x and C2x symmetries}}
\noindent
These space-group symmetries acts on the creation operators as
\begin{equation}
\begin{split}
    C_{3z}: & d^\dagger_{\vec{R},\eta,s} \rightarrow d^\dagger_{C_{3z}\vec{R},\eta+1,s}\\
    C_{2x}: & d^\dagger_{\mathbf{R},\eta,s} \rightarrow d^\dagger_{C_{2x}\mathbf{R},-\eta,s} \,,\\
 \end{split}
\end{equation}
with the identification $\eta \equiv \eta \text{ mod }3$.
\subsubsection{Particle-Hole symmetry}
\noindent
For a fermion operator at $\vec{R} = n\mathbf{a}_{M_1}+m\mathbf{a}_{M_2}$, under the action of
\begin{equation}
\hat{\mathcal{P}}:\hspace{0.5em}\hat{d}^\dagger_{C^\eta_{3z}\left(n\mathbf{a}_{M_1}+m\mathbf{a}_{M_2}\right),\eta,s}\rightarrow(-1)^{m}\hat{d}_{C^\eta_{3z}\left(n\mathbf{a}_{M_1}+m\mathbf{a}_{M_2}\right),\eta,s}
\end{equation}
and the simultaneous flip of the chemical potential $\mu\rightarrow-\mu$, the Hamiltonian of Eq.~\ref{eq:supp:Hamiltonian} is invariant. Hence, $\mathcal{P}$ takes the system from filling $\nu$ to filling $6-\nu$, and for half-filling, $\mathcal{P}$ is a symmetry. The sign factor $(-1)^m$ is there to ensure the kinetic term transforms correctly, and it crucially requires a bipartite hopping structure (see SI Sec.~\ref{section:supp:dqmc methods}).
\subsubsection{Time-Reversal Symmetry}
\noindent
The system is also invariant under the action of spin-$1/2$, anti-unitary time-reversal symmetry (TRS):
\begin{equation}
    \begin{split}
        \mathcal{T}=\mathbb{I}_{\text{valley}} \otimes \left(i\sigma^y\right)_{\text{spin}} \otimes \mathcal{K}:&\hspace{0.5em}\hat{d}^\dagger_{\vec{R},\eta,s} \rightarrow (i\sigma^y)_{ss'}\hat{d}^\dagger_{\vec{R},\eta,s'}\\
        :&\hspace{0.5em}i \rightarrow -i
    \end{split}
\end{equation}
The spin-summed density operators $\hat{n}_{\vec{R},\eta}$ are trivially invariant under $\mathcal{T}$ and the only requirement on the hopping is that $t_{\Delta\vec{R}} \in \mathbb{R}$.
\subsubsection{\texorpdfstring{$U(2)\otimes U(2)\otimes U(2)$ symmetry}{U(2) × U(2) × U(2) symmetry}}
\noindent
Beyond TRS, the system also possesses spin-$SU(2)$ rotation symmetry, and since the spin of each valley-$\eta$ electron operator can be rotated independently, there is an enlarged $SU(2)\otimes SU(2)\otimes SU(2)$ symmetry. Further, since valley-number is a good quantum number, there is a $U(1)$-charge conservation per valley, ultimately leading to the $U(2)\otimes U(2)\otimes U(2)$ symmetry.
Further, strictly in the case of one-dimensional hopping, i.e. when $t_\perp=0$, the system can be understood as $N_{\text{valley}} \times N_\text{spin}\times L=6L$ chains, each with a charge-$U(1)$ symmetry, coupled to each-other via density-density interactions. In this case, the system has a hugely enlarged internal symmetry of $U(1)^{\otimes 6L}$. This quasi-1D limit is not utilized in this work (besides the Exact Diagonalization discussion in SI Sec.~\ref{section:supp:benchmarks}) but it plays a crucial role in the complementary paper~\cite{Calugaru_2026}.
\subsection{Valley-Hund's Coupling}
Here we briefly comment on the most important valley-Hund's term generically present in the system, namely the \textit{local} term 
\begin{equation}
    \delta \hat{H} = J_{H} \sum_{\vec{R}}\sum_{\eta >\eta'}\hat{\mathbf{S}}^{(\eta)}_{\vec{R}}\cdot \hat{\mathbf{S}}^{(\eta')}_{\vec{R}} \,.
    \label{eq:supp:hunds}
\end{equation}
Such a term breaks the aforementioned valley-spin rotational $U(2)\otimes U(2)\otimes U(2)$ symmetry of the model by coupling the spin between the valleys, and hence in general it re-introduces frustration to the system as the triangular nature of the lattice becomes manifest. No matter the sign of $J_H$, such a term would frustrate the strong-coupling ground state found in the main text, which is two decoupled N\'eel AFM orders per valley. On the other hand, it is reasonable to assume that the intermediate-coupling picture remains largely intact upon the inclusion of such a term, as in that regime the valley-fluctuations also work against the formation of the N\'eel AFM order.
There are two physical processes contributing to the term in Eq.~\ref{eq:supp:hunds}. (1) A ferromagnetic $-|J_H|_{iv}$ contribution originating from the \textit{inter-valley} components of the microscopi Coulomb repulsion, which although suppressed by the moir\'e scale compared to $U$, is roughly comparable to $J\sim t^2/U$; and (2) and an antiferromagnetic $|J_H|_{ph}$ contribution originating from electron-phonon processes. As shown in Ref.~\cite{FAN26}, the M-point phonons are strongly coupled to electrons of the monolayer $\text{SnSe}_2$ system, which leads to them mediating an effective \textit{attractive} electron-electron interaction on the moir\'e lattice model.

Detailed expressions for $|J_H|_{iv},|J_H|_{ph}$ are provided in~\cite{Calugaru_2026}. While $|J_H|_{iv}$ can be estimated from existing DFT data, $|J_H|_{ph}$ would require knowledge of the coupling strength between monolayer electrons and M-point phonons. Without the coupling strength, we can only conclude that the overall scale of $J_H$ is expected to be small and of unclear sign, depending on which effect dominates. We have thus decided not to consider it here, leaving such an analysis for future work.

\subsection{Symmetry analysis of valley-fluctuations}
The valley-selective hopping processes, as well as the anisotropic local Hubbard interactions of the Hamiltonian (Eq.~\ref{eq:supp:Hamiltonian}) are the two $U(6)$ breaking effects which lead to the physical symmetry of the model being  $U(2)^{\otimes 3}$ (ignoring Hund's) as discussed in this Section. Qualitatively, however, it does seem that the system inherits some of the features of a $U(6)$ symmetric system, as discussed in the Main Text and in Sec.~\ref{section:supp:additional data}. Being concrete, for large interactions $U\gg t$ and small interaction anisotropy $\alpha \rightarrow 1^-$, the system has enhanced valley fluctuations leading to suppression of long-range magnetic order. In this subsection, we therefore investigate what is the appropriate observable to study regarding valley-fluctuations, and more generally, what are all the symmetry-inequivalent observables of the system.
\noindent
To answer this question, let us start by considering a system which indeed has the full $U(6)$ symmetry, using a combined index $a=(\eta,s)=1,\dots,6$ for the valley and spin. The electron operators $\hat{d}$ live in a vector space $\mathcal{V}$ and transform under the fundamental representation $\mathbf{6}$:
\begin{equation}
\hat{d}_a \rightarrow U_{ab}\hat{d}_b \hspace{0.5em} \text{with}\hspace{0.5em} U=e^{i\theta X},U \in U(6)\text{ and } X \in \mathfrak{u}(6) \,.
\end{equation}
The charge-neutral fermion bilinears $\hat{d}^\dagger_a \hat{d}_b$ live in $\mathcal{V}^* \otimes \mathcal{V}$ and transform under $\mathbf{6}^* \otimes \mathbf{6}$. Now, the vector space $\mathcal{V}^* \otimes \mathcal{V}$ can be decomposed into an identity element and a traceless subspace:
\begin{equation}
\hat{d}^\dagger_a \hat{d}_b = \frac{1}{6}\delta_{ab}\hat{d}^\dagger_c \hat{d}_c + \left(\hat{d}^\dagger_a \hat{d}_b - \frac{1}{6}\delta_{ab}\hat{d}^\dagger_c \hat{d}_c\right)
\end{equation}
where the identity element $\hat{N} = \hat{d}^\dagger_a \hat{d}_a$ is the total charge operator associated with the $U(1)$ part of the $U(6)$ symmetry. The $\hat{M}_{ab}=\hat{d}^\dagger_a \hat{d}_b - \frac{1}{6}\delta_{ab}\hat{d}^\dagger_c \hat{d}_c$ elements transform in the \textit{adjoint} representation of $SU(6)$. Crucially, this is spanned by the $6^2-1=35$ generators of the Lie algebra $\mathfrak{su}(6)$, labeled $\{T^I\}_{I=1}^{I=35}$ and hence any element of the adjoint can be expressed as a linear combination of the generators
\begin{equation}
\hat{M}_{ab} = \sum_{I} \hat{O}_I T^I \,.
\end{equation}
The coefficients of a generic element of the adjoint in the generator basis are exactly the fermion bilinears $\hat{O}^{I}\equiv \hat{d}^\dagger_a T^I_{ab} \hat{d}_b$. This is why these fermion bilinear operators naturally appear, at least from a mathematical point of view. Reinstating the momentum and imaginary time indices, we have:
\begin{equation}
\hat{O}^{I}(\mathbf{q},\tau) = \sum_{\mathbf{k}} \hat{d}^\dagger_{\mathbf{k}+\mathbf{q},\tau}T^{I}\hat{d}_{\mathbf{k},\tau}
\label{eq:supp:implementation:valley fluctuations generic}
\end{equation} 
 While the $\hat{O}^{I}$ bilinears have no free $a,b$ indices, they still transform under the adjoint representation since under the action of the symmetry:
\begin{equation}
\hat{O}^I(\mathbf{q},\tau) \rightarrow \sum_{\mathbf{k}} \hat{d}^\dagger_{\mathbf{k+q},\tau} U^\dagger T^I U  \hat{d}_{\mathbf{k},\tau}= \sum_{\mathbf{k}} \hat{d}^\dagger_{\mathbf{k+q},\tau} D^{IJ}_{\text{adj}} T^J  \hat{d}_{\mathbf{k},\tau} = D^{IJ}_{\text{adj}}(U)\hat{O}^J(\mathbf{q},\tau)\,.
\end{equation}
Now, extending the basis to include the total charge term: $I\in \{0,1,...,35\}$  the transformation matrix $D$ takes the following block diagonal form
\begin{equation}
D = \begin{pmatrix}
1 & 0 \\
0 & D_{\text{adj}}\,.
\label{eq:supp:implementation:D block diagonal}
\end{pmatrix}
\end{equation}
With this, the most general set of susceptibilities of the system for bilinear operators, can be written as
\begin{equation}
\chi^{IJ}(\mathbf{q}) = \int_{0}^{\beta}d\tau \hspace{0.5em} \langle \hat{O}^{I}(\mathbf{q},\tau) \hat{O}^{J}(-\mathbf{q},0) \rangle_{c} 
\end{equation}
which is constrained by the $U(6)$ symmetry through
\begin{equation}
\chi^{IJ}(\mathbf{q}) =  D^{IM}(U)\chi^{MN}(\mathbf{q})D^{JN}(U)\,.
\label{eq:supp:implementaton:susceptibility symmetry restriction}
\end{equation}
The block diagonal form of $D$ (Eq.~\ref{eq:supp:implementation:D block diagonal}) implies $\chi^{0I} = \chi^{I0} = 0$ for $I\neq 0$ and that the elements in the adjoint transform as
\begin{equation}
\chi^{IJ}(\mathbf{q}) =  D^{IM}_{\text{adj}}(U)\chi^{MN}(\mathbf{q})D^{JN}_{\text{adj}}(U) \hspace{0.5em} \text{ for } I,J,M,N>0 \,.
\end{equation}
According to Shur's lemma, since $D_{\text{adj}}$ is an \textit{irrep}  of $SU(6)$, the only matrix $\chi$ that commutes with $D_{\text{adj}}(U)$ for all $U \in SU(6)$ is the identity operator. Hence, there are only two unique susceptibilities in the $U(6)$ system:
\begin{equation}
\begin{split}
\chi^{00}(\mathbf{q}) &= \int_{0}^\beta d\tau \hspace{0.5em} \langle \hat{n}(\mathbf{q},\tau) \hat{n}(-\mathbf{q},0)\rangle_c \\
\chi^{IJ}(\mathbf{q}) &= \delta_{IJ} \times \int_{0}^\beta d\tau \hspace{0.5em} \langle \hat{O}^{I}(\mathbf{q},\tau) \hat{O}^{I}(-\mathbf{q},0)\rangle_c \hspace{0.5em}= \delta_{IJ} \chi^{\text{spin}}(\mathbf{q}) \text{ for } I,J>0 \,. \\
\end{split}
\label{eq:supp:implementation: susceptibilities in U(6) case}
\end{equation}
All possible $36^2$ susceptibilities collapse to two unique ones.: a charge and a flavor susceptibility.\par
Now,  returning to the model for twisted $\text{AA t-SnSe}_2$, it does not posses $U(6)$ symmetry, but rather $U(2)^{\otimes3}$ symmetry. Still, we would like to organize the susceptibilities keeping in mind the enlarged $U(6)$ symmetry explicitly broken by the aforementioned physical processes. The electron operators $\hat{d}_{\eta,s} = \hat{d}_{a}$ live in a vector space $\mathcal{V} =\mathcal{V}_1 \oplus \mathcal{V}_2 \oplus \mathcal{V}_3 $ with $\text{dim}(\mathcal{V}_i)=2$ and transform, not in $\mathbf{6}$ but rather in
\begin{equation}
\mathbf{6}\rightarrow (\mathbf{2},\mathbf{1},\mathbf{1}) \oplus (\mathbf{1},\mathbf{2},\mathbf{1})\oplus(\mathbf{1},\mathbf{1},\mathbf{2}) \,.
\end{equation}
This just formalises the fact that  electrons of valley-$\eta$ transform trivially under valley-$\eta' \left(\neq \eta\right)$ rotations. The fermion bilinears $\hat{d}^\dagger \hat{d}$ live in $\mathcal{V}^* \otimes \mathcal{V}$. This space can be decomposed as $\bigoplus_{\eta,\eta'=1}^3 \mathcal{V}^*_{\eta} \otimes \mathcal{V}_{\eta'}$ where each $\mathcal{V}^*_{\eta} \otimes \mathcal{V}_{\eta'}$ space contains bilinears annihilating an $\eta'$ electron and creating an $\eta$ electron. The ‘diagonal’ vector spaces $\mathcal{V}^*_{\eta} \otimes \mathcal{V}_{\eta}$ can be decomposed into a component that transforms under the identity (these are the valley charges) and components that transform under the adjoint of individual-valley $SU(2)$s. The six four-dimensional ‘off-diagonal’ vector spaces transform under some generic irrep. \par
Keeping track: Out of the $36$ bilinears $\hat{d}^\dagger T^I \hat{d}$, there are three charge bilinears transforming under the trivial $(1,1,1)$ representation, there are three sets of three bilinears transforming under the $(\textbf{\text{adj}}_2,1,1)$, $(1,\textbf{\text{adj}}_2,1)$ and $(1,1,\textbf{\text{adj}}_2)$ representations and there are also $6\times 4$ operators transforming under some generic irreps.\par
Starting with the valley-charge bilinears $\hat{n}_{\eta}(\mathbf{q},\tau)$, since they transform trivially under $U(2)\times U(2)\times U(2)$ their susceptibilities
 \begin{equation}
 \chi^{\eta \eta'}_{\text{nn}}(\mathbf{q}) = \int_{0}^{\beta}d\tau \hspace{0.5em} \langle \hat{n}_\eta(\mathbf{q},\tau) \hat{n}_{\eta'}(-\mathbf{q},0) \rangle_{c} 
 \end{equation}
are all symmetry-allowed and -- in the absence of any other symmetries -- are all inequivalent, leading to $9$ independent susceptibilities. However, the $C_{2x}$ and $C_{3z}$ space-group symmetries can further restrict the unique susceptibilities down to only the diagonal and the off-diagonal ones, $\chi^{\text{diag}}_{\text{nn}}(\mathbf{q}),\chi^{\text{off}}_{\text{nn}}(\mathbf{q})$.\par
Moving on to the spin correlators of the bilinears in  $(\textbf{\text{adj}}_2,1,1)$ and its permutations, the first thing to note is that correlators involving different valleys, such as $\langle \hat{\mathbf{S}}^{(\eta)}(\mathbf{q},\tau)\cdot \hat{\mathbf{S}}^{(\eta'\neq \eta)}(-\mathbf{q},0)\rangle$ vanish. Working with the same-valley spin-correlators and susceptibilities
\begin{equation}
 \chi^{\eta}_{\text{spin}}(\mathbf{q}) = \int_{0}^{\beta}d\tau \hspace{0.5em} \langle \hat{\mathbf{S}}^{(\eta)}(\mathbf{q},\tau) \cdot \hat{\mathbf{S}}^{(\eta)}(-\mathbf{q},0) \rangle_{c} 
 \end{equation}
it becomes obvious that, as above, the space-group symmetries reduce the naively three inequivalent susceptibilities to a single unique spin susceptibility $\chi_{\text{spin}}(\mathbf{q})$.\par
Finally, let us consider bilinears that mix valleys. Only combinations of bilinears that respect the conservation of valley-$U(1)$ numbers are symmetry-allowed. These correlators involve bilinears living in conjugate spaces $\mathcal{V}^*_{\eta} \otimes \mathcal{V}_{\eta'}$ and $\mathcal{V}^*_{\eta'} \otimes \mathcal{V}_{\eta}$.
There are $4^2 \times 6$ such correlators, which can be generically expressed as
\begin{equation}
C^{\eta \eta'}_{1234} = \langle O^{\eta\eta'}_{s_1 s_2} O^{\eta'\eta}_{s_3 s_4}\rangle_c  = \langle \hat{d}^\dagger_{\eta,s_1} \hat{d}_{\eta',s_2} \hat{d}^\dagger_{\eta',s_3} \hat{d}_{\eta,s_4} \rangle \,.
\end{equation}
Applying a spin-rotation only on valley $\eta$ electrons $\hat{d}_{\eta,s}\rightarrow U_{ss'}\hat{d}_{\eta,s'}$ constraints this correlator, as it implies
\begin{equation}
C^{\eta \eta'}_{1234} = U^\dagger_{11'}U_{44'}C^{\eta \eta'}_{1'234'} 
\end{equation}
for all $U \in SU(2)$. Hence, by Schur:  $C^{\eta \eta'}_{1234} =  \delta_{14} C^{\eta \eta'}_{1234}$. Repeating the argument for the $\eta'$ valley, we end up with a single unique correlator per pair $(\eta,\eta')$:
\begin{equation}
C^{\eta \eta'}_{1234} = \langle \hat{d}^\dagger_{\eta,s_1} \hat{d}_{\eta',s_2} \hat{d}^\dagger_{\eta',s_3} \hat{d}_{\eta,s_4} \rangle = \delta_{s_1,s_4}\delta_{s_2,s_3} \langle \hat{d}^\dagger_{\eta,s} \hat{d}_{\eta',s'} \hat{d}^\dagger_{\eta',s'} \hat{d}_{\eta,s} \rangle = \delta_{s_1,s_4}\delta_{s_2,s_3} C^{\eta \eta'}
\end{equation}
for $s_1,s_2 \in \{\uparrow,\downarrow\}$ and $\eta \neq \eta'$. Space-group symmetries reduce these six \textit{valley-coherent} correlators down to a single unique one. 
Thus, for the symmetries of the $\text{AA t-SnSe}_2$ model, there are four unique correlators of bilinears, and associated susceptibilities:
 \begin{equation}
 \begin{split}
 \chi^{\text{diag}}_{\text{nn}}(\mathbf{q}) &= \int_{0}^{\beta}d\tau \hspace{0.5em} \langle \hat{n}_\eta(\mathbf{q},\tau) \hat{n}_{\eta}(-\mathbf{q},0) \rangle_{c} \\
 \chi^{\text{off}}_{\text{nn}}(\mathbf{q}) &= \int_{0}^{\beta}d\tau \hspace{0.5em} \langle \hat{n}_\eta(\mathbf{q},\tau) \hat{n}_{\eta'\neq \eta}(-\mathbf{q},0) \rangle_{c} \\
 \chi_{\text{spin}}(\mathbf{q}) &= \int_{0}^{\beta}d\tau \hspace{0.5em} \langle \hat{\mathbf{S}}^{(\eta)}(\mathbf{q},\tau) \cdot \hat{\mathbf{S}}^{(\eta)}(-\mathbf{q},0) \rangle_{c} \\
 \chi_{\text{valley coh.}}(\mathbf{q}) &= \int_{0}^{\beta}d\tau \hspace{0.5em} \langle \hat{O}^{\eta\eta'}_{s s'} (\mathbf{q},\tau)\hat{O}^{\eta'\eta}_{s' s}(-\mathbf{q},0) \rangle_{c} \hspace{0.5em} \text{ for any } \eta \neq \eta' \hspace{0.5em} \text{ and any }s,s'\\
 \end{split}
 \label{eq:supp:implementation:symmetry inequivalent susceptibilities}
 \end{equation}
 This analysis, as mentioned previously, informs the choice of what observable to consider when discussing valley-fluctuations.
\subsection{Parameters}
In the main text, our approach to studying the  $\text{AA t-SnSe}_2$ model which parameters ($t_\perp/t,U/t,\alpha,V/t$), was to concentrate on the insulating versus itinerant behavior, which is qualitatively captured through $\alpha$ and the effective interaction strength $\tilde{U}/W = \left(U - V\right)/\left(2(t+t_\perp)\right)$ and thus study the `reduced' parameter space. To make the connection of our results with the material system clear, we provide a brief overview of realistic parameters for the projected AA-$\text{SnSe}_2$ Wannier Hamiltonian, modified from~\cite{Calugaru_2026}. There are many experimentally tunable -or not well known- parameters, such as the twist angle $\theta$, the interaction screening length $\xi$ and the dielectric constant $\epsilon$\footnote{The interaction is taken \textbf{}to be a Coulomb interaction in a dual-gated geometry: $V(\mathbf{q}) = \frac{e^2}{4\epsilon \epsilon_0}\frac{\tanh{(q\xi/2)}}{q/2}$ with $\epsilon_0$ the vacuum permittivity and $e$ the electron charge.}, and so the parameters presented in Table ~\ref{supp:model:tab:parameters_master} should be thought of as a representative sample of realistic parameters, chosen such that they lie in or near the intermediate-coupling regime.
\begin{table}[h]
	\centering
		\begin{tabular}{|l|c|c|c|c|c|c|c|l|l|l|l|l|}
			\hline
			No. & $\theta$ & $t/\si{\milli\electronvolt}$ & $t_{\perp}/t$ & $t'/t$ & $t''/t$ & $\epsilon$ & $\xi/\si{\nano\meter}$ & $U/t$ & $\alpha$ & $V/t$ & $|J_H|_{iv}/t$ & $\tilde{U}/W$ \\
			\hline
			$1$ & \multirow{3}{*}{\SI{9.43}{\degree}} & \multirow{3}{*}{$19.983$} & \multirow{3}{*}{$0.469$} & \multirow{3}{*}{$-0.086$} & \multirow{3}{*}{$0.233$} & \multirow{3}{*}{$12$} & $2.5$ & $4.083$ & $0.989$ & $0.803$ & $0.218$ & 1.116 \\
\cline{1-1}\cline{8-13}
$2$ &  &  &  &  &  &  & $5$ & $5.234$ & $0.990$ & $1.483$ & $0.218$ & 1.277\\
\cline{1-1}\cline{8-13}
$3$ &  &  &  &  &  &  & $10$ & $5.965$ & $0.991$ & $2.088$ & $0.218$ & 1.320 \\
\hline
$4$ & \multirow{3}{*}{\SI{7.34}{\degree}} & \multirow{3}{*}{$11.388$} & \multirow{3}{*}{$0.254$} & \multirow{3}{*}{$-0.083$} & \multirow{3}{*}{$0.202$} & \multirow{3}{*}{$12$} & $2.5$ & $6.324$ & $0.978$ & $0.727$ & $0.341$ & 2.232\\
\cline{1-1}\cline{8-13}
$5$ &  &  &  &  &  &  & $5$ & $8.277$ & $0.981$ & $1.612$ & $0.341$ & 2.657 \\
\cline{1-1}\cline{8-13}
$6$ &  &  &  &  &  &  & $10$ & $9.545$ & $0.983$ & $2.551$  & $0.341$ & 2.789\\
\hline
$7$ & \multirow{3}{*}{\SI{6.01}{\degree}} & \multirow{3}{*}{$5.871$} & \multirow{3}{*}{$0.112$} & \multirow{3}{*}{$-0.007$} & \multirow{3}{*}{$0.164$} & \multirow{3}{*}{$24$} & $2.5$ & $5.686$ & $0.974$ & $0.354$  & $0.312$& 2.397 \\
\cline{1-1}\cline{8-13}
$8$ &  &  &  &  &  &  & $5$ & $7.549$ & $0.978$ & $0.975$ & $0.312$ & 2.956\\
\cline{1-1}\cline{8-13}
$9$ &  &  &  &  &  &  & $10$ & $8.777$ & $0.98$ & $1.773$ & $0.312$ & 3.149\\
\hline
$10$ & \multirow{3}{*}{\SI{5.09}{\degree}} & \multirow{3}{*}{$4.152$} & \multirow{3}{*}{$0.057$} & \multirow{3}{*}{$-0.047$} & \multirow{3}{*}{$0.131$} & \multirow{3}{*}{$24$} & $2.5$ & $7.282$ & $0.969$ & $0.257$ & $0.387$ & 3.323\\
\cline{1-1}\cline{8-13}
$11$ &  &  &  &  &  &  & $5$ & $9.833$ & $0.974$ & $0.879$ & $0.387$ & 4.236\\
\cline{1-1}\cline{8-13}
$12$ &  &  &  &  &  &  & $10$ & $11.55$ & $0.977$ & $1.847$ & $0.387$& 4.590\\
\hline
$13$ & \multirow{3}{*}{\SI{4.41}{\degree}} & \multirow{3}{*}{$2.214$} & \multirow{3}{*}{$0.024$} & \multirow{3}{*}{$-0.132$} & \multirow{3}{*}{$0.088$} & \multirow{3}{*}{$48$} & $2.5$ & $6.470$ & $0.963$ & $0.114$ & $0.337$ & 3.104\\
\cline{1-1}\cline{8-13}
$14$ &  &  &  &  &  &  & $5$ & $8.827$ & $0.969$ & $0.517$ & $0.337$ & 4.058\\
\cline{1-1}\cline{8-13}
$15$ &  &  &  &  &  &  & $10$ & $10.43$ & $0.972$ & $1.286$ & $0.337$ & 4.465\\
\hline
$16$ & \multirow{3}{*}{\SI{3.89}{\degree}} & \multirow{3}{*}{$1.375$} & \multirow{3}{*}{$0.008$} & \multirow{3}{*}{$-0.191$} & \multirow{3}{*}{$0.063$} & \multirow{3}{*}{$72$} & $2.5$ & $6.323$ & $0.956$ & $0.061$ & $0.314$ & 3.106\\
\cline{1-1}\cline{8-13}
$17$ &  &  &  &  &  &  & $5$ & $8.765$ & $0.963$ & $0.357$ & $0.314$& 4.171\\
\cline{1-1}\cline{8-13}
$18$ &  &  &  &  &  &  & $10$ & $10.46$ & $0.969$ & $1.049$ & $0.314$& 4.668 \\
            \hline
		\end{tabular}\caption{Parameters of the projected $\text{AA t-SnSe}_2$ Wannier Hamiltonian studied in this work, modified from~\cite{Calugaru_2026}. For each twist angle $\theta$, we list the leading kinetic hopping amplitudes, as well as the interaction parameters as defined in the main text and above for different dielectric constants $\epsilon$ and screening lengths $\xi$. The different hopping processes are defined in Eq.~\ref{eq:supp:hoppings}}
	\label{supp:model:tab:parameters_master}
\end{table}
 \section{DQMC Implementation}
\label{section:supp:dqmc methods}
In this section we provide further information in the implementation of the DQMC algorithm for the model of $\text{AA t-SnSe}_2$ presented in the main text. We start off with a quick review of the basics of DQMC. Then, we describe how to re-write the interactions in the model in a way that is amenable to DQMC. There are multiple different ways to do this, each valid in different parameter regimes. We proceed by proving that the model is sign-problem-free in a large region of the parameter space, and finally, we define the observables we measure.
\subsection{Hubbard-Stratonovich decomposition and MC sampling of Hamiltonian}
In DQMC ---and more generally Auxiliary-Field methods--- the quartic fermion terms are expressed as fermion bilinears coupled to discrete auxiliary fields via the Hubbard-Stratonovich (HS) transformation. The Monte Carlo configuration space being sampled is precisely the configuration of auxiliary fields at each site (or bond, triangle etc) and imaginary time slice $\tau$. For a given auxiliary field configuration, the action is that of free fermions in some background potential and hence Wick's Theorem holds and all correlation functions can be computed exactly. This is one of the main advantage of Auxiliary-Field methods. \par
Consider the following expression of an interaction Hamiltonian as a sum of squares of fermion bilinears:

\begin{equation}
	\hat{H}_U = \sum_{\kappa} g_{\kappa}\hat{V}^2_{\kappa} \,.
	\label{eq:supp:interaction as sum of squares}
\end{equation}
After performing a Trotter decomposition of the partition function,
\begin{equation}
\begin{split}
\mathcal{Z} &= \text{Tr}\left(e^{-\beta \hat{H}}\right)\\
     		&=  \text{Tr}\left[\left(e^{-\Delta \tau \hat{T}}\prod_{\kappa} e^{-\Delta \tau g_\kappa\hat{V}					^2_{\kappa} }\right)^{L_\text{Trotter}}\right]+\mathcal{O}\left(\Delta \tau ^2\right)\,,\\
\end{split}
\label{eq:supp:Trotter decomposition}
\end{equation}
the interaction terms can be coupled to a background Hubbard-Stratonovich (HS) field, one field per $\tau$-slice and per interaction operator $\kappa$. The HS fields coupled to the fermion bilinears are discrete fields, which means the configuration space being sampled is discrete, offering a significant speedup compared to a continuous field. For operator $\kappa$ and time slice $\tau$, we have~\cite{Goth_2009}:
\begin{equation}
e^{-\Delta \tau g_\kappa\hat{V}^2_{\kappa}} = \frac{1}{4}\sum_{\ell=\pm1,\pm2}\gamma(\ell_{\kappa,\tau}) e^{i\sqrt{\Delta \tau g_\kappa}\eta(\ell_{\kappa,\tau})\hat{V}_{\kappa}}   +\mathcal{O}\left(\left(\Delta\tau g_\kappa\right)^4\right) \,.
\label{eqn:supp:HS}
\end{equation}
The $\gamma$ are positive and the $\eta$ are real:
\begin{equation}
    \gamma(\pm1)=1+\sqrt{2/3}, \qquad \gamma(\pm2)=1-\sqrt{2/3}, \qquad  \eta(\pm1)=\pm\sqrt{2(3-\sqrt{6})}, \qquad \eta(\pm2)=\pm\sqrt{2(3+\sqrt{6})}\,.
\end{equation}
With some algebraic manipulation, Eq.~\ref{eq:supp:Trotter decomposition} becomes:
\begin{equation}
    \begin{split}
        \mathcal{Z} &= \sum_{C=\{\ell_{\kappa,\tau}\}}\left(  \prod_{\tau}^{L_\text{Trotter}}\prod_{\kappa} \gamma(\ell_{\kappa,\tau})\right) \text{Tr}\left[ \prod_{\tau}^{L_\text{Trotter}} \left(e^{-\Delta \tau \hat{T}}\prod_{\kappa} e^{i\sqrt{\Delta \tau g_\kappa}\eta(\ell_{\kappa,\tau})\hat{V}_{\kappa} } \right)    \right] \\
        &=\sum_{C=\{\ell_{\kappa,\tau}\}}\underbrace{\left(  \prod_{\tau}^{L_\text{Trotter}}\prod_{\kappa} 						\gamma(\ell_{\kappa,\tau})\right)}_{\geq 0} \times W({\ell_{\kappa,\tau}})\,,\\
    \end{split}
    \label{eq:supp:Trotter decomposition 2}
\end{equation}
where $C$ is the configuration space of the H-S fields. Hence, proving the absence of the sign problem is equivalent to showing $W({\ell_{\kappa,\tau}})\geq 0$ for any given configuration. Before we do that, we go over two unique ways to decompose the interactions of our model in the form of Eq.~\ref{eq:supp:interaction as sum of squares}.

\subsection{Decompositions of the Interacting Hamiltonian}
The interacting part of the Hamiltonian in Eq.~\ref{eq:supp:Hamiltonian} is taken to be locally anisotropic, with the anisotropy captured by $\alpha$, and with an isotropic nearest-neighbor (N.N.) density-density term. Neither term as currently written is a sum of squares of bilinears. We first tackle the local term, by performing a change of basis in valley-space:
\begin{equation}
    \begin{split}
        \hat{Q}^{(c)} &= \frac{\hat{n}-3} {\sqrt{3}}\\
        \hat{Q}^{(v_1)}&=\frac{\hat{n}_{\eta=0} - \hat{n}_{\eta=1}}{\sqrt{2}} \\
        \hat{Q}^{(v_2)}&=\frac{\hat{n}_{\eta=0}+\hat{n}_{\eta=1}-2\hat{n}_{\eta=2}}{\sqrt{6}} \\
    \end{split}
	\label{eq: Change of basis}
\end{equation}
where the site index $\vec{R}$ was temporarily suppressed for clarity. The $\hat{Q}^{(c)}$ can be though of as a total charge operator whereas $\hat{Q}^{(v_i)}$ are charge-neutral but carry  valley-charge. In this basis, the interactions can be re-written as
\begin{equation}
\hat{H}_V = \frac{U}{2}\sum_{\vec{R}} \left[(1+2\alpha)[\hat{Q}^{(c)}_\vec{R}]^2 + (1-\alpha)\left([\hat{Q}^{(v_1)}_\vec{R}]^2 + [\hat{Q}^{(v_2)}_\vec{R}]^2\right)\right] 
			+ V\sum_{\tikz[baseline=-0.6ex]{\fill (0,0) circle (1.5pt); \draw (0,0) -- (0.4,0); \fill (0.4,0) circle (1.5pt);}}\left(\hat{n}_\vec{R}-3\right) \left(\hat{n}_{\vec{R}+\Delta\vec{R}}-3\right) \,.
\label{eq:supp:Triangle Interaction decomposition initial}
\end{equation}
As is clear by comparing with Eq.~\ref{eq:supp:interaction as sum of squares}, the local term is now in a form amenable to a HS decomposition. Next, we proceed by re-writing the non-local interactions, either as squares of bilinears on the bonds containing pairs of sites $\langle \vec{R}_1 ,\vec{R}_2 \rangle$ or on the triangles containing triples of sites $\langle \vec{R}_1, \vec{R}_2, \vec{R}_3\rangle$. Schematically:
\begin{equation}
    \begin{split}
        V\sum_{\tikz[baseline=-0.6ex]{\fill (0,0) circle (1.5pt); \draw (0,0) -- (0.4,0); \fill (0.4,0) circle (1.5pt);}} \hat{n}_{\vec{R}_1}\hat{n}_{\vec{R}_2}&=A\sum_{\tikz[baseline=-0.6ex]{\fill (0,0) circle (2pt);}} \hat{n}_{\vec{R}}^2+B\sum_{\tikz[baseline=-0.6ex]{\fill (0,0) circle (1.5pt); \draw (0,0) -- (0.4,0); \fill (0.4,0) circle (1.5pt);}} (\hat{n}_{\vec{R}_1}+x \hat{n}_{\vec{R}_2})^2  \\
                V\sum_{\tikz[baseline=-0.6ex]{\fill (0,0) circle (1.5pt); \draw (0,0) -- (0.4,0); \fill (0.4,0) circle (1.5pt);}} \hat{n}_{\vec{R}_1}\hat{n}_{\vec{R}_2}&=A'\sum_{\tikz[baseline=-0.6ex]{\fill (0,0) circle (2pt);}} \hat{n}_{\vec{R}}^2+B'\sum_{\tikz[baseline=-0.6ex]{\fill (-0.15,-0.15) circle (1.5pt); \draw (-0.15,-0.15) -- (0.15,-0.15); \fill (0.15,-0.15) circle (1.5pt); \draw (0.15,-0.15) -- (0,0.15); \fill (0,0.15) circle (1.5pt); \draw (0,0.15) -- (-0.15,-0.15);}} (\hat{n}_{\vec{R}_1}+\hat{n}_{\vec{R}_2}+\hat{n}_{\vec{R}_3})^2  \\
    \end{split}
\end{equation}
for some constants $A('),B('),x$. While the triangle decomposition will turn out to allow for a larger sign-problem-free parameter region, and all the main text results utilize it, it is instructive to proceed with both decompositions, as they lead to unique HS fields and hence, it provides us with an internal way to confirm the validity of our implementation.
\subsubsection{Bond decomposition}
Consider a system with local intra-valley repulsion $U$, anisotropy $\alpha$ and N.N. bond repulsion of strength $V$.  A term 
\begin{equation}
\frac{V}{2}\sum_{\tikz[baseline=-0.6ex]{\fill (0,0) circle (1.5pt); \draw (0,0) -- (0.4,0); \fill (0.4,0) circle (1.5pt);}}(\hat{n}_{\vec{R}_1} + \hat{n}_{\vec{R}_2})^2    
\label{eq:supp:bond interaction temp}
\end{equation}
contributes a N.N. term of amplitude $V$ per bond, as required, but it also contributes a local Hubbard term of amplitude $6V$ and anisotropy parameter $\alpha=1$, written in compact notation as $(6V,1)$. Hence, if we want to simulate a model with local interaction parameters $(U,\alpha)$ we need to supplement Eq.~\ref{eq:supp:bond interaction temp}  with a local term $(G,\alpha_G)$ such that:
\begin{equation*}
	(G,\alpha_G) \hspace{0.5cm} `` +"\hspace{0.5cm} (6V,1) \hspace{0.5cm} `` ="\hspace{0.5cm} (U,\alpha)  \,.
\end{equation*}    
The local interaction strengths add linearly, while the local anisotropies add like centers of mass, hence the required local term to be added is:

\begin{equation*}
\begin{split}
	G &= U - 6V \\
	\alpha_G &= \frac{\alpha U - 6V}{U-6V} \,.\\
\end{split}
\label{eq:supp:Bond Interaction decomposition initial}
\end{equation*}
Hence, from Eq.~\ref{eq:supp:Triangle Interaction decomposition initial} the interactions have been completely re-written as sums of squares:
\begin{equation}
\hat{H}_U = \frac{U'}{2}\sum_\vec{R} \left[(1+2\alpha')[\hat{Q}^{(c)}_\vec{R}]^2 + (1-\alpha')\left([\hat{Q}^{(v_1)}_\vec{R}]^2 + [\hat{Q}^{(v_2)}_\vec{R}]^2\right)\right] 
			+  \frac{g}{2}\sum_{\tikz[baseline=-0.6ex]{\fill (0,0) circle (1.5pt); \draw (0,0) -- (0.4,0); \fill (0.4,0) circle (1.5pt);}}\left(\hat{n}_{\vec{R}_1} + \hat{n}_{\vec{R}_2} \right)^2 
\label{eq:supp:Bond Interaction decomposition final}
\end{equation}
with $U'=U-6V,\alpha'=(\alpha U - 6V)/U', g= V$. Following arguments in Sec.~\ref{section:supp:dqmc methods:sign problem}, the sign-problem free parameter region is when all the coefficients are simultaneously positive, leading to:
\begin{equation}
\begin{split}
    0&\leq\alpha  \leq 1\\
    \frac{V}{U} &\leq \frac{2\alpha+1}{18} \,.\\
\end{split}
    \label{eq:supp:Bond sign order free}
\end{equation}
\subsubsection{Triangle Decomposition}
We take the same approach for the triangle decomposition, motivated by Ref.~\cite{Golor_2015} which considered a model on the honeycomb lattice. For a system with a local Hubbard term $(U,\alpha)$ and N.N. term $V$, the term 
\begin{equation}
\frac{g}{2}\sum_{\tikz[baseline=-0.6ex]{\fill (-0.15,-0.15) circle (1.5pt); \draw (-0.15,-0.15) -- (0.15,-0.15); \fill (0.15,-0.15) circle (1.5pt); \draw (0.15,-0.15) -- (0,0.15); \fill (0,0.15) circle (1.5pt); \draw (0,0.15) -- (-0.15,-0.15);}}(\hat{n}_{\vec{R}_1}+\hat{n}_{\vec{R}_2}+\hat{n}_{\vec{R}_3})^2
\label{eq:supp:tri interaction temp}
\end{equation}
contributes a NN term of strength $\frac{1}{2}\times2 \times 2 g = 2g$ on each bond, and hence $g=V/2$. It does, however, introduce an extra local Hubbard term $(6\times \frac{g}{2}\times 2,1)=(3V,1)$, since each site belongs to six triangles. Hence, to simulate a system with local term $(U,\alpha)$ we need to supplement Eq.~\ref{eq:supp:tri interaction temp} with a local term $(G,\alpha_G)$ such that:
\begin{equation*}
	(G,\alpha_G) \hspace{0.5cm} ``  +"\hspace{0.5cm} (3V,1)\hspace{0.5cm} `` ="\hspace{0.5cm} (U,\alpha) \,
\end{equation*}
which leads to 
\begin{equation}
\begin{split}
	G &= U - 3V \\
	\alpha_G &= \frac{\alpha U - 3V}{U-3V} \,.\\
\end{split}
\label{eq:supp:additional local term}
\end{equation}
Hence, from Eq.~\ref{eq:supp:Triangle Interaction decomposition initial} we have successfully re-written the interactions completely as sums of squares:
\begin{equation}
\hat{H}_U = \frac{U'}{2}\sum_{\vec{R}} \left[(1+2\alpha')[\hat{Q}^{(c)}_\vec{R}]^2 + (1-\alpha')\left([\hat{Q}^{(v_1)}_\vec{R}]^2 + [\hat{Q}^{(v_2)}_\vec{R}]^2\right)\right] 
			+  \frac{g}{2}\sum_{\tikz[baseline=-0.6ex]{\fill (-0.15,-0.15) circle (1.5pt); \draw (-0.15,-0.15) -- (0.15,-0.15); \fill (0.15,-0.15) circle (1.5pt); \draw (0.15,-0.15) -- (0,0.15); \fill (0,0.15) circle (1.5pt); \draw (0,0.15) -- (-0.15,-0.15);}}\left(\hat{n}_{\vec{R}_1}+ \hat{n}_{\vec{R}_2} + \hat{n}_{\vec{R}_3}\right)^2 
\label{eq:supp:Triangle Interaction decomposition final}
\end{equation}
with $U'=U-3V,\alpha'=(\alpha U - 3V)/U' ,g=V/2$.
Following arguments in Sec.~\ref{section:supp:dqmc methods:sign problem}, the sign-problem free parameter region is:
\begin{equation}
\begin{split}
    0&\leq\alpha  \leq 1\\
    \frac{V}{U} &\leq \frac{2\alpha+1}{9} \,.\\
\end{split}
    \label{eq:supp:Triangle sign order free}
\end{equation}
Comparing Eqs.~\ref{eq:supp:Bond sign order free},\ref{eq:supp:Triangle sign order free} it is clear that the triangle decomposition leads to a larger available sign-problem-free parameter regime and will be used exclusively in all the results (besides the benchmarking).
\subsection{Proof of absence of Sign-Problem for the model}
\label{section:supp:dqmc methods:sign problem}
We shall prove the absence of a sign problem in by utilizing the $SU(2)$ spin symmetry of the Hamiltonian\footnote{$U_{s}(1)$ is enough for this step, so one can add a spin-orbit coupling term in the model but no generic Rasha term, for example} to allow the weights in Eq.~\ref{eq:supp:Trotter decomposition 2} to be block-diagonalized for spin $s\in\{\uparrow,\downarrow\}$. We can decompose all operators appearing in the weights as
\begin{equation*}
\begin{split}
    \hat{H}_t&= \sum_s \hat{H}_{t,s} \\
    \hat{Q}^{(x)}_{\vec{R}} &= \sum_s \hat{Q}^{(x)}_{\vec{R},s} \hspace{0.5cm} \text{ for }x=c,v_1,v_2 \\
    \hat{V}_{\langle \vec{R}_1,\vec{R}_2,\vec{R}_3 \rangle} &= \sum_s \hat{V}_{\langle \vec{R}_1,\vec{R}_2,\vec{R}_3 \rangle,s} \\
    &\equiv\sum_s \left(\delta\hat{n}_{\vec{R}_1,s}+\delta\hat{n}_{\vec{R}_2,s}+\delta\hat{n}_{\vec{R}_3,s}\right)\,.\\
\end{split}    
\end{equation*}
Here, $\delta\hat{n}_{\vec{R},\eta,s}\equiv \hat{n}_{\vec{R},\eta,s}-1/2$ is the particle-hole symmetric density. This allows us to factorize the trace, since the different spin operators commute:
\begin{equation}\begin{split}
    W(\{\ell_{\kappa,\tau}\})  &= \prod_{s=\uparrow,\downarrow}W_s(\{\ell_{\kappa,\tau}\})\\
    W_s(\{\ell_{\kappa,\tau}\}) & = \text{Tr}\left[ \prod_{\tau}^{L_\text{Trotter}} \left(e^{-\Delta \tau \hat{T}_s}\prod_{\kappa} e^{i\sqrt{\Delta\tau g_{\kappa}}\eta(\ell_{\kappa,\tau})\hat{V}_{\kappa,s} } \right)    \right]\\
\end{split}
\label{eq:supp:factorized weight}
\end{equation}
where $\kappa$ encompasses both the local and the triangle interactions.\par 
Now, one can show that via a canonical anti-unitary particle-hole transformation $\hat{d}_{\vec{R},\eta,\uparrow}\rightarrow \text{sgn}(\vec{R},\eta)\hat{d}^\dagger_{\vec{R},\eta,\downarrow}$ that  $W^*_\uparrow(\{\ell_{\kappa,\tau}\})=W_\downarrow(\{\ell_{\kappa,\tau}\})$, which then means that $W(\{\ell_{\kappa,\tau}\}) = |W_\uparrow(\{\ell_{\kappa,\tau}\})|^2 \geq 0$. \newline
To see this, we first clarify the properties of the sign function $\text{sgn}(\vec{R},\eta)$: This function provides one covering of the triangular lattice for each valley $\eta$, such that if sites $\vec{R},\vec{R}'$ are connected via a hopping process of valley $\eta$ electron: $\text{sgn}(\vec{R},\eta)\text{sgn}(\vec{R}',\eta)=-1$. This relies crucially on the hopping being bipartite, despite the superficially non-bipartite triangular geometry. Being explicit, for $\vec{R}=C^{\eta}_{3z} \left(n\mathbf{a}_{M_1} + m\mathbf{a}_{M_2}\right)$, the sign function is $\text{sgn}(\vec{R},\eta) = (-1)^{m}$
Thus, applying the canonical transformation:
\begin{equation}
    \begin{split}
        \hat{H}_{t,\uparrow} &= \sum_{\vec{R},\Delta \vec{R}}\hat{d}^\dagger_{\vec{R},\eta,\uparrow}t^{\eta}_{\Delta \vec{R}}\hat{d}_{\vec{R}+\Delta \vec{R},\eta,\uparrow} \\
 & \rightarrow \sum_{\vec{R},\Delta \vec{R}}-\hat{d}_{\vec{R},\eta,\downarrow}\left(t^{\eta}_{\Delta \vec{R}}\right)^*\hat{d}^{\dagger}_{\vec{R}+\Delta \vec{R},\eta,\downarrow} \\
 & = \sum_{\vec{R},\Delta \vec{R}}\hat{d}^\dagger_{\vec{R}+\Delta \vec{R},\eta,\downarrow}\left(t^{\eta}_{\Delta \vec{R}}\right)^*\hat{d}_{\vec{R},\eta,\downarrow} \\
 & = \hat{H}_{t,\downarrow} \,,\\
    \end{split}
\end{equation}
where  $\left(t^{\eta}_{-\Delta \vec{R}}\right)^* = t^{\eta}_{\Delta \vec{R}}$ due to hermiticity. Similarly, for a single triangle interaction bilinear:
\begin{equation}
    \begin{split}
        i\hat{V}_{\langle \vec{R}_1,\vec{R}_2,\vec{R}_3 \rangle,\uparrow} &=i\left[ (\hat{n}_{\vec{R}_1,\uparrow}-3/2)+(\hat{n}_{\vec{R}_2,\uparrow} - 3/2) + \hat{n}_{\vec{R}_3,\uparrow}-3/2)\right] \\
 &\rightarrow -i\left[(3 - \hat{n}_{\vec{R}_1,\downarrow}-3/2)+(3 - \hat{n}_{\vec{R}_2,\downarrow} - 3/2) +(3 - \hat{n}_{\vec{R}_3,\downarrow} - 3/2)\right] \\
 & = i\hat{V}_{\langle \vec{R}_1,\vec{R}_2,\vec{R}_3 \rangle,\downarrow}\\
    \end{split}
\end{equation}
with the local interaction bilinears behaving the same way. Thus, as long as  
\begin{equation}
    \frac{U'}{2}\left(1+2\alpha'\right)\geq0 ,\qquad\frac{U'}{2}\left(1-\alpha'\right)\geq0,\qquad \frac{g}{2}\geq 0\,,
\end{equation}
in Eq.~\ref{eq:supp:Triangle Interaction decomposition final},
it is clear that $W^*_\uparrow(\{\ell_{\kappa,\tau}\})=W_\downarrow(\{\ell_{\kappa,\tau}\})$. this ensures that the weight of each Auxiliary Field configuration is positive: $W(\{\ell_{\kappa,\tau}\}) = |W_{\uparrow}(\{\ell_{\kappa,\tau}\})|^2 $ and leads to the sign-problem-free regime of Eq.~\ref{eq:supp:Triangle sign order free}.

\subsection{Observables}
A major advantage of DQMC is that one can in principle evaluate all observables with relative ease compared to e.g. world-line Monte Carlo methods. While more details can be found in Refs.~\cite{ALF_article,Assaad_2007} the basic idea is that, for an observable $\hat{O}$, the expectation value is calculated via:
\begin{equation*}
    \langle \hat{O}\rangle = \frac{\text{Tr}[e^{-\beta \hat{H}}\hat{O}]}{\text{Tr}[e^{-\beta \hat{H}}]} = \sum_{C}P(C)\langle \langle\hat{O}\rangle \rangle_C
\end{equation*}
where $C$ is a configuration of HS fields, $P(C)$ is the weight of that configuration, and $\langle \langle\hat{O}\rangle \rangle_C$ is the expectation value of the observable for those fixed HS fields. As is typical, the distribution $P(C)$ is not known a priori, and is sampled using, in this case, a simple Metropolis-Hastings algorithm. Since for a given configuration $C$, the electron problem is non-interacting, Wick's theorem holds and $\langle \langle\hat{O}\rangle \rangle_C$ can be evaluated, no matter how many fermion operators it involves. Hence, the core observable in DQMC is the single particle Green's function
\begin{equation}
 \langle \mathcal{T}\hat{d}_{\vec{R}+\Delta \vec{R},\eta,s}(\tau)\hat{d}^\dagger_{\vec{R},\eta,s}(0) \rangle
\end{equation}
where we have assumed $\vec{R}$ and $\tau$ translation symmetry, as well as the $SU(2)^{\otimes 3}$ symmetry which renders only the diagonal $\eta,s$ elements as non-trivial.
Given a generic local operator $\hat{O}_{\vec{R}}$ , the connected 2-point correlator is
\begin{equation}
    \mathcal{S}_{\hat{O}}(\mathbf{q},\tau) = \langle \hat{O}_{\mathbf{q}}(\tau)\hat{O}_{-\mathbf{q}}(0)\rangle_c = \frac{1}{L^2}\sum_{\vec{R} ,\vec{R}'}e^{-i\mathbf{q}\cdot (\vec{R} - \vec{R}')}\left(\langle \hat{O}_{\vec{R}}(\tau)\hat{O}_{\vec{R}'}(0)\rangle - \langle \hat{O}_{\vec{R}}(0)\rangle \langle \hat{O}_{\vec{R}'}(0)\rangle\right)
\end{equation}
for an $L\times L$ lattice. The correlator \textit{susceptibility} is
\begin{equation}
    \chi_{\hat{O}}(\mathbf{q}) = \int_{0}^{\beta}d\tau \mathcal{S}_{\hat{O}}(\mathbf{q},\tau)\,.
\end{equation}
Here, The Fourier transform is defined as
\begin{equation}
\hat{O}_{\mathbf{k}} = \frac{1}{\sqrt{L^2}}\sum_{\vec{R}}e^{-i\mathbf{k}\cdot \vec{R}}\hat{O}_{\vec{R}} \,.
\end{equation}
Throughout this work, we mainly utilize momentum-resolved correlation functions, as well as \textit{local} correlation functions, which are just $\frac{1}{L^2}\sum_{\mathbf{q}}\mathcal{S}_{\mathcal{O}}(\mathbf{q},\tau)$.
\subsubsection{Spin correlators}
The local spin operator of valley-$\eta$ electrons is defined as
\begin{equation*} \hat{\mathbf{S}}^{(\eta)}_{\mathbf{R}}(\tau)=\frac{1}{2}\sum_{ss'}\hat{d}^\dagger_{\mathbf{R},\eta,s}(\tau)\bm{\sigma}_{ss'}\hat{d}_{\mathbf{R},\eta,s'} (\tau)
\end{equation*}
with $\bm{\sigma}$ the Pauli spin-$1/2$ matrices. The fact that the system is symmetric under \textit{independent} rotations in spin-space for valleys $\eta \neq \eta'$ implies that $\langle \hat{\mathbf{S}}^{(\eta)}_{\vec{R}}(\tau) \cdot\hat{\mathbf{S}}^{(\eta')}_{\vec{R}'}(\tau')\rangle =0$ which just reflects the intuitive fact that the spins of electrons belonging to different valleys are uncorrelated. Hence, we can restrict our attention to the momentum-resolved spin correlators, as well as the spin susceptibility, defined as:
\begin{equation}
\begin{split}
    \mathcal{S}^{(\eta)}(\mathbf{q},\tau)&=\langle \hat{\mathbf{S}}^{\eta}_{\mathbf{q}}(\tau)\cdot\hat{\mathbf{S}}^{\eta}_{\mathbf{-q}}(0) \rangle\\
    \chi^{(\eta)}_{\text{spin}}(\mathbf{q}) &=\int_0^{\beta}d\tau  \mathcal{S}^{(\eta)}(\mathbf{q},\tau)\,.\\
\end{split}
\label{eq:supp:implementation:spin}
\end{equation}
As mentioned in Section~\ref{section:supp:model}, the spin correlators for different valleys all encode the same information, due to the space-group symmetries $C_{2x},C_{3z}$. Hence, without loss of generality, we will concentrate on $\eta=0$ valley spin observables. The static $\tau=0$ correlators are useful for the structure factor peaks, but the time-displaced correlators contain additional information about the spin dynamics, such as the presence of Goldstone modes in the system.
\subsubsection{Density correlators}
Another family of observables which are crucial to our work, are fully resolved density operators $\hat{O}_{\vec{R},\eta,s}(\tau) = \hat{n}_{\vec{R},\eta,s}(\tau)$ and we will mainly be interested in their local, static correlations
\begin{equation}
\langle \hat{n}_{\vec{R},\eta,s}\hat{n}_{\vec{R},\eta',s'} \rangle_c  =  \langle \hat{n}_{\mathbf{0},\eta,s}\hat{n}_{\mathbf{0},\eta',s'} \rangle_c = \langle \hat{n}_{\mathbf{0},\eta,s}\hat{n}_{\mathbf{0},\eta',s'} \rangle - \langle \hat{n}_{\mathbf{0},\eta,s}\rangle^2 \,.
\end{equation}
Through these operators, one can access the \textit{local moment}, \textit{double-occupancy} and the \textit{inter-valley correlations}, defined in Eqs.~\ref{eq:supp:additonal:ZZ},\ref{eq:supp:additonal:C&D},\ref{eq:supp:additonal:C_connected}, which are crucial in understanding the nature of local fluctuations.
\subsubsection{Green's functions}
While the Green's functions were already defined above, here we use the more compact notation, directly in momentum-space:
\begin{equation}
\mathcal{G}^{(\eta)}(\mathbf{k},\tau) = \langle \mathcal{T}\hat{d}_{\mathbf{k},\eta,s}(\tau)\hat{d}^\dagger_{\mathbf{k},\eta,s}(0) \rangle \,.
\label{eq:supp:implementation:green}
\end{equation} 
The Green's function provides information about the single-particle excitation spectrum of the system and so is very important for distinguishing metallic and insulating behaviors.
\subsubsection{Valley fluctuations}
Another quantity which proves crucial in understanding the physics of $\text{AA t-SnSe}_2$ is a measure of how electrons of different valleys are correlated to each other. Following the discussion regarding the symmetry-inequivalent correlators of the model in Sec.~\ref{section:supp:model}, we introduce the \textit{valley-coherence} operator
\begin{equation}
    \hat{O}^{\left(\eta,\eta'\right)}_{(s,s')}(\mathbf{k},\tau) = \hat{d}^\dagger_{\mathbf{k},\eta,s}(\tau)\hat{d}_{\mathbf{k},\eta',s'}(\tau) \hspace{1em}\text{ for } s,s'\in \{ \uparrow,\downarrow \} \hspace{0.5em}\text{ and } \eta \neq \eta'
\end{equation}
where the choice of spin indices is arbitrary due to the individual spin-rotation symmetries in each valley. The associated 2-pt correlator is 
\begin{equation}
    \mathcal{F}(\mathbf{k},\tau) = \frac{1}{6}\sum_{\eta\neq\eta'}\langle \hat{d}^\dagger_{\mathbf{k},\eta,s}(\tau)\hat{d}_{\mathbf{k},\eta',s'}(\tau)\hat{d}^\dagger_{-\mathbf{k},\eta',s'}\hat{d}_{-\mathbf{k},\eta,s}\rangle \,.
    \label{eq:supp:implementation:valley flucts}
\end{equation}
This correlator preserves the valley-$U(1)$ symmetries and captures the degree of coherence of inter-valley fluctuations. As in the case of the spin correlators, space-group symmetries lead to all six different valley-coherence channels $(\eta,\eta')$ being equivalent up to a rotation in momentum. 
\subsubsection{Analytic continuation of imaginary-time operators}
The Monte Carlo algorithm is much more naturally developed in imaginary time $\tau$, with the data then able to be transformed to imaginary-time Matsubara frequencies $i\omega_n$. Experimentally accessible quantities, however, like the response functions, are associated with \textit{real} frequencies $\omega$. Analytic continuation is the process of bringing the observables from the imaginary to the real axis. This process is ill-defined: small changes in the input (imaginary $\tau$) data due to \textit{e.g.} MC noise, lead to large changes in the output (real $\omega$) data. The general approach to mitigate this, via the so called Maximum-Entropy methods (see e.g. Refs.~\cite{Bryan_1990,JARRELL_1996,Levy_2017}), is to strike a balance between fitting the data and maximizing `entropy', a quantity that imposes a sense of smoothness to the output and penalizes overfitting. In this work, we analytically continue the Green's function and the valley-fluctuations using the pre-build MaxEnt code in ALF~\cite{ALF_article,ALF_codebase}. Below we very briefly summarize the basics of analytic continuation using a generic (fermionic or bosonic) operator $\mathcal{O}(\tau)$ and its analytically continued form $O(\omega)$. The two are related via
\begin{equation}
    \mathcal{O}(\tau) = \int_{-\infty}^{+\infty}d\omega K^{\pm}_{\tau}O(\omega)
    \label{eq:supp:implementation:ana cont defintion}
\end{equation}
with $K^{\pm}_{\tau} = \frac{1}{\pi}\frac{e^{-\tau \omega}}{1\pm e^{-\beta \omega}}$ and $+(-)$corresponding to fermionic (bosonic) operators. The kernel has the property $K^\pm_\tau(\omega) = \pm K^{\pm}_{\beta - \tau}(-\omega)$. Analytic continuation methods such as MaxEnt aim at inverting Eq.~\ref{eq:supp:implementation:ana cont defintion}.
\par
One may improve the quality of the input data (and thus of the analytically continued output data) by taking advantages of the symmetries of the observables. Taking as an example the single-particle Green's function $\mathcal{G}^{(\eta)}(\mathbf{k},\tau)$ and the spectral function $A^{(\eta)}(\mathbf{k},\omega)$: Under the particle-hole operation $\mathcal{P}:\mathcal{G}^{(\eta)}(\mathbf{k},\tau) \rightarrow \mathcal{G}^{(\eta)}(\mathbf{Q}^{(\eta)} - \mathbf{k},\beta-\tau)$ and since $\mathcal{P}$ is a symmetry: $A^{(\eta)}(\mathbf{k},\omega) = A^{(\eta)}(\mathbf{Q}^{(\eta)} - \mathbf{k},-\omega)$, we can use the improved estimator:
\begin{equation}
    \mathcal{G}^{(\eta)}_{\text{symm}}(\mathbf{k},\tau) = \frac{1}{2}\left(\mathcal{G}^{(\eta)}(\mathbf{k},\tau) + \mathcal{G}^{(\eta)}(\mathbf{Q}^{(\eta)} - \mathbf{k}, \beta -\tau) \right)= \int_{-\infty}^{\infty} d\omega K^+_\tau(\omega)A^{(\eta)}(\mathbf{k},\omega)
\end{equation}
which would reduce the noise by essentially doubling the data. Using the improved estimator $\mathcal{G}^{(\eta)}_{\text{symm}}(\mathbf{k},\tau)$ would require a good amount of scaffolding to the MaxEnt code in ALF since it is $\mathbf{k}$-agnostic. The next best thing, which however treats the noise sub-optimally but at least enforces the symmetries of the system, is to symmetrize after the fact:
\begin{equation}
    \tilde{A}^{(\eta)}(\mathbf{k},\omega) = \frac{1}{2}\left(A^{(\eta)}(\mathbf{k},\omega) + A^{(\eta)}(\mathbf{Q}^{(\eta)} -\mathbf{k},-\omega)\right)
\end{equation}
We can also utilize the $C_{3z}$-symmetry to further improve the data:
\begin{equation}
    \tilde{\tilde{A}}^{(\eta)}(\mathbf{k},\omega) = \frac{1}{3}\left(\tilde{A}^{(\eta)}(\mathbf{k},\omega) + \tilde{A}^{(\eta+1)}(C_{3z}\mathbf{k},\omega) + \tilde{A}^{(\eta+2)}(C^2_{3z}\mathbf{k},\omega)\right)
\end{equation}
For the case of the valley-coherence correlator, since the underlying object is bosonic and thus particle-hole symmetric, one may use an improved kernel
\begin{equation}
    \mathcal{F}(\mathbf{k},\tau) = \int_{0}^{\infty} d\omega \left(K^{-}_\tau(\omega)+K^{-}_{\tau}(-\omega)\right)A^v(\mathbf{k},\omega) \,.
\end{equation}
\par
Often, we will be interested in extracting some information about response functions without performing the full analytical continuation with its potential pitfalls. Specifically, there is an accurate way to extract the $\omega=0$ behavior of any function $\mathcal{O}(\omega)$ directly from the $\tau=\beta/2$ value of $\mathcal{O}(\tau)$. This for example, applied to the single-particle Green's function can give us the $\omega=0$ density of states, which is a direct measure of the presence and momentum dependence of low-energy quasiparticles. Let's consider first the case of a \textit{fermionic observable},labeled by the subscript $f$: At $\tau=\beta/2$, Eq.~\ref{eq:supp:implementation:ana cont defintion} can be written as
\begin{equation}
\begin{split}
    \mathcal{O}_f(\tau=\beta/2) &= \frac{1}{\pi}\int_{-\infty}^{\infty} d\omega \mathcal{O}_{f} (\omega)\times \frac{1}{2\cosh{(\beta \omega/2)}}\\
    &\approx \mathcal{O}_{f}(\omega=0) \times \int_{-\infty}^{\infty}d\omega\frac{1}{2\pi \cosh{(\beta\omega/2)}}\\
    &=\frac{1}{\beta} \mathcal{O}_{f}(\omega=0)
\end{split}
\end{equation}
The $\cosh{(\beta \omega/2)}$ acts as a low-frequency filter and so the only assumption here is that $\mathcal{O}_{f}(\omega) \approx \mathcal{O}_{f}(\omega=0)$ in a window of width $\sim 1/\beta$ within $\omega=0$. A similar calculation can be done for bosonic operators (denoted with subscript $b$): 
\begin{equation}
\begin{split}
    \mathcal{O}_b(\tau=\beta/2) &= \frac{1}{\pi}\int_{-\infty}^{\infty}d\omega \frac{\mathcal{O}_{b}(\omega)}{\omega} \times \frac{\omega}{2\sinh{(\beta \omega/2)}}\\
    &\approx \frac{\mathcal{O}_{b}}{\omega}|_{\omega=0} \times \int_{0}^{\infty} d\omega \frac{\omega}{\pi \sinh{(\beta\omega/2)}}\\
    &=\frac{\pi}{\beta^2} \frac{\mathcal{O}_{b}}{\omega}|_{\omega=0}\,.
\end{split}
\end{equation}
Here, we run into the complication that $K^{-}_{\tau}(\omega)$ is singular at $\omega=0$, and that further, the dynamical susceptibility $\mathcal{O}^b(\omega)$ is \textit{odd} in $\omega$ and hence not positive definite, which thus required the redefinitions of the Kernel and the quantity being analytically continued. But the key assumption here is that again, it requires the bosonic susceptibility to have reached it low-$\omega$ behavior (which in this case is $\mathcal{O}^b(\omega) \sim \omega$) by $|\omega|\lesssim 1/\beta $. Discussion of these proxies, as well as their applications can be found in Refs.~\cite{ALF_article,JARRELL_1996,Wang_2020,Trivedi_1996}. For our purposes, re-instating the $\mathbf{q}$ labels, we can gain access to the low-frequency single-particle and valley-coherence excitation spectrum
\begin{equation}
\begin{split}
    A^{(\eta)}(\mathbf{k},\omega=0) &\approx \beta \mathcal{G}^{(\eta)}(\mathbf{k},\tau = \beta/2)\\
    &\\
     A^v(\mathbf{k},\omega=0) &\approx \frac{\beta^2}{\pi} \mathcal{F}(\mathbf{k},\tau = \beta/2)\,.
\end{split}
\end{equation}
with the $\mathcal{G},\mathcal{F}$ observables defined in Eqs.~\ref{eq:supp:implementation:green},\ref{eq:supp:implementation:valley flucts}. \section{Benchmarks}
\label{section:supp:benchmarks}
In this section, we carefully benchmark our results in two different ways:  (1) By comparing to Exact Diagonalization (ED) data for small system sizes and (2) by using different Determinant Quantum Monte Carlo (DQMC) implementations of the model, as explained in SI Sec.~\ref{section:supp:dqmc methods}. Then, we analyze the Monte Carlo data to ensure that (a) the bins used for the error estimation are not auto-correlated (b) that their distribution is sufficiently close to a normal distribution $\mathcal{N}(\mu,\sigma^2)$ and (c) that the symmetries of the Hamiltonian (see SI Sec.\ref{section:supp:model}) ---those that are not `hard-coded' into the DQMC algorithm--- are reflected in the data. All these benchmarks and checks ensure the correctness and statistical significance of the numerical results presented.

Beyond this we can compare with another benchmark: as mentioned in the main text, in the companion paper~\cite{Calugaru_2026}, we simulate the $\text{AA t-SnSe}_2$ system at arbitrary filling $\nu$ for the quasi-one-dimensional case where $t_\perp=t'=t''=0$. This can be done with Stochastic Series Expansion~\cite{Sandvik_1991,Sandvik_1992,Sandvik_2010,XU15} which is a QMC method usually reserved for 1D fermionic systems. For technical reasons elaborated in~\cite{Calugaru_2026}, the SSE and DQMC methods can be compared at $\nu=3$ (half-filling) and $t_\perp=0$ only on a system with \textit{open boundary conditions}. This analysis is performed in the S.I. of~\cite{Calugaru_2026} and it displays excellent agreement between the two methods, providing another non-trivial benchmark for both methods.
\subsection{Benchmarking DQMC with ED}
\begin{figure}
    \centering
    \includegraphics[width=\linewidth]{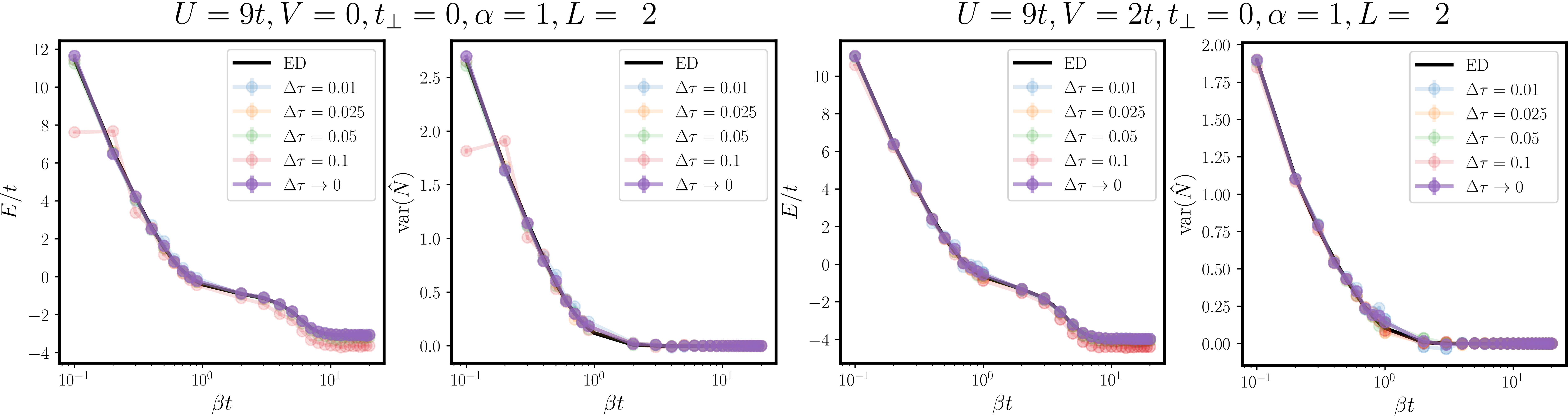}
    \caption{\textbf{ED and DQMC Benchmarking:} Calculating the energy $E/t$ and the charge variance $\text{var}(\hat{N})=\langle \hat{N}^2\rangle - \langle\hat{N} \rangle^2$ for two different sets of parameters. ED is the black line and DQMC are the other lines, with the purple line being the extrapolation to $\Delta \tau \rightarrow 0$. }
    \label{fig:supp:bench:ED comparison}
\end{figure}
Due to the three valleys, the M-point model has a very large local Hilbert space dimension: $|\mathcal{H}_\mathbf{R}|=2^{6}$. This makes even a very small $2\times 2$ system quite difficult to do ED on since it has a dimension $2^{24} \sim 10^7$. However, in the case when $t_\perp=0$, the system is comprised of $N_\text{spin}\times N_\text{valley} \times L^2$ chains, each one with an associated U$(1)$ conserved charge. Utilizing these conservation laws, we perform ED in each symmetry sector c which is characterized by a tuple of $N_\text{spin}\times N_\text{valley} \times L^2$ integers, each taking values from $0$ to $L$. Further, one may utilize time-reversal-symmetry, $C_{3z}$, and $T_{\mathbf{a}_1}$, $T_{\mathbf{a}_2}$ translations to group sectors $c$ into equivalence classes:
\begin{equation}
    \tilde{\mathbf{c}} = \{c| \exists g \in G \text{ such that } g c g^{-1} = \tilde{c}\}
\end{equation}
with $\tilde{c}$ the `representative' sector of an equivalence class $\mathbf{\tilde{c}}$ and $G$ the group generated by the aforementioned symmetries. Sectors $c \sim c'$ lying in the same equivalence class, have identical spectrum, and symmetry-related eigenstates. This dramatically simplifies the calculations: For the $2\times 2$ system, there are a total of $531441$ sectors, grouped into $23115$ classes. Diagonalizing the system in the representative sector of each class provides the full information of the system. This allows for very quick full diagonalization of the $2\times 2$ system, in less than 10 minutes and 2GB of RAM on a single CPU node. However, things become very costly, very quickly. For an $L \times L$ system:
\begin{equation}
    \begin{split}
        \#_{c} &=(L+1)^{6L}\\
        \#_{\tilde{c}} &\gtrsim \#_{\tilde{c}}^{\text{optimal}}=\frac{(L+1)^{6L}}{|G|} = \frac{(L+1)^{6L}}{6L^2}\\
        \text{max}\{|\mathcal{H}^c|\} &= \begin{pmatrix}
            L \\ L/2
        \end{pmatrix}^{6L}\\
    \end{split}
\end{equation} 
with $\#_{c(\tilde{c})} $ the number of (representative) sectors, and $\text{max}\{\mathcal{H}^c\}$ the dimension of the largest Hilbert space among these sectors. Already for a $3\times 3$ system, $ \text{max}\{|\mathcal{H}^c|\}\sim 4 \times 10^8$ meaning sophisticated Lanczos/Sparse Matrix constructions need to be implemented, and a full high-$T$ exact diagonalization becomes prohibitively expensive. As a comparison,  while the $2\times 2$ system is approximately equivalent to a $12$-site 2D Hubbard model the $3 \times 3$ system is equivalent to a $27$-site Hubbard model. So while this approach of utilizing the U$(1)$ symmetries of the model is quite useful, it requires further modifications if one wants to tackle larger systems. This is especially true if one wants to compute non-diagonal (in the occupation basis) observables, as such calculations require a complicated mapping between the exponentially-many symmetry sectors. Even for the $2 \times 2$ system, calculating the Green's function $\mathcal{G}^{\eta}_{\mathbf{R}}(\tau)$  is quite arduous. For our purposes however, it is enough to compare the results for the $2\times 2$ system at $\nu=3$, i.e. half-filling, for some simple static observables. In Fig. \ref{fig:supp:bench:ED comparison} we compare the energy and charge fluctuations  between ED and DQMC for two parameter values. As can be seen, once we take into account the Trotter error due to the finite $\Delta \tau$, the DQMC data agrees perfectly with the ED data. \par

\subsection{Internal DQMC Benchmarking }
\begin{figure}\centering
    \includegraphics[width=0.75\linewidth]{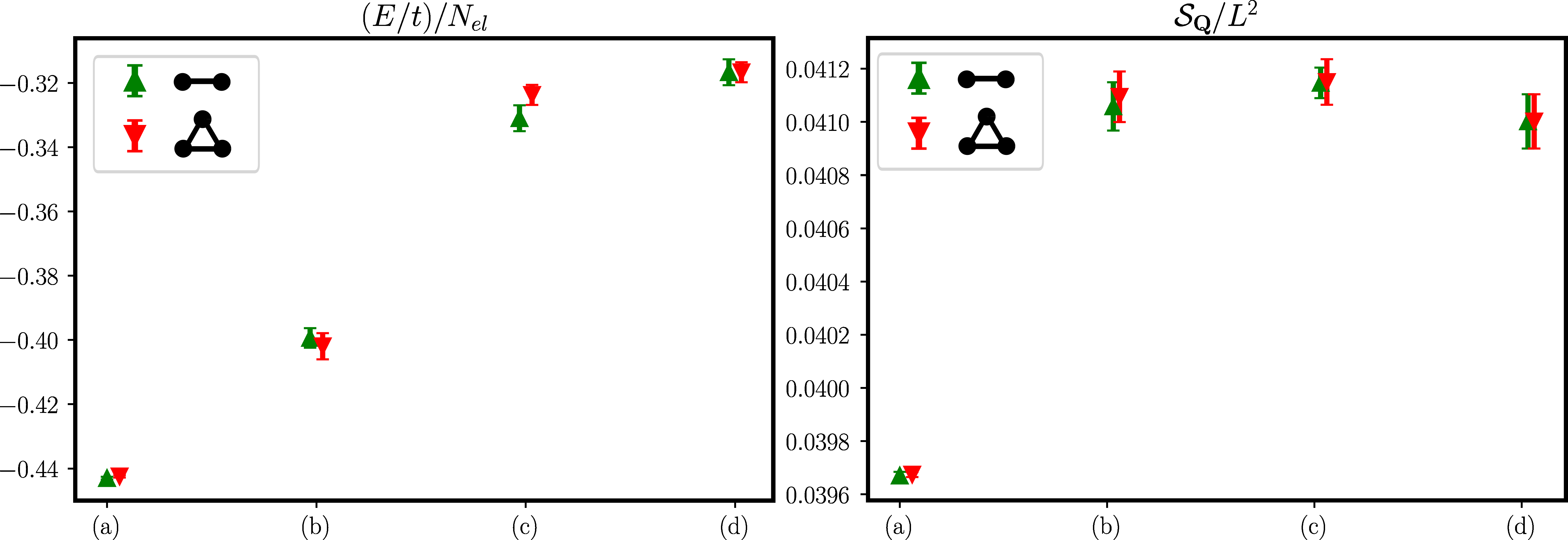}
    \caption{\textbf{Comparison of different DQMC implementations:} In this figure we are showing two different observables: Energy per particle and the magnetic structure factor at its peak, using two different DQMC implementations: The `bond' and the `triangle' implementations, as specified in Sec. \ref{section:supp:dqmc methods}. We compare four sets of parameters at $L=8,U=9t,\beta=5t$ and $(V,\alpha)$ equal to (a): $(1.5t,1)$, (b):$(1.4t,0.9)$,(c):$(0.4t,0)$ and (d):$(0,0)$}
    \label{fig:supp:bench:internal DQMC}
\end{figure}
Another verification of the numerics is to utilize the different DQMC implementations of the model: the `bond' and `triangle' decompositions of the interactions, as described in Sec.~\ref{section:supp:dqmc methods}. Since each implementation utilizes a different set of auxiliary fields and hence a different sample space of configurations, their results being identical (up to statistical noise) is a non-trivial check on our numerical implementation. The results of this check are shown in Fig. \ref{fig:supp:bench:internal DQMC}.

\subsection{Quality of the Monte Carlo data}
To correctly interpret the Monte Carlo data and end up with estimates for the mean and error of the observables $\langle\hat{O}\rangle$, we must ensure the assumptions underlying the data analysis are being met. The two key assumptions are that the bins are independent, and that their number $N_{\text{bins}}$ is large enough for the central limit theorem (CLT)to apply, and the sample mean of the observable approaches $\sim \mathcal{N}(\mu,\sigma^2)$~\cite{Krauth_2006}. 
\subsubsection{Autocorrelation time}
\begin{figure}
    \centering
    \includegraphics[width=0.75\linewidth]{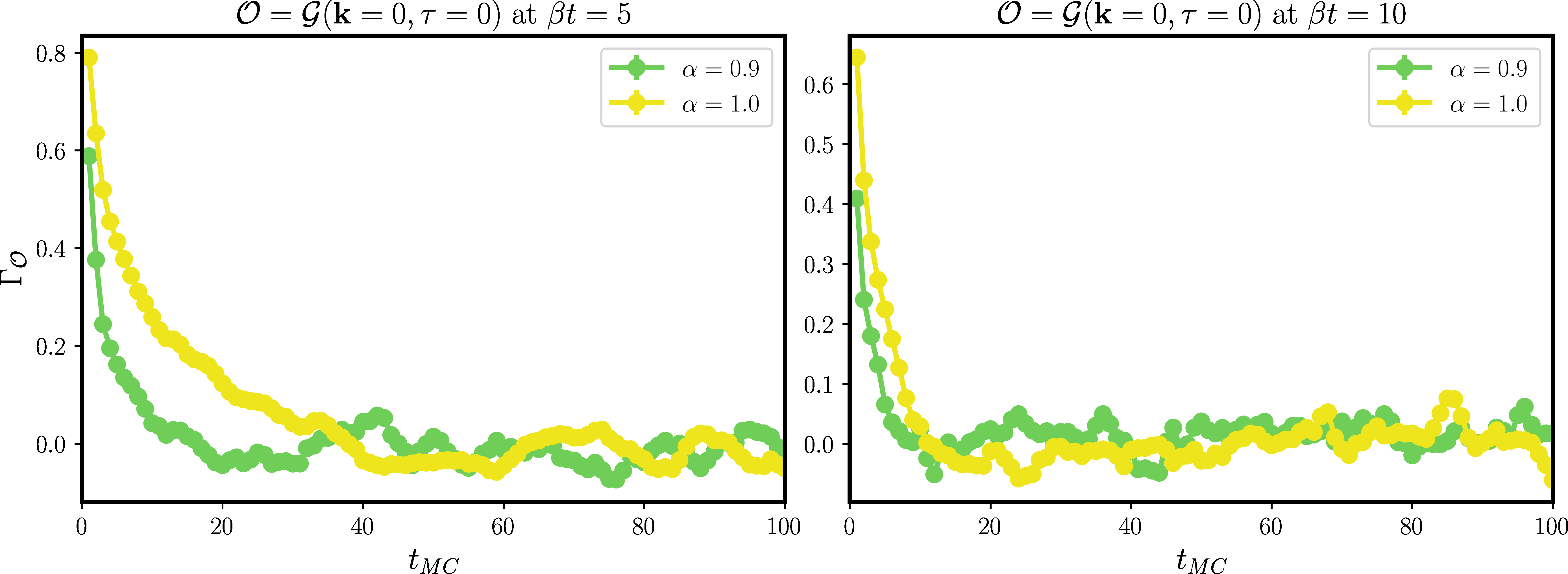}
    \caption{The autocorrelation of the Monte Carlo data for the Green's function observable. Parameters are  $L=12,t_\perp=0.25t,U=9t,V=0t$ and $N_{\text{sweeps}}=1,N_{\text{bins}}=2^{12}$. The data remains correlated in $t_{MC}$ due to $N_{\text{bins}}$ being quite small. The faster decay seen in larger $\beta t$ is because the `time-steps' for each sweep of the algorithm scale with $\beta$ }
    \label{fig:supp:bench:autocorrelation}
\end{figure}

The autocorrelation of the data quantifies how correlated the Monte Carlo bins are. To generate one bin from another along the Markov chain, the ALF algorithm loops through all auxiliary fields $N_\text{sweep}$ times, proposing local spin-flip updates. The number of auxiliary fields scales with $L^2 \times \beta \sim L^2 \times N_\tau$. For sufficiently small $N_\text{sweep}$, the $i$'th  and the $(i+t)$'th bin will  be correlated, which is captured through
\begin{equation}
    \Gamma_{\hat{\mathcal{O}}}(t) = \sum_{t_0=1}^{N_\text{Bins}-t}\frac{(\langle \hat{\mathcal{O}}(t_0+t)\rangle - \langle\hat{\mathcal{O}}\rangle)(\langle \hat{\mathcal{O}}(t)\rangle - \langle\hat{\mathcal{O}}\rangle)}{(\langle \hat{\mathcal{O}}(t_0)\rangle - \langle\hat{\mathcal{O}}\rangle)^2} \,.
\end{equation}
This is expected to decay as $\sim e^{-t/\tau_{\text{Auto}}}$ as $t\rightarrow \infty$.  The autocorrelation time for a particular choice of  observable $\mathcal{O}$ is shown in Fig.~\ref{fig:supp:bench:autocorrelation}. For all the data presented in this work, we choose $N_\text{sweeps}=128 \gg \tau_{\text{Auto}}$ between collected bins, which should be more than enough to guarantee the bins are statistically independent.
\subsubsection{Distribution of MC Observables}
\begin{figure}\centering
    \includegraphics[scale=0.4]{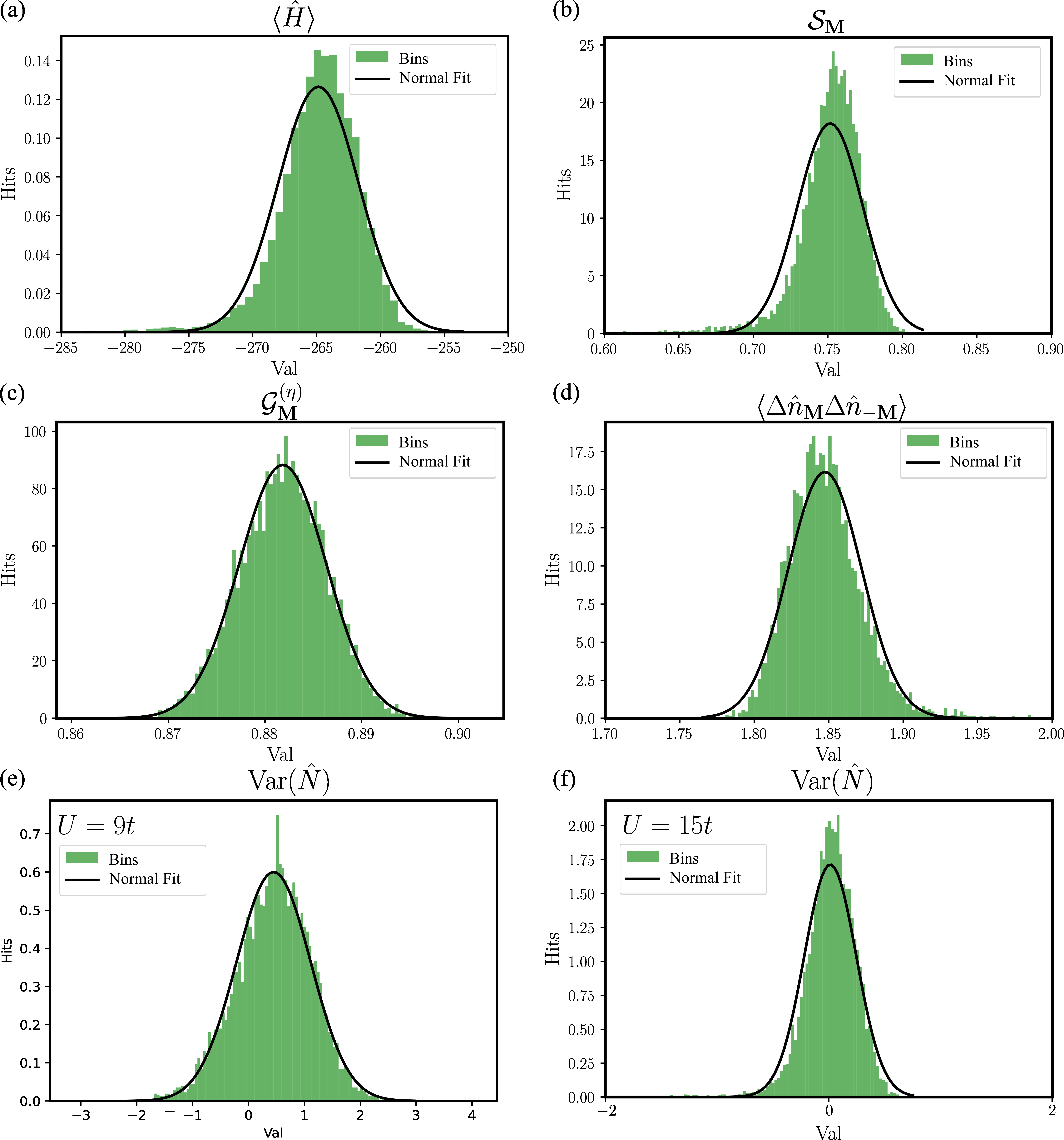}
    \caption{\textbf{Is the collected MC data normally-distributed?} The figures (a)-(e) show different observables (energy, spin structure factor, Green's function, valley-charge imbalance and charge variance) for the parameters $L=12,t_\perp=0.25t,U=9t,V=0,\alpha=1,\beta=20t$ while in (f), $U=15t$. This data is with $\Delta \tau = 0.2t$ and $N_\text{bins}=100\times 2^6$ and $N_{\text{sweeps}}=2^7$. }
    \label{fig:supp:bench:MC data bins}
\end{figure}

The other key assumption, used in deriving the sample mean and error, is that the $N_{\text{bins}}$ is large enough for the CLT to hold. We can check this explicitly by plotting histograms of all the samples generated by the Markov chain Monte Carlo. As it can be seen in Fig.~ \ref{fig:supp:bench:MC data bins}, the observables' bins are distributed quite close to a normal $\mathcal{N}(\mu,\sigma^2)$ distribution, despite some having slightly heavier tails.
\subsubsection{Does the MC data respect the system's symmetries?}
\begin{figure}
    \centering
    \includegraphics[scale=0.4]{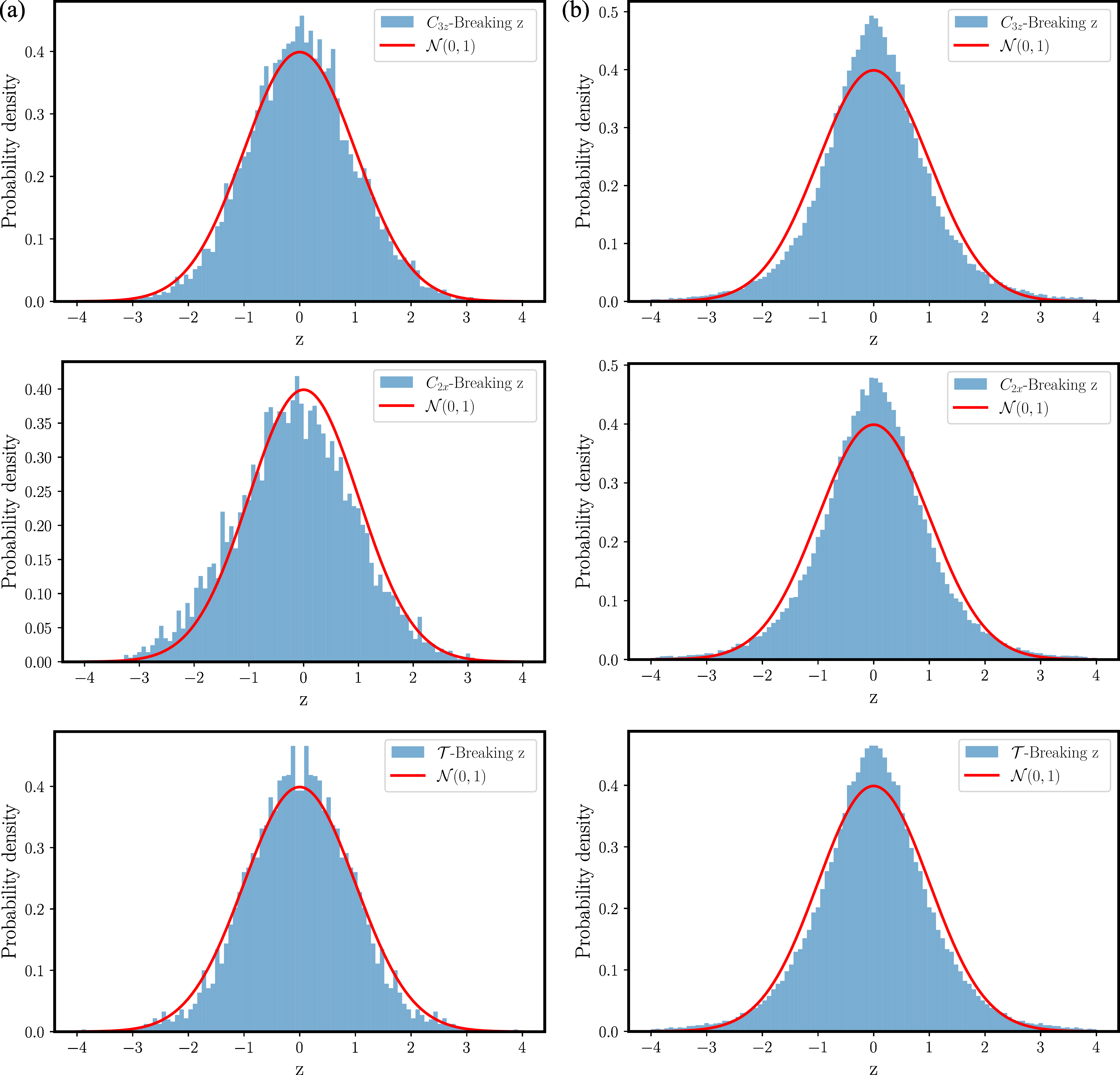}
    \caption{\textbf{Does the MC data respect the system's symmetries?} The left column (a) is for $L=12,t_\perp=0.25t,U=9t,V=0,\alpha=0.9,\beta=10t$ and the right (b) column is for $L=12,t_\perp=0.25t,U=6t,V=0,\alpha=1,\beta=100t$ , the two serving as representatives of all the simulations. We see that indeed, the samples of $z$'s as defined in Eq.\ref{eq:supp:zs} are quite close to normally-distributed, meaning the implicit symmetries of the model are indeed respected by the simulation.}
    \label{fig:supp:bench:symmetry bins}
\end{figure}

Certain symmetries like the $SU(2)$ spin symmetry, and translation symmetry are enforced explicitly in the DQMC algorithm, and are used to improve the error estimates of observables and speed up the algorithm. Other symmetries, like the $C_{3z}$ and $C_{2x}$ symmetries are not enforced explicitly, and so we would like to see if they are reflected in the measured observables ( up to statistical errors). Consider a set of tuples of observables that are equal to each other according to the symmetries of the system: $\{\text{pairs}\} = \{(\mathcal{O}_1,\mathcal{O}_2)\}$. Each tuple may be, for example: $(\mathcal{G}^{(\eta)}(\mathbf{k},\tau),\mathcal{G}^{(\eta)}(\mathbf{Q}^{(\eta)}-\mathbf{k},\beta - \tau))$for each $\eta,\mathbf{k},\tau$. We can calculate a new rescaled observable as follows:
\begin{equation}
    z_\text{pair} = \frac{\mathcal{O}_1 - \mathcal{O}_2}{\sqrt{\sigma_{\mathcal{O}_1}^2 + \sigma_{\mathcal{O}_2}^2}} \hspace{0.5cm}\text{for pair}=(\mathcal{O}_1,O_2) \in \{\text{pairs}\}
    \label{eq:supp:zs}
\end{equation}
If the MC distribution's samples approximate a normal distribution, then the sample distribution should approach $\sim \mathcal{N}(0,1)$. It should be noted that $z$ ignores the covariance between measurements, small autocorrelation effects, \textit{etc} but it is none the less instructive. In Fig.~\ref{fig:supp:bench:symmetry bins} we show the distribution of the samples $z$'s for different symmetries and that indeed, they are respected up to statistical error.
 \section{Additional Monte Carlo Data}
\label{section:supp:additional data}
In this section, we present additional Monte Carlo data to further bolster the statements made in the main text, as well as to make some additional comments on some other aspects of the model. We begin by exploring the magnetic properties of the half-filled system: We demonstrate that, while throughout the parameter regime explored in this work, the magnetic correlations are peaked at the M-points, whether the system actually develops long-range N\'eel Anti-ferromagnetic (AFM) order at  $T=0$ and $L \rightarrow \infty$, is a much subtler question, which is at the heart of the interesting physics present in the system. \par
We also present data on the charge compressibility $\kappa = \frac{\partial \nu}{\partial \mu}$. This serves as a simple observable capturing the ease with which electrons can be added to the system. In certain parameter regimes, we find the system remaining compressible down to temperatures much lower that the localization scale due to the Hubbard term $T^* \sim U$, hinting at the existence of low-energy quasiparticles whose energy scale is \textit{not} associated with a doublon/holon cost as it would be in the more standard Hubbard models.  To investigate these low-energy processes in detail, we turn our attention to local fluctuations, as well as to momentum- and frequency- resolved charge and valley fluctuations. \par
Throughout this Section, we also present results with different values of $V$ (the nearest-neighbor interaction) and $t_\perp$ from those presented in the main text. As argued previously, the physics of the system at generic interaction and hopping strength is qualitatively captured by a dimensionless `interaction strength' parameter $\tilde{U}/W = (U-V)/2(t+t_\perp)$. Here we demonstrate this explicitly, and we also briefly comment also some quantitative effects turning on $V,t_\perp$ brings.
\subsection{Magnetic Correlations}
\begin{figure}[h]
    \centering
    \includegraphics[width=\linewidth]{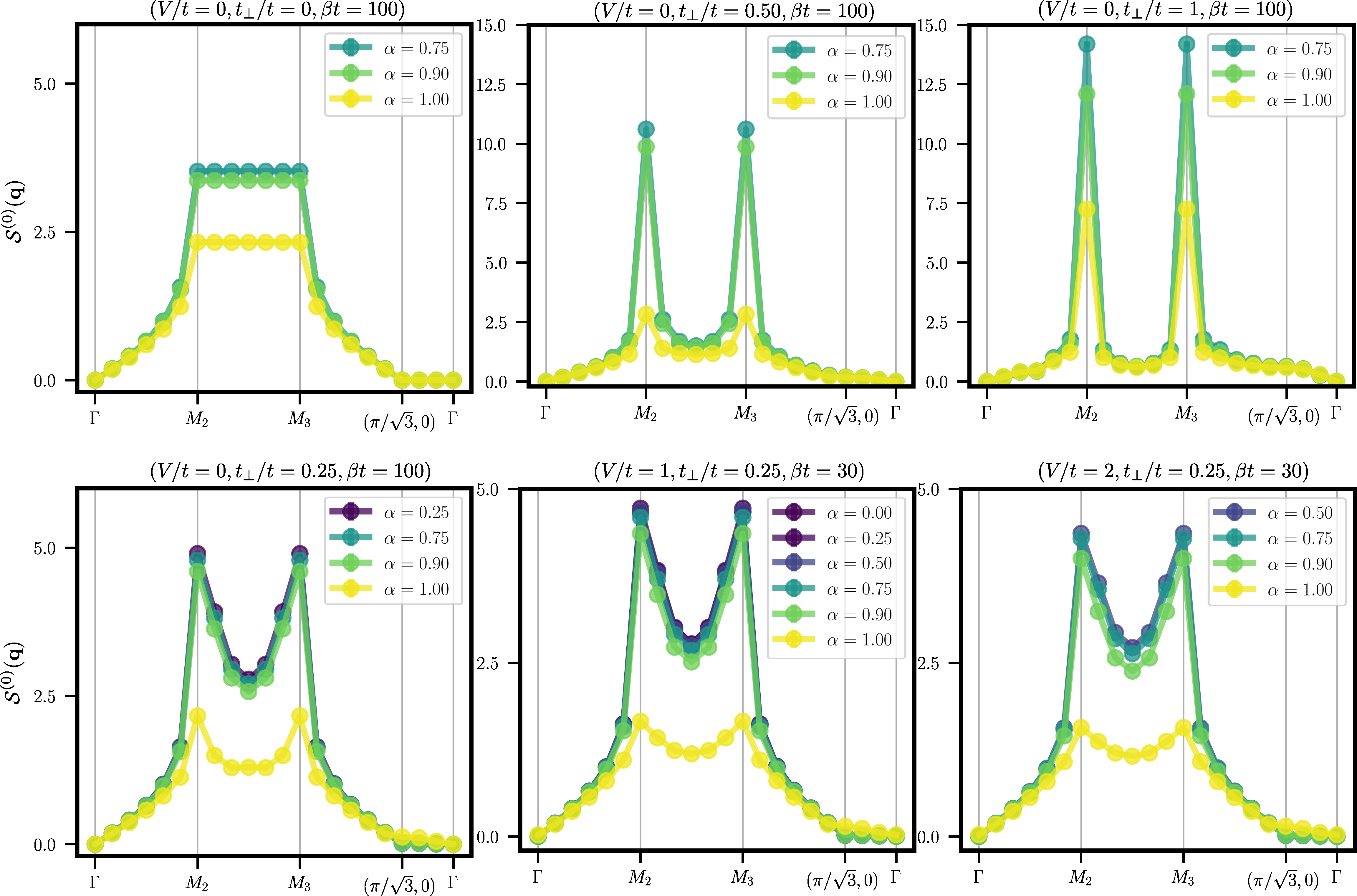}
    \caption{The magnetic structure factor for  $\eta=0$ valley electrons in the Brillouin zone, for six representative parameter sets at $U/t=9$, for a $12 \times 12$ system.}
    \label{fig:supp:magnetic correlations: S(q) in BZ}
\end{figure}
One of the key observables to study in this system, is the static spin correlations $\langle \mathbf{S}^{(\eta)}_{\vec{R}} \cdot \mathbf{S}^{(\eta)}_{\vec{R} + \Delta \vec{R}} \rangle$, for $\mathbf{S}^{(\eta)}_\vec{R}  =\frac{1}{2} \sum_{s,s'}d^\dagger_{\vec{R},\eta,s}\bm{\sigma}_{ss'}c_{\vec{R},\eta,s'}$, as well as its Fourier transform $\mathcal{S}^{(\eta)}(\mathbf{q})$, as defined in Eq.~\ref{eq:supp:implementation:spin}. It is an observable related to many experimental probes. From a theoretical perspective, the magnetic ordering in this system is quite unique: The moir\'e lattice is triangular, which would frustrate a collinear AFM order. However, the \textit{non-symmorphic} $\tilde{M}_z$ symmetry~\cite{Calugaru_2025} reduces the translation symmetry and leads to the electrons of each valley hopping on  frustration-free rectangular sublattices ($t'=0$ in Fig.~\ref{fig:main:lattice diagram}).Further, since electrons of different valleys $\eta$ are only sensitive to other valleys' densities, the spin structure the system develops should be completely valley-decoupled. The precise statement is that the individual spin-rotation symmetries of each valley render $\langle \hat{\mathbf{S}}^{(\eta)}_{\vec{R}}\cdot\hat{\mathbf{S}}^{(\eta')}_{\vec{R}'}\rangle$ spin-diagonal. In a sense, the spin degrees of freedom have no knowledge of the triangular (and hence frustrated) nature of the lattice. In combination with the small but non-zero inter-chain hopping $t_\perp$, the system is in principle allowed to form true long-range AFM order at $T=0$, in accordance with Mermin-Wagner-Hohenberg~\cite{Mermin_66,Hohenberg_67}. This is to be contrasted to the ‘chain' limit of $t_\perp=0$ studied in the companion paper~\cite{Calugaru_2026} in which case the system never develops a magnetic order. Similar cases of weakly-coupled chains have been explored numerically in ~\cite{Raczkowski_2012,Raczkowski_2013,Raczkowski_2015}. \par
The symmetry analysis performed in Sec.~\ref{section:supp:model} shows that there is a single unique spin-spin correlator, with all the others being symmetry-related. For example, $C_{3z}$ implies
\begin{equation}
    \begin{split}
    \langle \mathbf{S}^{(\eta)}_{\vec{R}} \cdot \mathbf{S}^{(\eta)}_{\vec{R} + \Delta \vec{R}} \rangle &=\langle \mathbf{S}^{(\eta)}_{C_{3z}\vec{R}} \cdot \mathbf{S}^{(\eta)}_{C_{3z}\vec{R} + C_{3z}\Delta \vec{R}} \rangle\\
    \mathcal{S}^{(\eta)}(\mathbf{q}) &= \mathcal{S}^{(\eta+1)}(C_{3z}\mathbf{q}) \,.
    \end{split}
    \label{eq:supp:additional data:spin structure}
\end{equation}
Thus, we will only concentrate on the valley-$0$ spin-correlations, with the $\eta=1,2$ ones being identical up to a $C_{3z}$ rotation (and noise).\par
As a final comment: the lack on any \textit{inter-valley} spin-correlations implies that, if the system were to order, the orientation of the spin-order (e.g. the direction in $\mathbb{R}^3$ defining a N\'eel AFM order) would in general be different in each valley, as there is no term locking them together, such as a local Hund's $\hat{\mathbf{S}}^{(\eta)}_{\vec{R}}\cdot \hat{\mathbf{S}}^{(\eta'\neq \eta)}_{\vec{R}}$. This, combined with the fact that even within a given valley the hopping is restricted to two independent sublattices, means that at this level, any magnetic order would have $3 \times 2 = 6$ independent orientations, which only couple once additional terms are included. Upon reinstating smaller terms that are present in the realistic $\text{AA t-SnSe}_2$ system, this behavior breaks down and in general the different spin orientations get coupled to eachother, leading to ‘locking', although in that case, whether the system even orders should be evaluated carefully. This situation is discussed in the main text, as well as in the SI Secs.~\ref{section:supp:model},~\ref{section:supp:ObD}.
\subsubsection{Structure factor in the Brillouin Zone}
\begin{figure}
    \centering
    \includegraphics[width=\linewidth]{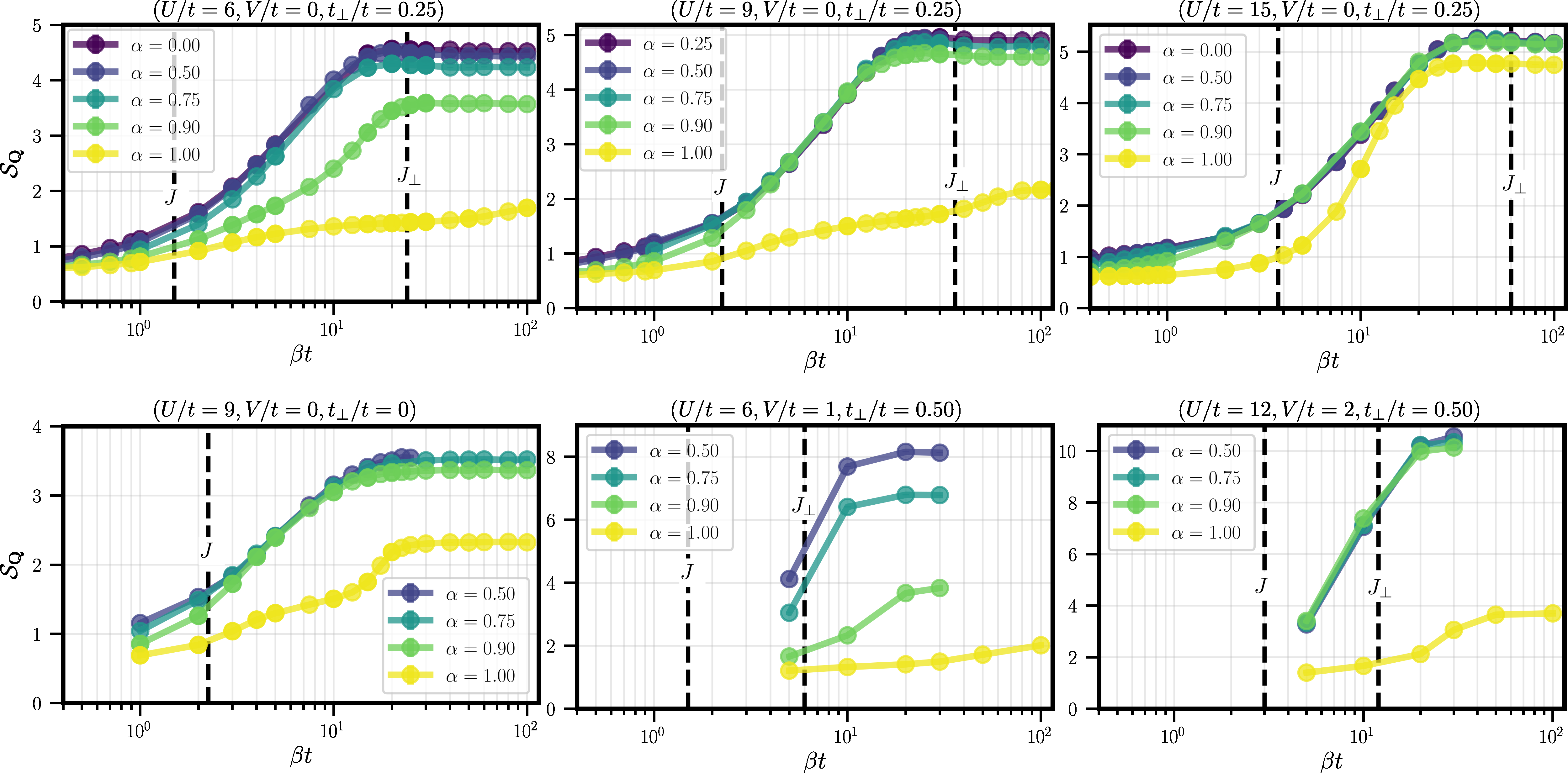}
    \caption{How the magnetic structure factor peak evolves with cooling the system (increasing $\beta$), for a range of parameters. The relevant super-exchange energy scales when $T=J,J_\perp$ are also shown, for $J=4t^2/(U-V)$ and $J_\perp = 4t^2_\perp/(U-V)$ }
    \label{fig:supp: magnetic correlations: S(Q) vs beta}
\end{figure}
We begin by investigating the momentum-space structure for the spin correlators $\mathcal{S}^{(\eta)}(\mathbf{q})$, shown for a range of parameters, in Fig.~\ref{fig:supp:magnetic correlations: S(q) in BZ}. Concentrating on $\eta=0$-valley electrons, we see clear peaks at two of the M points($M_2,M_3$) with no $M_1$ peak. This can be understood by considering the \textit{effective rectangular} Brillouin Zone that electrons of each valley $\eta$ hop on, as shown in Fig.~\ref{fig:main:lattice diagram}. For $\eta=0$ (blue) electrons, the $M_2,M_3$ points coincide and are the BZ-corner points. In analogy to the square-lattice Hubbard model, finding magnetic peaks at the corners is hardly surprising and is associated with N\'eel AFM correlations. We denote these special corner momenta as $\mathbf{Q}^{(\eta)} = C_{3z}^{\eta}\left(\pi/\sqrt{3},\pi\right)$. Besides the peaks, one can also see the one-dimensional magnetic correlations in the ‘chain' limit of$t_\perp=0$ as the $2\text{D}$ Brillouin Zone collapses to a $1\text{D}$ line. Further, comparing plots with same local interaction parameters but different $V$, we notice the general trend of a modest decrease in spin-correlations as $V/U$ increases.

We can also investigate how the value of the spin-correlations at the peaks $\mathbf{q} = \mathbf{Q}^{(\eta)}$ depends on temperature, which shown in Fig.~\ref{fig:supp: magnetic correlations: S(Q) vs beta}. Invoking again the familiar case of the square-lattice half-filled Hubbard model in a Mott-insulating phase, we would expect a sudden increase in correlations below the super-exchange scale $J=4t^2/\left(U-V\right)$ as electrons are localized and their spins can coherently anti-align~\cite{Varney_2009}, eventually saturating to their ground state value. In this system, such behavior is indeed observed, but only when $\tilde{U}/W$ is sufficiently large and $\alpha$ is sufficiently small. In other cases however, the anisotropy $\alpha$ seems to play two roles: (1) as $\alpha \rightarrow 1$, correlations decrease, as seen already in Fig.~\ref{fig:supp:magnetic correlations: S(q) in BZ} and (2) for $\alpha \rightarrow 1$, the correlations are sensitive to a very low-energy scale ($\Delta< J_\perp,J,\dots$), seen as a low-$T$ rise in $\mathcal{S}$. This scale is quite unexpected as it cannot be understood by a strong-coupling spin model picture. As we will soon demonstrate, this scale is also system-size dependent.
\subsubsection{Long Range Antiferromagnetic order}
\begin{figure}[h]
    \centering
    \includegraphics[width=\linewidth]{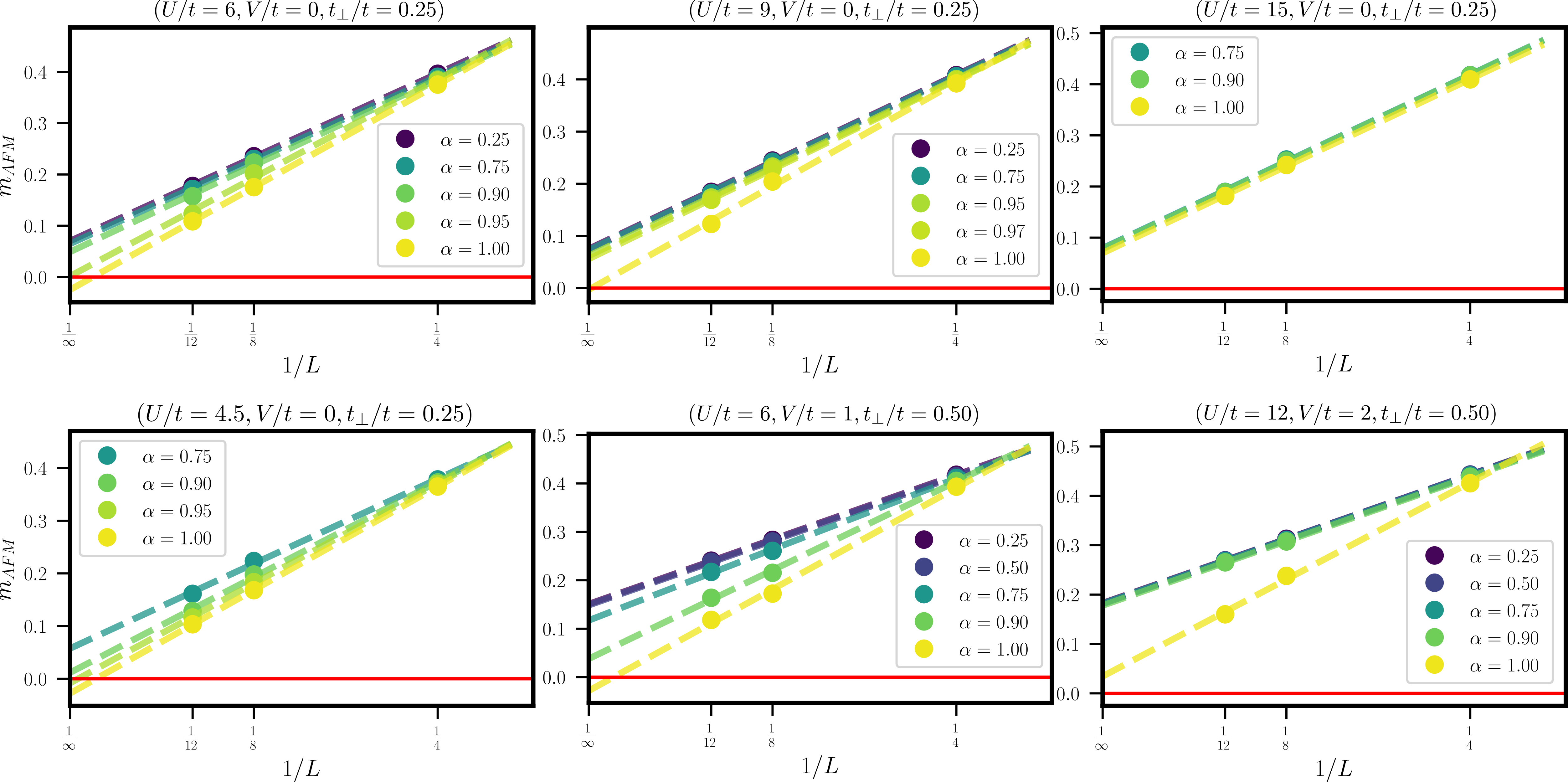}
    \caption{Scaling results for the AFM staggered moment (the square root of the correlations) using the fit $\mathcal{S}_{\vec{Q}}/L^2\equiv m_{\text{AFM}}(L)=  m_{\infty} + a/L$.}
    \label{fig:supp: magnetic correlations: LRO}
\end{figure}
\begin{figure}
    \centering
    \includegraphics[width=0.75\linewidth]{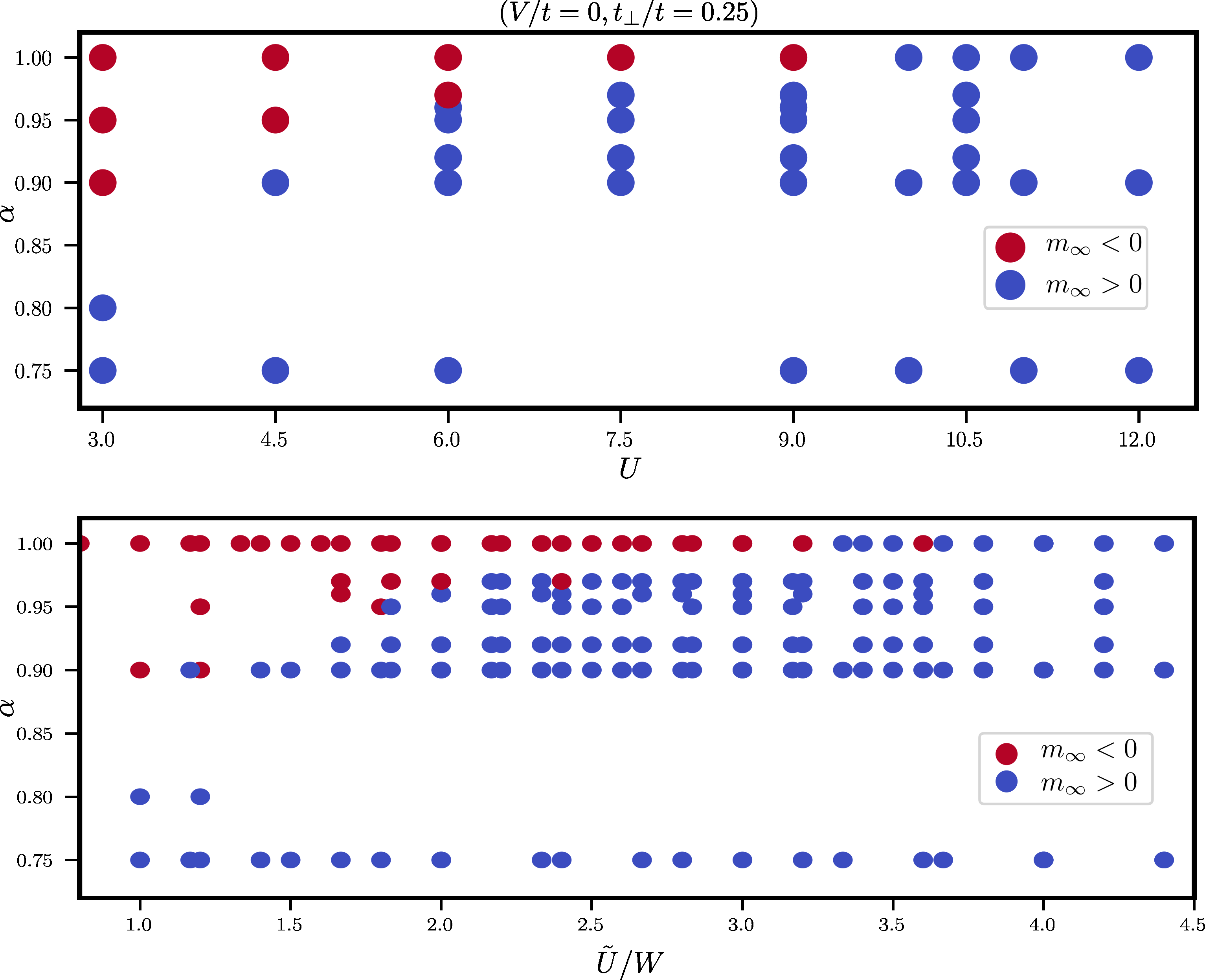}
    \caption{The extrapolated $L\rightarrow \infty$ value of the staggered magnetization $m_\infty$ from the spin-correlations at $\mathbf{Q}^{(\eta)}$.This is shown both for a single value of $V,t$ (\textbf{top panel}) and using the coupling-strength parameter $\tilde{U}/W$ which combines multiple measurements with different $U,V,t_\perp$ (\textbf{bottom panel}). The output is a binary red/blue for absence/presence of LRO.}
    \label{fig:supp: magnetic correlations: LRO phase diagram}
\end{figure}

As seen in Fig.~\ref{fig:supp:magnetic correlations: S(q) in BZ}, the system displays peaks at the M-points of the moir\'e Brillouin zone for most parameter values. However, whether these give rise to a genuine long range order(LRO) in the thermodynamic limit ($L\rightarrow \infty$) is a different question, one that depends on how the value of these peaks rises with increasing system-size. For larger system sizes, the maximum correlation length $\xi_{\text{AFM}}$ rises with $L$, and so for the system to genuinely order, one would expect 
\begin{equation}
    \mathcal{S}^{(\eta)}(\mathbf{Q}^{(\eta)}) \sim \frac{1}{L^2}\sum_{\vec{R},\vec{R}+\Delta \vec{R}}e^{-i\mathbf{Q}^{(\eta)}\cdot\Delta R}\langle \mathbf{S}^{(\eta)}_{\vec{R}} \cdot \mathbf{S}^{(\eta)}_{\vec{R} + \Delta \vec{R}} \rangle \propto \frac{L^4}{L^2}
\end{equation}
i.e. $\mathcal{S}(\mathbf{Q}) \equiv \mathcal{S}_{\mathbf{Q}}$ should grow linearly with the number of sites. The finite-size correction to this, obtained via spin-wave theory~\cite{Varney_2009,Huse_1988}, is
\begin{equation}
    \frac{\mathcal{S}(\mathbf{Q})} {L^2} = m^2_{\infty}  + \frac{a}{L}+\frac{b}{L^2}+\mathcal{O}(L^{-5/2})\,.
    \label{eq:supp:magnetic ordering:scaling}
\end{equation}
In our case, we extract the $m_\infty$ parameter using $L\times L$ systems, with  $L=4,8,12$. $m_\infty \leq0$ signifies the absence of the LRO.\footnote{While strictly speaking, $m_\infty <0$ is unphysical, it signals that the ansatz of an ordered state is the wrong one.} The fit and $m_\infty$ extrapolation is shown for a variety of  parameters in Fig.~\ref{fig:supp: magnetic correlations: LRO}. This reflects what was already suggested in Fig.~\ref{fig:supp: magnetic correlations: S(Q) vs beta}: that the magnetic correlations, for not-too-strong coupling strength $\tilde{U}/W$, are very sensitive to the interaction anisotropy $\alpha$, which drastically suppresses the correlations near $\alpha=1$. This behavior is shown again in Fig.~\ref{fig:supp: magnetic correlations: LRO phase diagram} which displays the presence/absence of an AFM long-range-order for different system parameters, demonstrating the role of $\alpha$ in suppressing the formation of AFM long range order. It should be noted that the scaling behavior of Eq.~\ref{eq:supp:magnetic ordering:scaling} does not apply to the ‘chain' limit of $t_\perp=0$.

\subsection{Charge compressibility}
 \begin{figure}
     \centering
     \includegraphics[width=\linewidth]{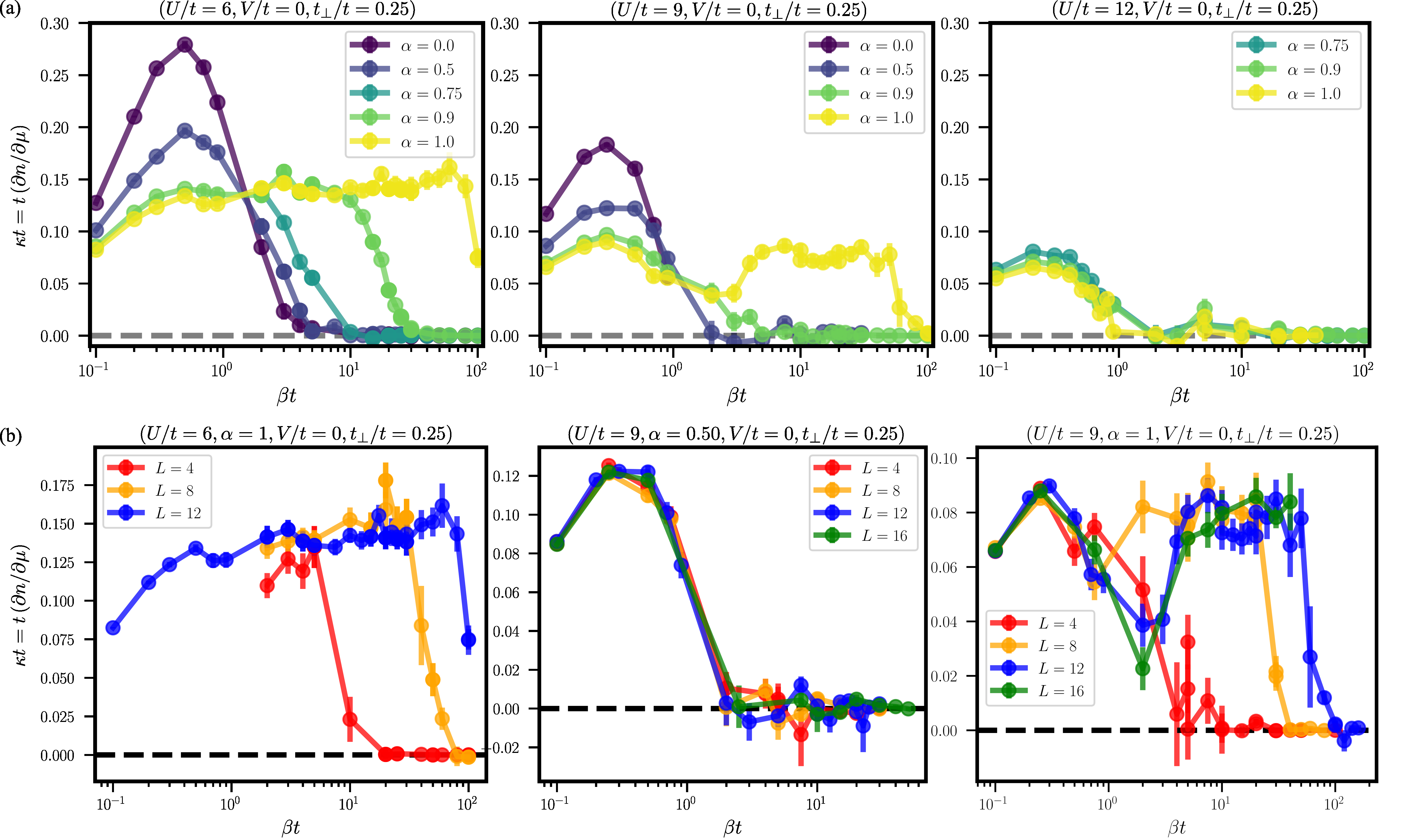}
     \caption{\textbf{Charge compressibility $\kappa$} for multiple parameters (a) at fixed system size of $12 \times12$ (b) for different system sizes.}
     \label{fig:supp:charge fluctuations:compressibility}
 \end{figure}
In an effort to understand the role of the interaction anisotropy $\alpha$ on the magnetic correlations, we proceed by studying the  charge fluctuations in the system. A useful measure of this, is the total charge variance, and related, the charge compressibility
 \begin{equation}
 \begin{split}
     \text{var}{(\hat{N})} &=\langle\hat{N}^2 \rangle - \langle \hat{N}\rangle^2 \\
     \kappa &= \frac{\beta \text{var}(\hat{N})}{L^2} \,,\
 \end{split}
 \end{equation}
 for $\hat{N}=\sum_{\mathbf{R},\eta,s}\hat{d}^\dagger_{\mathbf{R},\eta,s}\hat{d}_{\mathbf{R},\eta,s}$, the total number operator. The charge compressibility captures how susceptible the system is to the addition or removal of charge, and hence distinguishes gapped and gapless systems. For an insulating system with single-particle gap $\Delta_c$, the charge compressibility as a function of temperature $T$ has a Schottky peak at $T \sim \Delta_c$ and decays exponentially like $\sim e^{-T/\Delta_c}$ as the system is further cooled down. Further, if the insulating gap is set by the Hubbard $U$, as it is commonly understood to happen in Mott-insulating systems, then finite size effects do not strongly affect the compressibility since $\Delta_c \sim U \gg 1/L$.  
By contrast, in a metallic system the charge gap $\Delta_c$ vanishes in the thermodynamic limit. At any finite size $L$, $\lim_{T\rightarrow0}\kappa(L)$ depends on whether the discretized BZ contains $\mathbf{k}$ which are on the Fermi surface. If it does not, which is the generic situation for an interacting system, there will be a finite size gap associated with a $|\delta \mathbf{k}|\sim 1/L$. Thus, once the temperature is below $T \lesssim 1/L$ the system appears incompressible ($\kappa \approx 0$). Importantly, in this case, the temperature at which the compressibility gets suppressed shifts with system size. 
\footnote{If the finite-size system does contain a momentum lying on the FS, then $\lim_{T\rightarrow 0}\kappa(L)=\infty$, and the order of limits matters if the system is to acquire a finite compressibility for $T=0,L=\infty$. This latter behavior is not observed in the intermediate- and strong-coupling regimes of our system.}
 \par
 The compressibility $\kappa$ is shown in Fig.~\ref{fig:supp:charge fluctuations:compressibility}. For a range of parameters, the system seems to unambiguously be in the strong-coupling insulating regime, akin to a Mott insulator, with a single peak at high-$T$ and then a rapid decay to $\kappa\rightarrow0$. In these cases, the $\kappa \text{ vs }\beta$ plot looks identical for any system size (\textit{e.g.} Fig.~\ref{fig:supp:charge fluctuations:compressibility} (b) middle plot). For other parameter values, the system seems to lie in an intermediate-coupling regime, which displays a peak at high-$T$ associated with a $\Delta \sim U$ scale, but it also displays a second peak (or a flat `shoulder' of high-compressibility) down to very low temperatures, associated with another scale, the same one identified in the spin-correlations shown in Fig.\ref{fig:supp: magnetic correlations: S(Q) vs beta}. Qualitatively, the intermediate-coupling regime occurs when $\alpha$ is near $1$, and coupling is strong, but not too strong ($1\lesssim\tilde{U}/W \lesssim 4$). The fact that the second energy scale is also system-size dependent (Fig.~\ref{fig:supp:charge fluctuations:compressibility} (b)) supports the idea that in the intermediate-coupling regime, there seems to be a coexistence between \textit{localized} and \textit{itinerant} degrees of freedom. Such a regime cannot be found in a square-lattice half-filled $U(2)$-Hubbard model, unless AFM nesting is suppressed, and is typical of the intermediate-coupling regime of larger-$N$ $SU(N)$ models, despite the system never having $U(6)$ symmetry.

\subsection{Local fluctuations}
As demonstrated in the previous sections, the intermediate-coupling regime seems to be characterized by a suppression of AFM correlations, and the presence of both localized and itinerant charge degrees of freedom. To shed light on the nature of the low-energy degrees of freedom in this regime, we can focus on the local fluctuations, disentangling them from $\text{var}(\hat{N})$ which also includes non-local charge fluctuations:
\begin{equation}
    \frac{\text{var}(\hat{N})}{L^2} = \text{var}(\hat{n}_\vec{R}) + \sum_{\vec{R} \neq \vec{R}'}\text{cov}(\hat{n}_\vec{R},\hat{n}_{\vec{R}'})\,.
\end{equation}
For the following discussion which is confined to local operators, we drop the site subscript for simplicity: $\hat{d}_{\vec{R},\eta,s}\to \hat{d}_{\eta,s}$. For a system with $U(2)^{\otimes 3}$ symmetry as well as a valley-permutation symmetry$S_3$\footnote{This is the effect of $C_{3z}$ symmetry at the local level.}, there are only two unique quartic correlators that determine the local properties :
\begin{equation}
\begin{split}
     C &= \langle\hat{n}_{\eta,\uparrow}\hat{n}_{\eta' \neq \eta,\uparrow} \rangle\\
      D &= \langle \hat{n}_{\eta,\uparrow}\hat{n}_{\eta,\downarrow} \rangle \,.\\
\end{split}
\label{eq:supp:additonal:C&D}
\end{equation}
For example, the (square of the) local moment can be written as:
\begin{equation}
    \langle Z^{(\eta)}Z^{(\eta)} \rangle =\frac{1}{4}\left(\langle \hat{n}_{\eta}\rangle -2D\right) = \frac{1}{4}-\frac{D}{2}\,,
     \label{eq:supp:additonal:ZZ}
\end{equation}
for $Z^{(\eta)} = \frac{1}{2}\sum_{s,s'}\hat{d}^\dagger_{\eta,s}\sigma^z_{ss'}\hat{d}_{\eta,s'}$. Another useful quantity, which captures how correlated the different valleys are, is $C^{\text{conn.}}$:
\begin{equation}
    C^{\text{conn.}} = 4C-\langle n_\eta \rangle \langle n_\eta' \rangle = 4C - 1 \,.
    \label{eq:supp:additonal:C_connected}
\end{equation}
Further, we can write the local charge fluctuations $\hat{n}=\sum_{\eta,s}\hat{n}_{\eta,s}$, as well as local, charge-neutral valley fluctuations  $\delta \hat{n}_{+}=\left(\hat{n}_{\eta=1}-\hat{n}_{\eta=2}\right)/\sqrt{2}$ and $\delta \hat{n}_{-}=\left(\hat{n}_{\eta=1}+\hat{n}_{\eta=2}-2\hat{n}_{\eta=3}\right)/\sqrt{6}$  as functions of  $C$ and $D$:
\begin{equation}
    \begin{split}
        \text{var}(\hat{n})&= \langle \hat{n}\rangle\left(1-\langle \hat{n}\rangle \right) + 6\left(D+4C\right) = 6(D+4C-1)\\
        \text{var}(\delta\hat{n}_+)&=\text{var}(\delta\hat{n}_-)=\left(\frac{\langle \hat{n}\rangle}{3}+2 D-4C\right) =(1+2D-4C) \,.\\
    \end{split}
    \label{eq:supp:additonal:delta_n}
\end{equation}
It is also instructive to consider the values $C$ and $D$ take in the atomic limit ($U/t \rightarrow \infty$) for a $U(2N_f)$-symmetric  Mott insulator: $C=D=(N_f-1)/2(2N_f-1)$ which is $0.2$ for $N_f=N_\eta = 3$. It should be noted however, that the $\text{AA t-SnSe}_2$ system is never $U(6)$ symmetric, since even for isotropic interactions ($\alpha=1$) the hopping terms are valley-selective and hence always break $U(6)$.
\begin{figure}
    \centering
    \includegraphics[width=\linewidth]{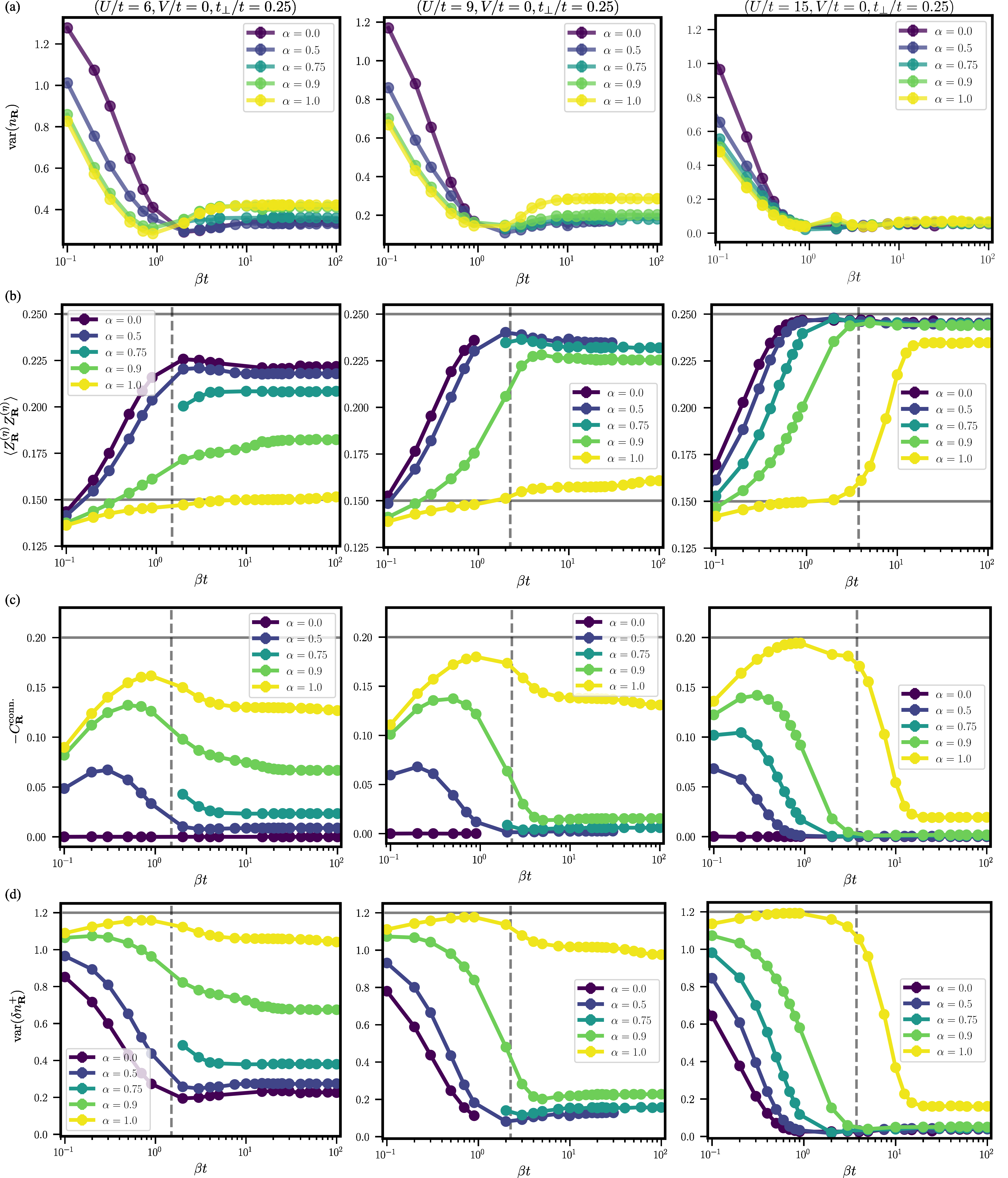}
    \caption{\textbf{Local observables in the intermediate- and strong-coupling regime:} Each row of figures displays a different local observable, with each column being a different set of parameters. Roughly, the first two columns should be understood as being in the intermediate-coupling regime for large-enough $\alpha$. For the last column, the system is in the strong-coupling regime for all $\alpha$. The dashed vertical lines denote the $T=J$ scale and the horizontal lines the $U(2)$ and $U(6)$ atomic limits of a given observable. The observables are defined in Eqs.~\ref{eq:supp:additonal:ZZ},~\ref{eq:supp:additonal:C_connected} and~\ref{eq:supp:additonal:delta_n}.}
    \label{fig:supp:additional data:local fluctuations}
\end{figure}
\begin{figure}[h]
    \centering
    \includegraphics[width=0.75\linewidth]{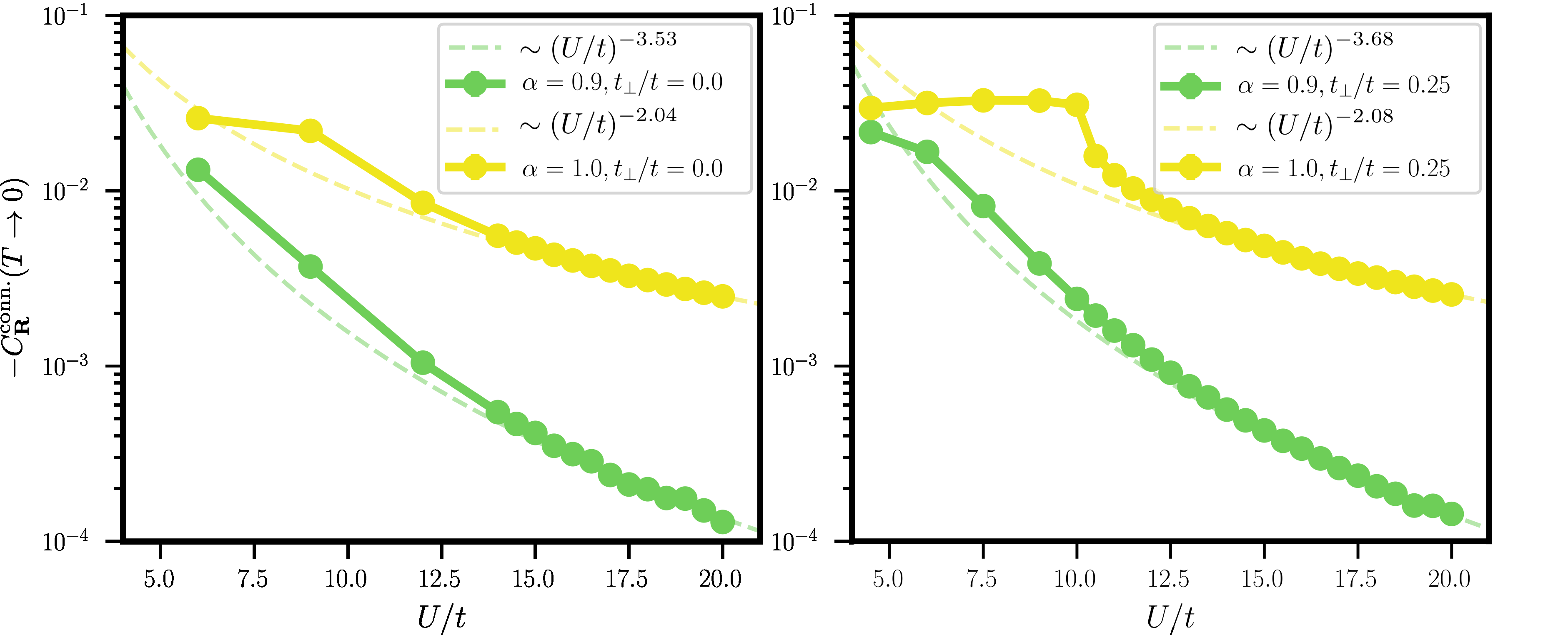}
    \caption{The connected part of the $C$ observable (Eq.~\ref{eq:supp:additonal:C_connected}) and its power-law fit deep inside the strong-coupling regime. The decay of the valley-couplings for $\alpha=1$ is very close to $\sim (U/t)^{-2}$ and seems largely independent of $t_\perp$. If the system is tuned away from the isotropic-interaction point, the decay is much faster.}
    \label{fig:supp:additional data: large-U valley-coupling}
\end{figure}
These local observables, shown in Fig.~\ref{fig:supp:additional data:local fluctuations}, help us understand the sensitivity of the system on $\alpha$: The $U(6)$-breaking scales are $(1-\alpha)U,t^2/U,t_\perp^2/U$, corresponding to the local anisotropy, and the valley-selective hoppings. The intermediate-coupling regime is a regime where the local anisotropy \textit{is not the dominant} $U(6)$-breaking process: $(1-\alpha)U \lesssim J$ or $ U/t \lesssim1/\sqrt{1-\alpha}$. In this regime, the near-degeneracy of states at the local level, dramatically changes the behavior of the system.
Fig.~\ref{fig:supp:additional data:local fluctuations}(a) shows that even in the intermediate-coupling regime, there is a well-formed local charge throughout, with only small local charge variance. So, states containing more/less than $\nu=3$ electrons per site, are heavily suppressed. But, local configurations that are `valley-imbalanced' such as $(210)$, are only slightly penalized, compared to the `valley-singlet' $(111)$ states. So in this regime, there seems to be a proliferation of such valley fluctuations, as shown by the suppression of the local moment and the enhanced charge-neutral valley fluctuations shown in Fig.~\ref{fig:supp:additional data:local fluctuations} in the first two columns.
\noindent
Turning our attention to the strong-coupling behavior, the effects of the anisotropy $\alpha$ are less dramatic compared to the intermediate-coupling regime, but they are nonetheless very interesting: First, we concentrate in the temperature regime $\text{max}(J,(1-\alpha)U)\lesssim T\lesssim U$ where the dominant $U(6)$-breaking scale remains incoherent due to thermal fluctuations. As a result, the local observables behave as if they are $U(6)$-symmetric and approach the values they would take in a $U(6)$ Mott insulator. This can be seen as the $\alpha=1$ curves plateauing at the gray horizontal lines in the third column. Second, there are remnants of the valley-fluctuations even in the ground state: As can be seen from Fig.~\ref{fig:supp:additional data:local fluctuations} (b)-(d) in the last column, at very low-$T$, there is a \textit{suppression} of the local moment for the $\alpha=1$ isotropic case, alongside enhanced inter-valley coupling and valley-neutral fluctuations. This local moment suppression cannot be solely explained from the double-occupancy $D$,  in the way that it can in the $\alpha \neq 1$ case: The $\alpha = 1$ behavior seems to be due to some higher-order processes, as elucidated in Sec.~\ref{section:supp:perturbation theory}. We can delve deeper into the strong-coupling regime by tracking $C^{\text{conn.}}$ in the ground state, i.e. how coupled the valleys remain, as $\tilde{U}/W$ (or simply $U/t$) increases. This is shown in Fig.~\ref{fig:supp:additional data: large-U valley-coupling}. Numerical data seems to suggest that for $\alpha=1$, the decay is $C^{\text{conn.}}\sim(U/t)^2$ and that it is $t_\perp$-independent.

\subsection{Time displaced, momentum-resolved observables}
While the spin-correlations and local, static observables shed some light on the low-energy mechanisms at play when the interactions are dominant and nearly-isotropic, to fully understand the physics of the system we can investigate the full, time-displaced, momentum-resolved correlation functions $\mathcal{S}_{\hat{\mathcal{O}}}(\mathbf{q},\tau)$ associated with charge excitations and valley-fluctuations. These will provide insight into the dynamics of the low-energy degrees of freedom. We probe the low-energy spectral properties of the system using three closely related methods. The first method involves analyzing the late-time behavior of the $\mathcal{S}_{\hat{\mathcal{O}}}(\mathbf{q},\tau)$ correlation functions
\begin{equation}
\begin{split}
    \mathcal{S}_{\hat{\mathcal{O}}}(\mathbf{q},\tau) = \langle \hat{O}_{\mathbf{q}}(\tau) \hat{O}_{-\mathbf{q}}\rangle &= \frac{1}{Z}\text{Tr}\left[e^{-(\beta-\tau)\hat{H}}\hat{O}_{\mathbf{q}}e^{-\tau\hat{H}}\hat{O}_{-\mathbf{q}}\right]\\
    &= \frac{1}{Z}\sum_{ij} e^{-\frac{\beta}{2}(E_i + E_j)}e^{-(\tau - \beta/2)(E_i - E_j)}[O_{\mathbf{q}}]_{ij}[O_{-\mathbf{q}}]_{ji}\\
\end{split}
\label{eq:supp:additional:obs late time}
\end{equation}
where $i,j$ are some generic basis states. Now, consider $\beta$ large enough to project the system to its ground state. For $\tau -\beta/2<0$, the terms where $i\equiv0$, i.e. when $i$ is the ground state, dominate the sum of Eq.~\ref{eq:supp:additional:obs late time}
\begin{equation}
\begin{split}
    \mathcal{S}_{\hat{\mathcal{O}}}(\mathbf{q},\tau<\beta/2)  
    &\approx \frac{1}{Z}\sum_{j} e^{-\frac{\beta}{2}(E_i + E_j)}e^{-(\tau - \beta/2)(E_i - E_j)}[O_{\mathbf{q}}]_{ij}[O_{-\mathbf{q}}]_{ji}\\
    &\approx e^{\beta E_0}\sum_{j} e^{-\beta E_0}e^{-\tau(E_j - E_0)}[O_{\mathbf{q}}]_{0j}[O_{-\mathbf{q}}]_{j0} \,,\\
\end{split}
\label{eq:supp:additional:decay}
\end{equation}
where the ground state was assumed to be non-degenerate. Now, among all the states that can be reached from the ground state $| 0\rangle$ upon the action of $\hat{O}_{-\mathbf{q}}$, we can find the one with the minimum gap, indexed by $j^*$:
\begin{equation}
    \Delta_{-,j^*}(\mathbf{q}) = \text{min}_{\langle j|\hat{O}_{-\mathbf{q}}|0\rangle \neq 0}\{E_j - E_0\} \,,
\end{equation}
with an associated weight $Z_{-,\hat{O}_{-\mathbf{q}}}=[O_{\mathbf{q}}]_{0j^*}[O_{-\mathbf{q}}]_{j^*0}$. For $\tau<\beta/2$ but also much larger than the gap between $\Delta_{j^*}(\mathbf{q})$ and the next excited state's energy, the $\sum_j$ in Eq.~\ref{eq:supp:additional:decay} is dominated by $j^*$:
\begin{equation}
    \mathcal{S}_{\hat{\mathcal{O}}}(\mathbf{q},\tau<\beta/2)  \approx Z_{-,\hat{O}_{-\mathbf{q}}} \times e^{-\Delta_{-,j^*}(\mathbf{q})\tau}\,.
\end{equation}
A similar calculation for $\tau>\beta/2$ leads to the unified expression
\begin{equation}
    \mathcal S_{\hat{\mathcal O}}(\mathbf q,\tau)\sim
\begin{cases}
Z_{-,\hat{O}_{-\mathbf{q}}}(\mathbf q)e^{-\Delta_{-,j^*}(\mathbf{q})\tau}, & 0<\tau<\beta/2,\\[4pt]
Z_{+,\hat{O}_{+\mathbf{q}}}(\mathbf q)e^{-\Delta_{+,j^*}(\mathbf q)(\beta-\tau)}, & \beta/2<\tau<\beta \,.
\end{cases}
\end{equation}
Hence, a clear exponential decay of $\mathcal{S}_{\hat{\mathcal{O}}}(\mathbf{q},\tau)$ in $\tau$ can reveal the energy gap associated with $\hat{O}$ excitations of the ground state. For the single-particle Green's function, the  $\Delta_{\pm,j^*}(\vec{q})$ gaps acquire a particularly clean interpretation, being the cost of adding or removing an electron with momentum $\mathbf{q}$ from the ground state. The total single-particle excitation gap can then be defined as
\begin{equation}
    \Delta_{\vec{q},\eta,s} \equiv \Delta E^{+}_{\vec{q},\eta,s} + \Delta E^{-}_{\vec{q},\eta,s}
\end{equation}
with $\Delta E^{\pm}_{\vec{q},\eta,s} = \text{min}_{\langle j|\hat{d}^{(\dagger)}_{\vec{q},\eta,s}|0\rangle}\{E_j - E_0\}$.\par
Another, less direct but perhaps more intuitive way to extract the spectrum, is via \textit{analytic continuation}, giving access to the real frequency dynamical information. Finally, one can utilize proxies to extract information about the $\omega=0$ behavior of the real-frequency quantities, which bypasses the difficulties inherent to analytical continuation. Both of these methods were covered in Sec.~\ref{section:supp:dqmc methods}.

\subsubsection{Single-particle Green's function}
\begin{figure}
    \centering
    \includegraphics[width=\linewidth]{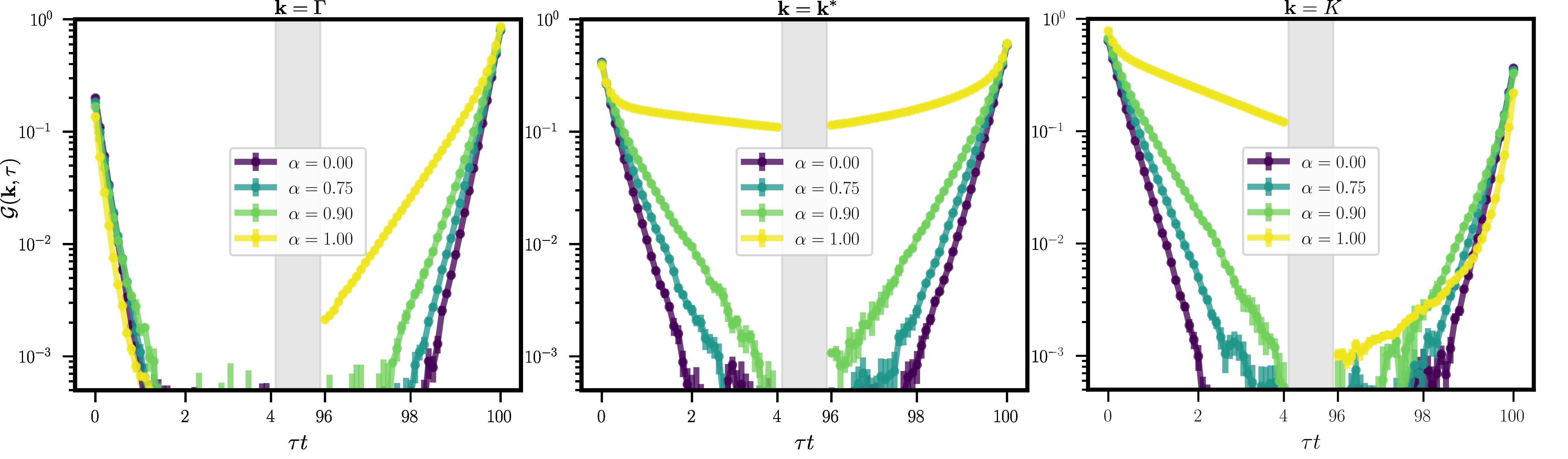}
    \caption{Imaginary-time decay of single-particle Green's functions at very low temperatures $T=0.01t (\beta t=100)$. The decay is shown for $U=9t,V=0,t_\perp=0.25t,L=12$ and a range of anisotropies $\alpha$, at three different Brillouin zone momenta. For $\alpha=1$, the three momenta should be thought of as lying respectively inside, (nearly) on, and outside the Fermi surface.}
    \label{fig:supp:dynamical:greens function time-decay}
\end{figure}
Here we study the single-particle Green's function $\mathcal{G}^{(\eta)}(\mathbf{k},\tau)$ and its spectral function $A^{(\eta)}(\mathbf{k},\omega)$ (See SI Sec.~\ref{section:supp:dqmc methods} for definitions).  The Green's function is spin-symmetric and valley-diagonal due to the $U(2)^{\otimes3}$ valley symmetry. $\mathcal{G}^{(\eta)}(\mathbf{k},\tau)$ provides information about the dispersion of single-particle charge excitations. We begin by studying the $\tau$-decay of the Green's function which is shown in Fig.~\ref{fig:supp:dynamical:greens function time-decay}. For an insulator with both an addition and subtraction gap $\Delta_\pm >0$, the expected behaviour would be that $\mathcal{G}(\tau)$ decays exponentially with a decay rate given by the gaps. Further, for a Mott insulator, these gaps should be very large and proportional to $U$. This is indeed what is observed in  Fig.~\ref{fig:supp:dynamical:greens function time-decay} for $\alpha\lesssim 0.90$. The $\mathcal{G}(\tau)$ decay when the system is in the intermediate regime however, deviates significantly from this picture: For different momenta $\mathbf{k}$, the tails show a two-mode behaviour, involving a scale of order $U/t$ which decays by early times $\tau$, as well as a smaller scale which dominates at later times. For certain momenta, that second energy scale is extremely small. These results are consistent with two kinds of charged excitations: localized, gapped electrons as in the Mott-insulating regimes, as well as itinerant, (nearly) gapless quasiparticles with a well-defined Fermi Surface. Momenta inside (outside) the Fermi Surface probe the energy to create a quasi-hole (electron) excitation, while excitations associated with momenta on, or near, the Fermi surface are nearly gapless!\par
\begin{figure}[h!]
    \centering
    \includegraphics[width=0.95\linewidth]{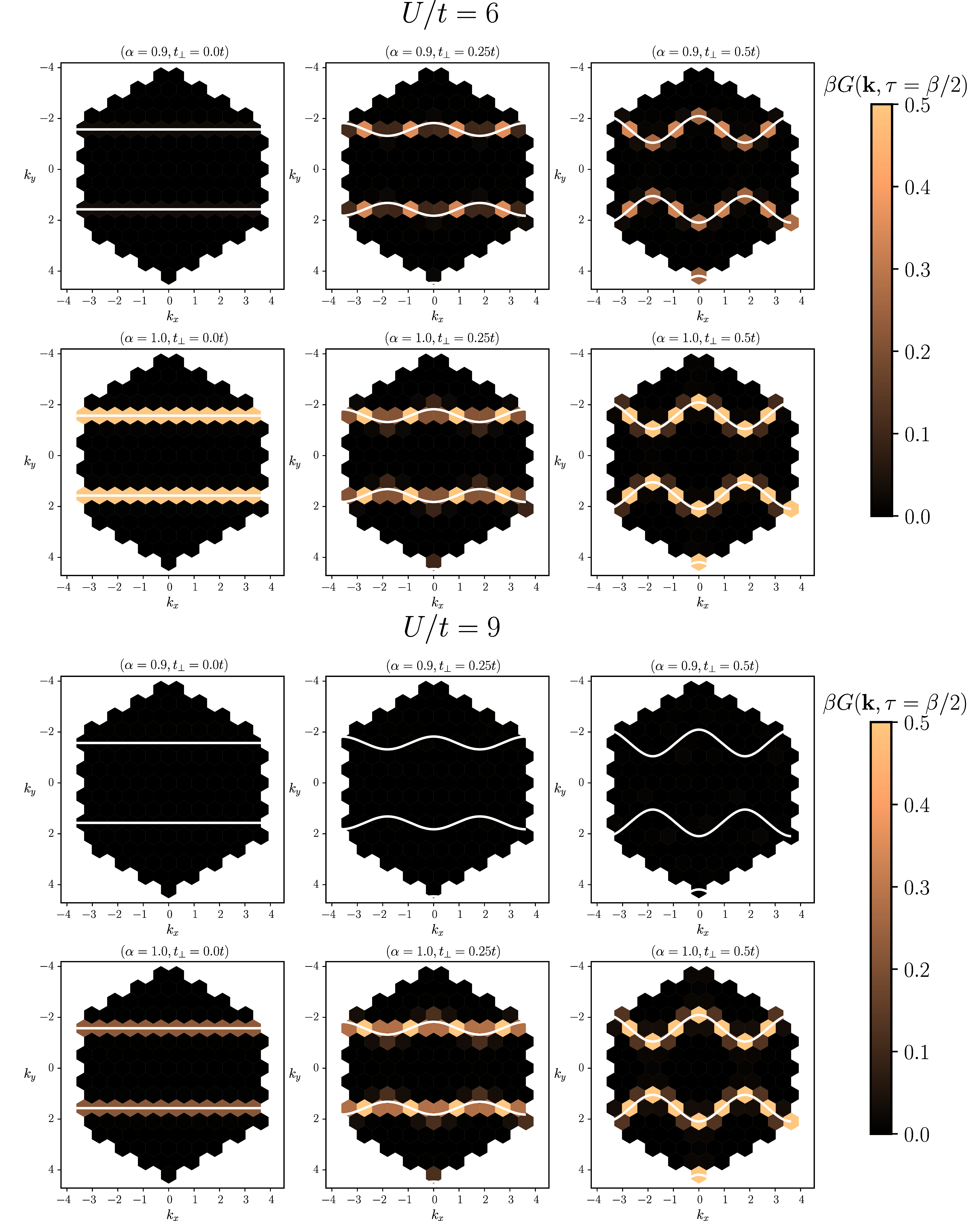}
    \caption{A proxy for the low-frequency spectral weight of valley-$\eta=0$ electrons $A^{(0)}(\mathbf{k},\omega=0)$. We simulate the model for $V=0,L=12$ at $\beta=20t$ and different $U,t_\perp$. A lit-up Brillouin zone indicates the existence of spectral weight within $\sim 0.05t$ of $\omega=0$ and hence may be characterized as metallic. The white lines trace the non-interacting Fermi surface.}
    \label{fig:supp:dynamical:spectral weight}
\end{figure}
\begin{figure}
    \centering
    \includegraphics[width=\linewidth]{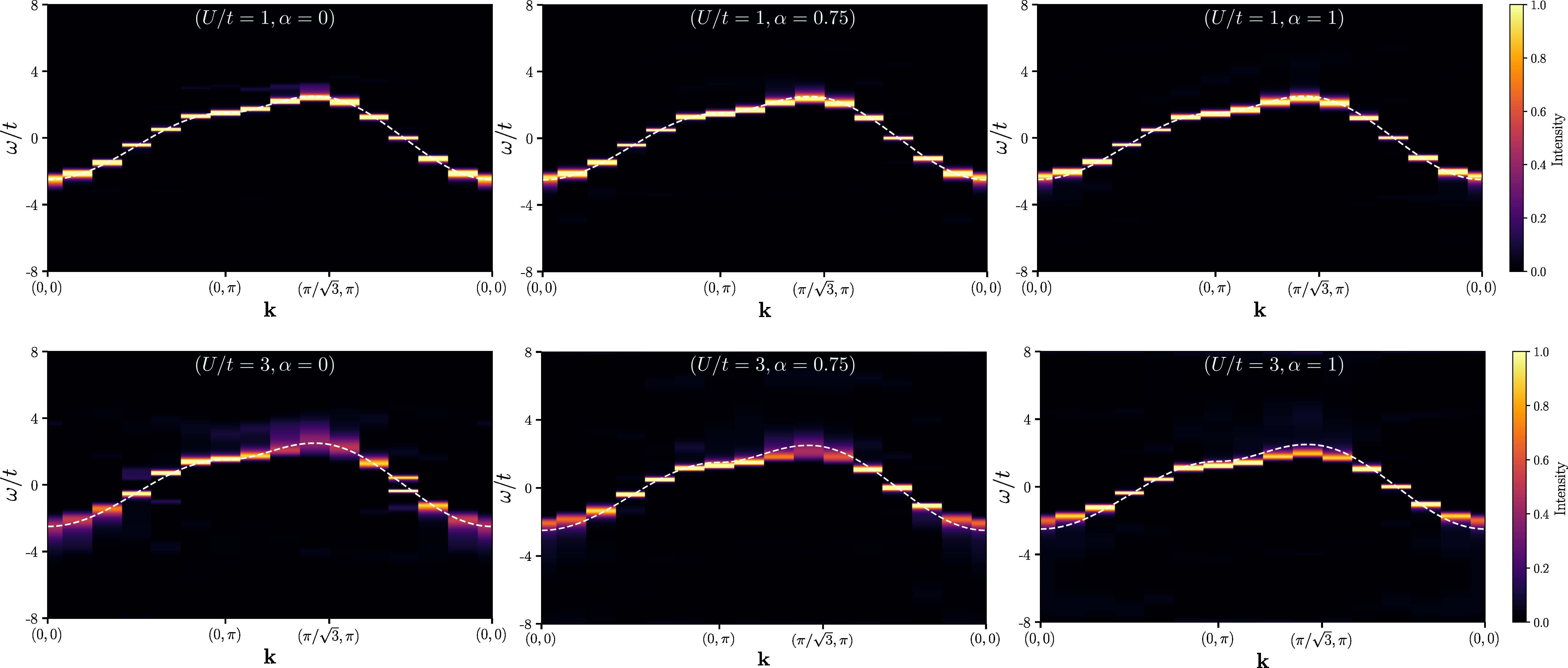}
    \caption{A collection of spectral functions for different parameter values with $V=0,t_\perp=0.25t$ at $\beta t = 20$, showing the systems both in the weak-, intermediate- and strong- coupling regimes. In cases where there is a clear gap, the data was extrapolated from a $12\times12$ to a larger system.}
    \label{fig:supp: weak U spectral function}
\end{figure}
\begin{figure}
    \centering
    \includegraphics[width=\linewidth]{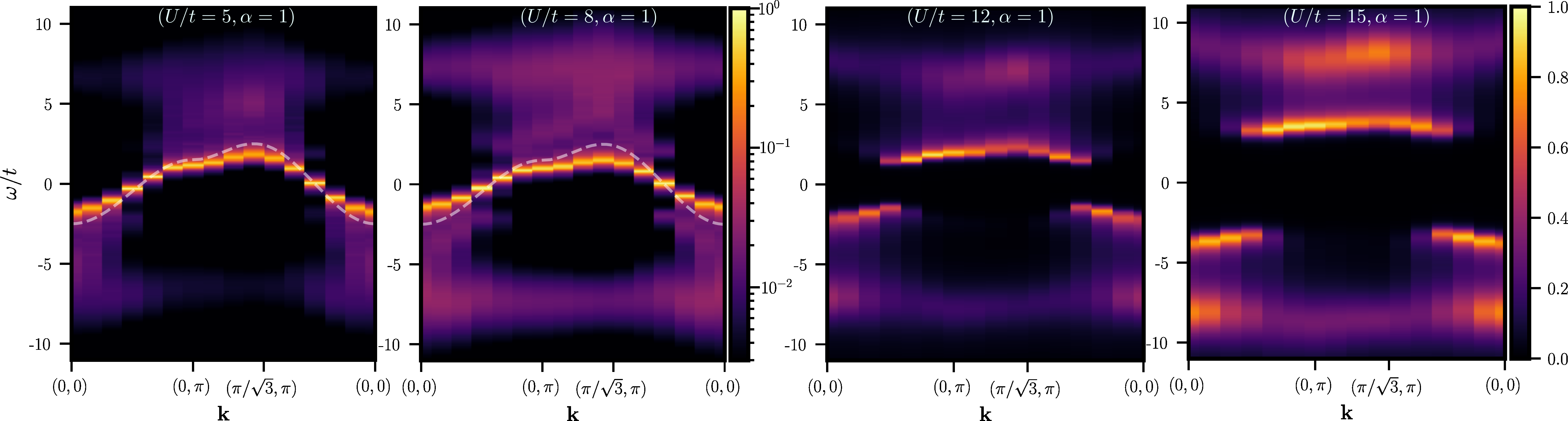}
    \caption{Electronic spectral function $A^{(\eta)}(\mathbf{k},\omega) = -\frac{1}{\pi}\text{Im}G^{(\eta)}(\mathbf{k},\omega)$ for $\eta=0$-valley electrons with $\alpha=1,T=0.05t$ and $V=0,t_\perp=0.25t$. The dashed lines correspond to the non-interacting bandstructure.}
    \label{fig:supp:dynamical:ana cont spectral function 1}
\end{figure}

This picture is further corroborated by studying the proxy $A^{(\eta)}(\mathbf{k},\omega=0) \approx \beta \mathcal{G}^{(\eta)}(\mathbf{k},\tau=\beta/2)$ for the low-energy spectral weight. For a Mott-insulating phase, there are no low-energy excitations and so $A^{(\eta)}(\mathbf{k},\omega=0) \approx 0$ while in a scenario where localized, gapped modes coexist with gapless, itinerant modes, $A^{(\eta)}(\mathbf{k},\omega=0)$ should be non-zero, tracing the Fermi Surface! This is exactly what is observed in Fig.~\ref{fig:supp:dynamical:spectral weight}. It is quite unexpected that well-defined, gapless quasiparticles are present for such large interaction strengths $\tilde{U}/W\approx4$.
Finally, we can perform the analytic continuation to get the entire spectral function $A^{(\eta)}(\mathbf{k},\omega)$  allowing us to directly observe the `two-mode' itinerant and localized behavior at intermediate-coupling. The spectral function results are shown in Fig.~\ref{fig:supp: weak U spectral function} \& Fig.~\ref{fig:supp:dynamical:ana cont spectral function 1}.  In the weak-coupling regime $\tilde{U}/W \lesssim1$, the spectral function is very weakly dependent on $\alpha$, largely following the non-interacting bands. As interaction is increased, the spectral function acquires a gap. For small $\alpha$, a clear gap develops and the spectral weight rearranges into two bands, with nearly zero spectral weight at $\omega=0$. These bands evolve into the familiar Mott bands upon further increasing interactions. For $\alpha \approx 1$, as interactions are increased, the system enters the intermediate-coupling regime: there is an accumulation of spectral weight at `high' energies; these are the localized degrees of freedom ---usually though of as doublon/ holon excitations--- but the spectral weight remains mostly near the non-interacting bands, carried by the strongly-renormalized quasiparticles. For sufficiently large $\tilde{U}/W$ the quasiparticles acquire a gap but they still retain a lot of spectral weight. As  $\tilde{U}/W$ is further increased,  the gap increases, and simultaneously, spectral weight is depleted from the quasiparticles and into the remote bands. Still, even for $\tilde{U}/W=6$ the spectral function is quite different between $\alpha=1$ and $\alpha < 1$ as in the latter case, the Mott bands are quite well defined, while in the former, the presence (of the now gapped) quasiparticles remains visible.
\subsubsection{Valley fluctuations}
\begin{figure}[h!]
    \centering
	\includegraphics[width=0.9\linewidth]{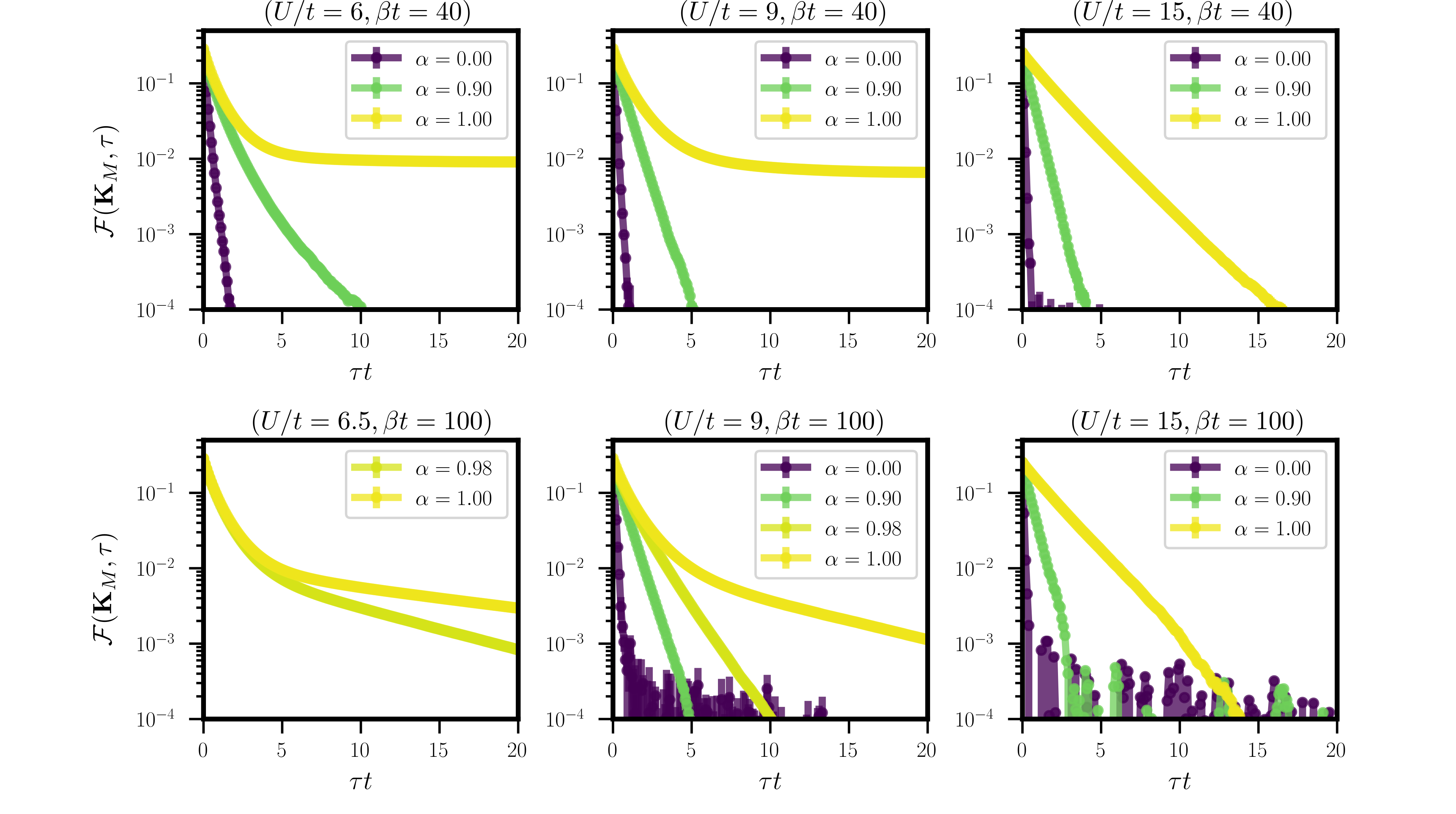}    
    \caption{The imaginary-time decay of the valley-coherence modes $\mathcal{F}(\mathbf{k},\tau)$ at the M points of the Brillouin Zone. The plot is for $V=0,t_\perp=0.25t,L=12$.}
    \label{fig:supp:strong coupling: valley fluctuations tau decay}
\end{figure}
In order to clarify the nature of the itinerant quasiparticles present in the intermediate-coupling regime, we now turn our attention to the valley-fluctuations. These were defined in Eq.~\ref{eq:supp:implementation:valley flucts} but are repeated here for ease of presentation:
\begin{equation*}
    \mathcal{F}(\mathbf{k},\tau) = \frac{1}{6}\sum_{\eta\neq\eta'}\langle \hat{d}^\dagger_{\mathbf{k},\eta,s}(\tau)\hat{d}_{\mathbf{k},\eta',s'}(\tau)\hat{d}^\dagger_{-\mathbf{k},\eta',s'}\hat{d}_{-\mathbf{k},\eta,s}\rangle \,.
\end{equation*}
\begin{figure}
    \centering
    \includegraphics[width=0.9\linewidth]{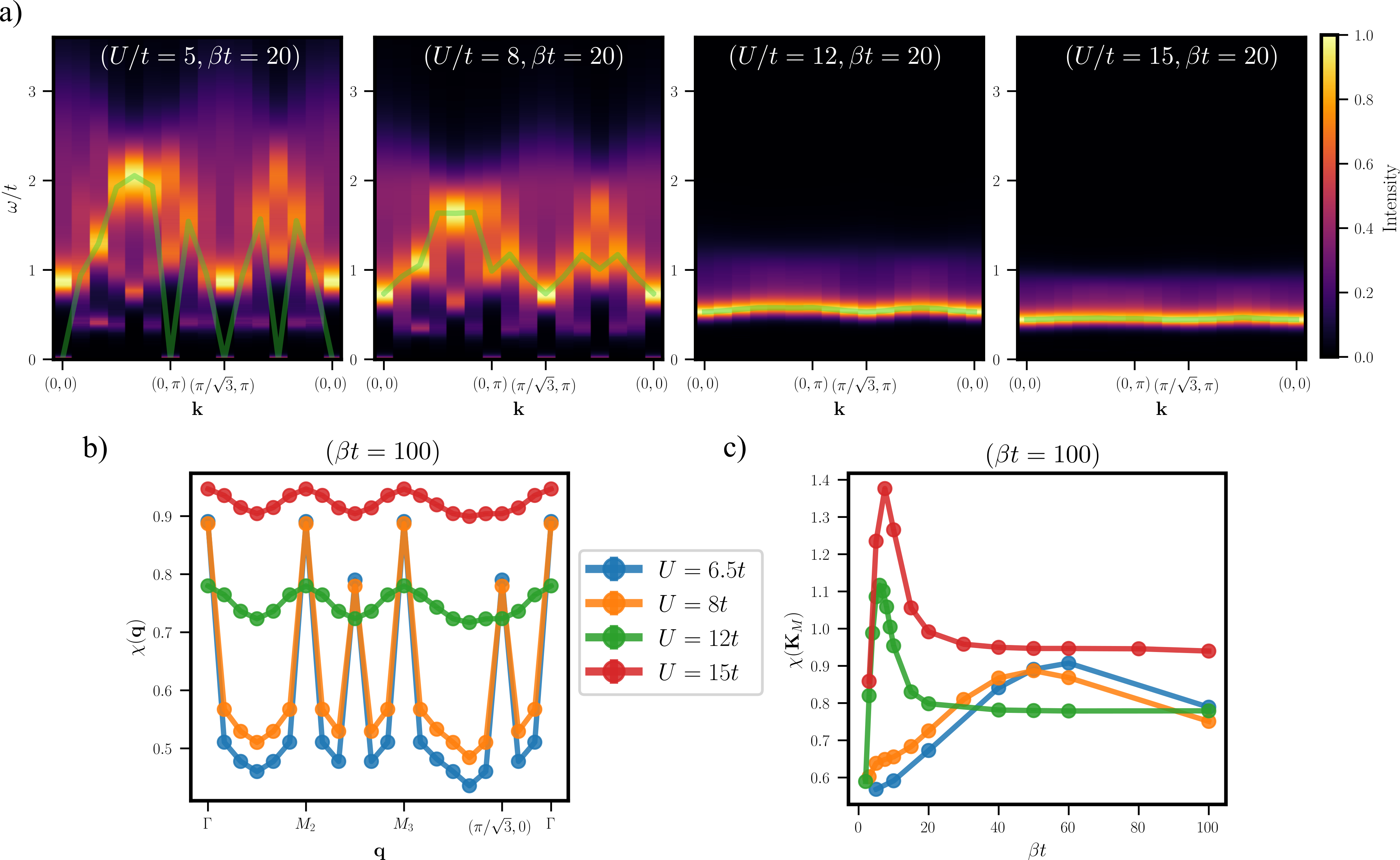}
    \caption{\textbf{Valley Fluctuation spectrum and susceptibility:}(a)Analytically continued valley fluctuations $F(\mathbf{k},\omega)$ at temperatures $T=0.05t$. The faint green line tracks the $\omega^*$ frequency at which $F(\mathbf{k},\omega^*)$ is maximum. (b)\& (c) Display of the valley-coherence susceptibility $\chi_{\text{valley coh.}}$ as defined in Eq.~\ref{eq:supp:implementation:symmetry inequivalent susceptibilities}, both in the BZ at $\beta t=40$ and how its peak value, at the M points evolves with cooling down the system. All the figures correspond to systems with $\alpha=1,V=0,t_\perp=0.25t,L=12$}
    \label{fig:supp:valley-fluctuation analytic continuation}
\end{figure}
The analytically-continued quantity is denoted as $F(\mathbf{k},\omega)$. This correlator captures the degree of valley-coherence in the system: it can be thought of as describing the propagation of valley-excitons which are charge-neutral modes carrying valley charge. While there are no processes in the Hamiltonian that change the valley of an electron, the correlator can probe the dynamics of such excitations. We are interested both in the energy cost of such modes, and their momentum-space dispersion.  We begin by considering their imaginary-time $\tau$ decay, shown in Fig.~\ref{fig:supp:strong coupling: valley fluctuations tau decay}. For bosonic excitations such as these valley-coherence modes, $\mathcal{F}(\mathbf{k},\tau)=\mathcal{F}(\mathbf{k},\beta-\tau)$ and so we restrict the discussion to $0\leq\tau\leq\beta/2$. For $U/t$ large enough and $\alpha$ small enough for the system to be in the strong-coupling regime, the valley-coherence modes behave as expected: for all momenta $\mathbf{k}$ they decay with a decay rate $\sim e^{-\Delta \tau}$ associated with a `large' gap of order $U$ and this behavior is roughly temperature-independent for cold enough systems. If, on the other hand, the system has $U/t$ and $\alpha$ such that it lies in the intermediate-coupling regime, there are drastic departures from the above description: The valley-coherence modes display a `two-mode ' decay, just as the single-particle excitations did, with them being (nearly) gapless at the M point of the Brillouin Zone. While this is indeed observed at low-$T$, upon further lowering the temperature, the second (small) gap can be resolved. This `emergent' gap can be understood as the temperature getting small enough to resolve the very low-energy process which sets the gap for these modes. This picture is further justified through the full analytically continued spectral function for these modes which is shown in Fig.~\ref{fig:supp:valley-fluctuation analytic continuation}. The fact that the valley-coherence modes mirror the single-particle excitations and become nearly gapless in the same parameter regime, hints at the two observations being related. Further, the valley-coherence modes being nearly gapless begs the following question: Does the system actually order in this channel? A valley-coherent order would be highly unusual, as it would naturally be associated with Spontaneous Symmetry Breaking of a $U(3)$ valley-rotation symmetry, which is of course not present in the system. To answer this, we can look at the valley-coherence susceptibility $\chi_{\text{valley coh.}}(\mathbf{q})$, shown in Fig.~\ref{fig:supp:valley-fluctuation analytic continuation} (b)\&(c). For systems in the strong-coupling regime, the valley-coherence susceptibility is mostly uniform in momentum-space which corresponds to local valley-fluctuations in real space. Further, if a strongly-coupled system cools down, the susceptibility rapidly rises as if it were to diverge and signal the advent of log-range order. But, once the system reaches $J\sim t^2/U$, the susceptibility peaks, decreases and then plateaus. For systems at intermediate-coupling, the susceptibility is greatly enhanced at the M, $\Gamma$ and other non-high-symmetry points of the Brillouin Zone. But, similarly to the strong-coupling regime, the susceptibility rises as the system is cooled until it reaches a  peak and then decreases. Everything considered, the data supports a picture where in the strong-coupling regime the valley-coherence fluctuations are local and gapped, but they remain present in the system down to the ground state (as evidenced by the $\chi_{\text{valley coh.}}$ plateau). On the other hand, for systems in the intermediate-coupling regime, the valley-coherence fluctuations are coherent over at least some short ranges, facilitated by the gapless charged modes also present in the system. In both cases, for $alpha\rightarrow 1^-$, the system seems to be proximate to some $U(6)$ physics, as seen by the tendency to form a SSB valley-coherent order. This is a consequence of the interactions dominating over the hopping, and replacing it as the dominant $U(6)$-breaking scale.

\subsubsection{On the nature of the low-energy quasiparticles}
\begin{figure}
    \centering
    \includegraphics[width=0.9\linewidth]{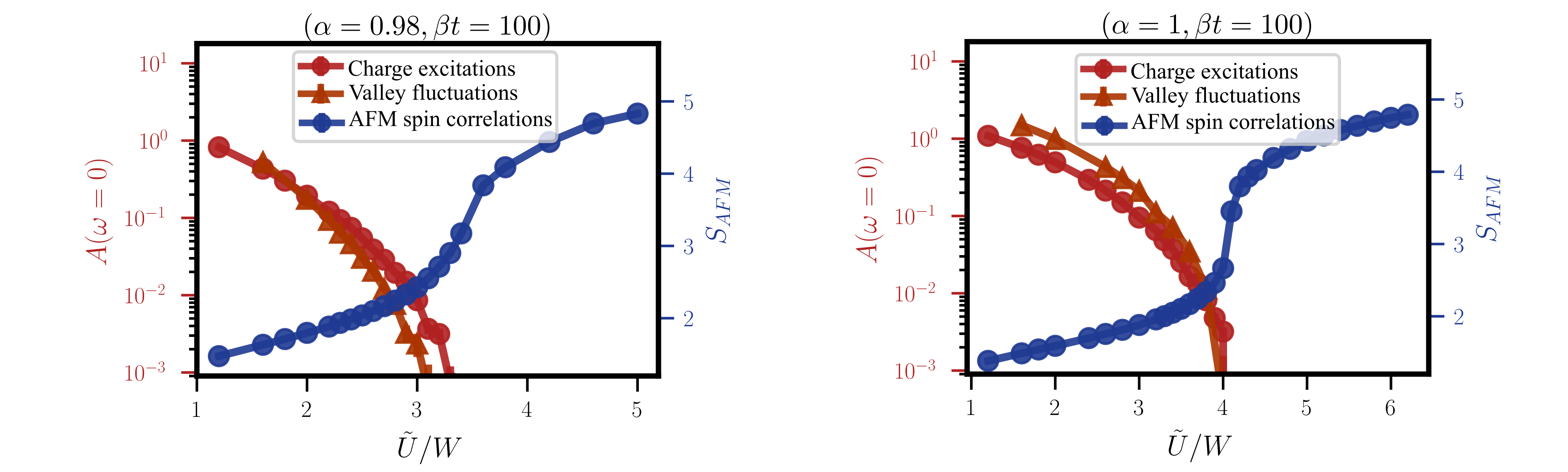}
    \caption{The competition between charge/valley fluctuations with the AFM correlations and the crossover between the intermediate- and strong-coupling regimes, as the interaction strength $\tilde{U}/W$ increases. The calculations are at very low temperatures $T=0.01t$, for both a realistic value of the anisotropy $\alpha=0.98$, and the isotropic point $\alpha=1$. The data is for $V=0,t_\perp=0.25t,L=12$.}
    \label{fig:supp:qp nature}
\end{figure}
Tying everything together, the intermediate-coupling regime can be characterized by the simultaneous presence of the following phenomena: (1) The AFM spin correlations are weak (compared to same $\tilde{U}/W$ coupling but smaller $\alpha$, for example) and the state does not seem to develop long range order (2) The system remains compressible down to very low-temperatures, but the low-energy charged excitations \textit{are not} the familiar holon/doublon excitons, which are remain suppressed by a scale $\sim U$. (3) Valley-coherence fluctuations have some short-range coherence and are nearly gapless.  These features, we believe, are intertwined: The quasiparticles should be thought of as collective degrees of freedom comprised of electrons and valley-fluctuations, which are allowed due to the local degeneracy near the isotropic point $\alpha=1$. These quasiparticles impede the development of coherence of long-range AFM correlations. For an interaction strong enough to gap out the quasiparticles, the local moments can finally increase their AFM correlations over longer distances and the system develops long-range AFM order. This competition between charge/valley fluctuations and AFM correlations can be seen directly in Fig.~\ref{fig:supp:qp nature}. Further, this intermediate-coupling window should be present for realistic twisted $\text{AA t-SnSe}_2$ materials with $\alpha \approx 0.96-0.98$, as long as there is a regime where the interactions dominate over the kinetic terms, but not so much so that the leading $U(6)$-breaking scale is $(1-\alpha)U$: An estimate for the intermediate-coupling  regime is $1\ll U/t \lesssim 1/\sqrt{1-\alpha}$. For $\alpha=0.98$, we see from Fig.~\ref{fig:supp:qp nature} that the intermediate- to strong- transition happens at $\tilde{U}/W\approx 3-3.5$. As seen from the Parameter Table~\ref{supp:model:tab:parameters_master}, many material realizations of twisted $\text{AA t-SnSe}_2$ should lie in this regime.

\section{Weak-coupling physics}
\label{section:supp:weak_coupling}
In this section, we briefly comment on the behavior of the model for $\text{AA t-SnSe}_2$ in the weak-coupling regime. Overall, we observe a weaker dependence of the physics on the interaction-anisotropy $\alpha$, and for $\tilde{U}/W\lesssim1$, the system remains metallic down to very low temperatures, despite it having a peak in the spin-spin correlations at the M points.\par
These results should be compared to the more familiar square-lattice half-filled Hubbard model: There, DQMC results find an Insulating state with antiferromagnetic long range order (AFM LRO) already for $U/t=2$ ($\tilde{U}/W=0.5$ in our notation) \cite{Varney_2009}, while RPA predicts a Slater Insulating state for any $U>0$ due to a Fermi Surface instability. There are, however, two crucial aspects of the M-point moir\'e model which significantly alter the picture: the strongly-anisotropic hopping ($t_\perp < t$) and the existence (and interplay) of the three valleys which  are coupled via interactions. Already for an anisotropic Hubbard model  the physics deviate from the isotropic case: Within RPA, the temperature $T^*(U)$ below which the Slater insulator forms is exponentially suppressed compared to the isotropic case, due to the van Hove singularities no longer being at the level of the Fermi surface. In Refs.~\cite{Raczkowski_2012,Raczkowski_2013,Raczkowski_2015}, weakly-coupled Hubbard chains were found to still form AFM LRO at sufficiently low $T$.
\par
\begin{figure}
    \centering
    \includegraphics[width=0.75\linewidth]{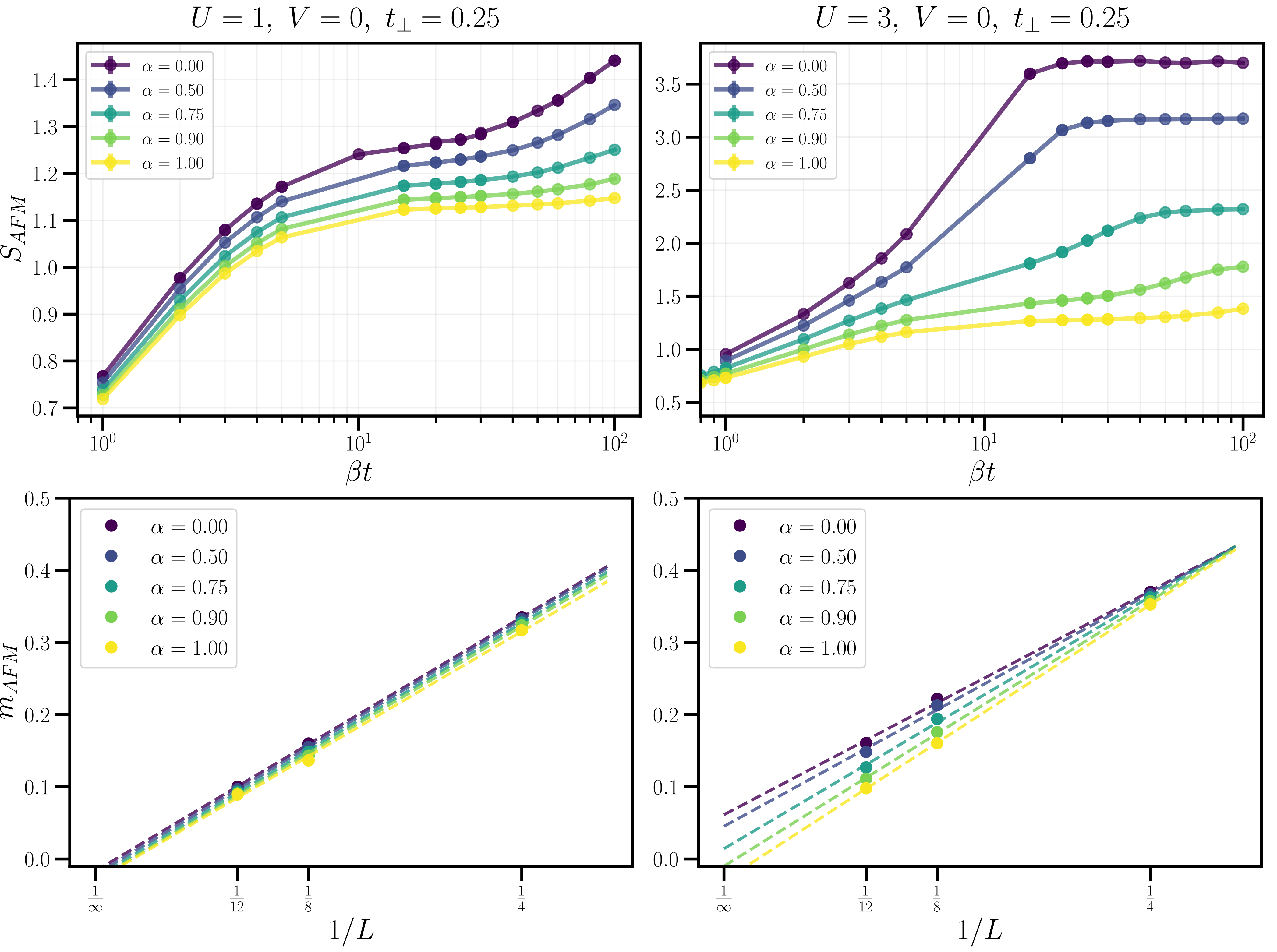}
    \caption{DQMC results in the weak-coupling regime. \textbf{Top}: The Spin Correlations $\mathcal{S}_{\mathbf{Q}}$ at the M-points as a function of inverse temperature $\beta$ for an $12 \times 12$ system. \textbf{Bottom}: The extrapolated staggered moment of magnetization, showing presence or absence of AFM long-range order for different values of the interaction anisotropy $\alpha$.}
    \label{fig:supp:weak coupling DQMC}
\end{figure}
\begin{figure}
    \centering
    \includegraphics[width=0.8\linewidth]{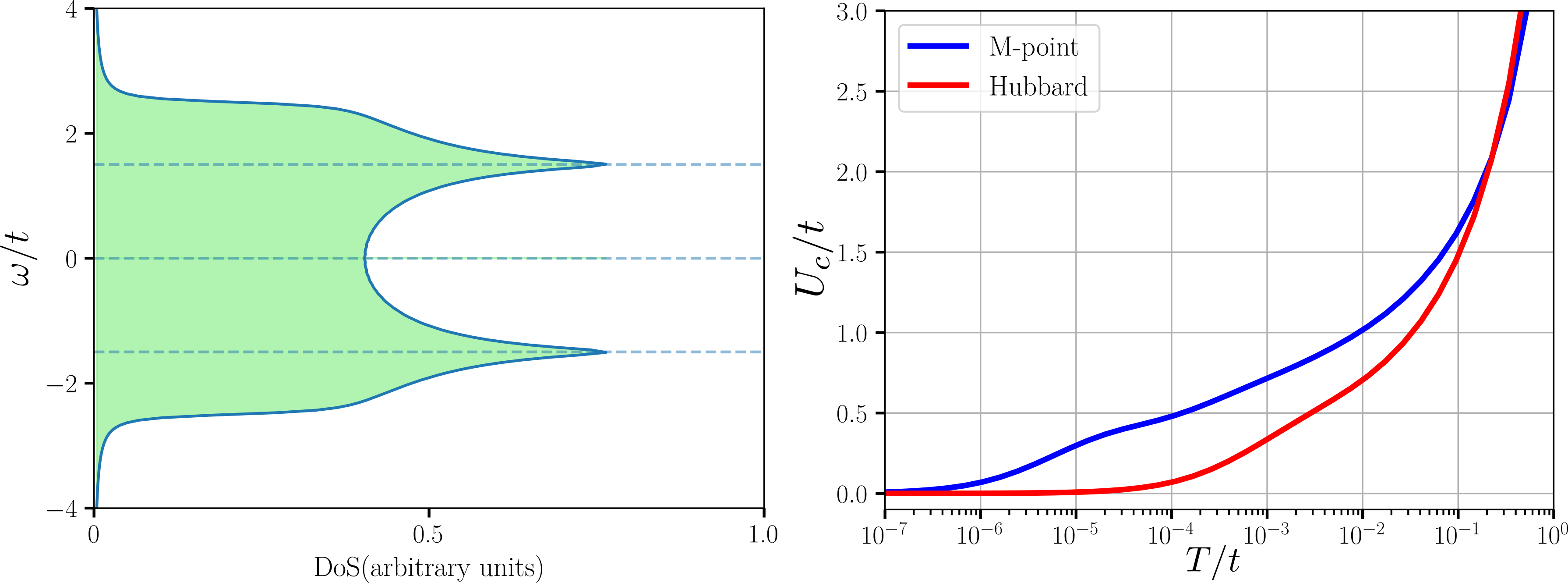}
    \caption{\textbf{Left}: The Density of States of the non-interacting system of Eq.~\ref{eq:supp:weak coupling:hamiltonian}. Notably, the vHs are away from $\omega=0$. \textbf{Right}: A comparison between the critical temperature $T_{\text{SDW}}^*$ within RPA for the square lattice Hubbard model and the $\text{AA t-SnSe}_2$ model, showing the suppression of $T_{\text{SDW}}^*$ in the latter. In both figures,$t_\perp=0.25t$ for the $\text{AA t-SnSe}_2$ model.}
    \label{fig:supp:RPA}
\end{figure}

As demonstrated in the rest of this work, the system is highly sensitive on the local interaction anisotropy $\alpha$ in the intermediate- and strong-coupling regimes. That is because, when the interaction terms dominate over the kinetic terms, the system  becomes very sensitive to the degeneracy of the local problem as $\alpha \rightarrow 1^-$, and valley-fluctuations play an important role. On the other hand, for the weak-coupling regime, physics is dominated by the kinetic term which breaks $U(6)$ due to it being valley-diagonal.
Thus, the local anisotropy plays a much less important role in the physics. This is reflected in the DQMC results of Fig.~\ref{fig:supp:weak coupling DQMC}, where for $\tilde{U}/W=0.4$, the $\alpha$ dependence on the spin correlations is much less dramatic compared to, for example Fig.~\ref{fig:supp: magnetic correlations: S(Q) vs beta}, and the scale at which a slater AFM forms is at least below $T=0.01t$ for all anisotropies $\alpha$. As $\tilde{U}/W$ increases, the system becomes more sensitive to $\alpha$ and already by $\tilde{U}/W=1.2$, AFM LRO can be detected at $T=0.01t$ only for $\alpha<0.90$.
We can qualitatively reproduce this behavior, through a very simple RPA calculation. Taking $V=0$, the Hamiltonian of the model is
\begin{equation}
\begin{split}
        \hat{H} &= \hat{H}_t + \sum_{\vec{R}} \frac{U}{2}\left(\sum_{\eta}\hat{n}_{\vec{R},\eta}^2 + 2\alpha\sum_{\eta > \eta'}\hat{n}_{\vec{R},\eta}\hat{n}_{\vec{R},\eta'}\right)\\
        \hat{H}_t &= \sum_{\mathbf{k},\eta,s}\epsilon^{(\eta)}(\mathbf{k})\hat{d}^\dagger_{\mathbf{k},\eta,s}\hat{d}_{\mathbf{k},\eta,s} \,,
\end{split}
\label{eq:supp:weak coupling:hamiltonian}
\end{equation}
with \begin{equation}
    \begin{split}
        \epsilon^{(0)}(\mathbf{k})&= -2t\cos{(k_y)}-2t_\perp \cos{(\sqrt{3}k_x)} - \mu\\
        \epsilon^{(\eta)}(\mathbf{k})&=\epsilon^{(0)}(C^{\eta}_{3z}\mathbf{k}) \,.\\
    \end{split}
\end{equation}
At half-filling ($\mu=0$), the non-interacting Fermi surfaces satisfy a nesting condition:
\begin{equation}
   \epsilon^{(\eta)}(\mathbf{k}) = -  \epsilon^{(\eta)}(\mathbf{k+Q^{(\eta)}})
\end{equation}
with $\mathbf{Q}^{\eta} = C^{\eta}_{3z}\left(\pi/\sqrt{3},\pi\right)$. This nesting vector is nothing but the AFM ordering vector encountered in the DQMC calculations in the main text. As mentioned above, the van Hove singularities are no longer at $\epsilon=0$, but at $\epsilon=\pm2|t-t_\perp|$, as seen in Fig.~\ref{fig:supp:RPA}). The absence of the singularity at $\epsilon=0$ will result in the suppression of the ordering temperature for the spin-density-wave (SDW) instability.

Working with the Green's function as defined in Eq.~\ref{eq:supp:implementation:green}, Fourier-transformed to Matsubara frequencies, the non-interacting bubble is
\begin{equation}
    [\Pi^{0}(\mathbf{q},i\nu)]_{(\eta,s),(\eta',s')}=T\sum_{\omega_n}\sum_{\mathbf{k}}\mathcal{G}^{(\eta)}_{s}(\mathbf{k+q},i(\omega_n+\nu))\mathcal{G}^{(\eta')}_{s'}(\mathbf{k},i\omega_n)=T\sum_{\omega_n}\sum_{\mathbf{k}}\mathcal{G}^{(\eta)}(\mathbf{k+q},i(\omega_n+\nu))\mathcal{G}^{(\eta')}(\mathbf{k},i\omega_n)
\end{equation}
which leads to a susceptibility
\begin{equation}
    [\chi^{0}(\mathbf{q},i\nu)]_{\eta_1,\eta_2;\eta_3,\eta_4} = -\frac{1}{N}\sum_{\mathbf{k}}\frac{f(\epsilon^{(\eta_1)}(\mathbf{k})) - f(\epsilon^{(\eta_2)}(\mathbf{k+q}))}{i\nu + (\epsilon^{(\eta_2)}(\mathbf{k}))-\epsilon^{(\eta_2)}(\mathbf{k+q}))} \delta_{\eta_1,\eta_3}\delta_{\eta_2,\eta_4}:= [\tilde{\chi}^{0}(\mathbf{q},i\nu)]_{\eta_1,\eta_2}\delta_{\eta_1,\eta_3}\delta_{\eta_2,\eta_4}\,.
    \label{eq:app:bare susc}
\end{equation}
The RPA spin-susceptibility, in matrix form, is~\cite{Kemper_2010,bjornson_21}:
\begin{equation}
    \chi^{\text{RPA}}_s(\mathbf{q},i\nu) = \chi^0(\mathbf{q},i\nu)\left(1-\hat{U}\chi^0(\mathbf{q},i\nu)\right)^{-1}
\end{equation} 
with this being a matrix equation in the valley-basis $(00,11,22,01,10,02,20,12,21)$. We have $\hat{U}=U\text{diag}\left(1,1,1,\alpha,\alpha,\dots,\alpha\right)$
and $\chi^{0} = \text{diag}(\tilde{\chi}^{0}_{\eta\eta'})$. Thus, the RPA spin-susceptibility can be re-written as
\begin{equation}
\begin{split}
    [\chi^{\text{RPA}}_s(\mathbf{q},i\nu)]_{\eta \eta} &=\frac{2[\tilde\chi^0(\mathbf{q},i\nu)]_{\eta,\eta}}{1-2U[\tilde\chi^0(\mathbf{q},i\nu)]_{\eta,\eta}}\\
    [\chi^{\text{RPA}}_s(\mathbf{q},i\nu)]_{\eta \eta'} &=\frac{2[\tilde\chi^0(\mathbf{q},i\nu)]_{\eta,\eta'}}{1-2\alpha U[\tilde\chi^0(\mathbf{q},i\nu)]_{\eta,\eta'}} \hspace{0.25cm}(\eta \neq \eta')\,.\\
\end{split}   
\label{eq:app:RPA}
\end{equation}

The first susceptibility in Eq.~\ref{eq:app:RPA} corresponds to the familiar spin-spin susceptibility, whereas the second one corresponds an \textit{inter-valley coherent}(IVC) spin susceptibility. Their respective forms, alongside Eq.~\ref{eq:app:bare susc} should make it clear that the Spin Density Wave (SDW) instability in the spin susceptibility is enhanced by the nesting, which is not the case in the spin IVC susceptibility. Thus, for $U>0$ the system becomes unstable in the SDW channel at a critical temperature $T_{\text{SDW}}^*(U)>0$  which is exponentially larger than the IVC critical temperature. Note that at this level of approximation, there are no valley-mixing diagrams in the $\eta\eta$ channel, leading to $T_{\text{SDW}}^*$ being independent of $\alpha$. While this is an artifact of the approximation, it does demonstrate the insensitivity of the system to $\alpha$ in the weak-coupling regime.
\par
As mentioned before, upon the inclusion of anisotropy ($t_\perp\neq t$), the vHS no longer coincide with the Fermi surface at half-filling. This leads to a critical temperature $T^*(U)$ that is much smaller that for the $t=t_\perp$ case of the Hubbard model. Hence, for small $U$, the temperature required to observe the insulator is probably too small to be achievable with DQMC, which is seen in Fig.~\ref{fig:supp:weak coupling DQMC} as the spin-correlations not saturating even down to $T=0.01t$. 
 \section{Strong-coupling effective model}
\label{section:supp:perturbation theory}
What happens in the ground-state of the strong-coupling regime, when the electrons have been localized on the lattice sites and all charge excitations are suppressed? As hinted at in the DQMC data, even in this regime, the local degeneracy brought on by $\alpha \approx 1$ affects the physics. To be specific, we have already seen in the DQMC data that for $\tilde{U}/W \gtrsim 5$ and low temperatures, the system has a clear gap to all charge excitations, and comfortably develops long-range AFM order. However, as seen in Figs.~\ref{fig:supp:additional data:local fluctuations} and in the main text, if $\alpha \approx 1$, there remains a degree of inter-valley correlation and a depletion of the local moment. In this section, we start off by deriving an effective spin model for the system for the $U\gg t,t_\perp$ regime (taking $V=0$ for simplicity). The resulting spin model makes the origin of the N\'eel AFM order explicit, and it also partially identifies that something interesting happens at $\alpha \rightarrow 1 ^-$. It should be noted that such a model has already been derived in Ref.~\cite{Li_2025} . We follow a similar approach here, placing particular emphasis on the local anisotropy. After constructing the spin model, we show that at least to second order in $t/U$, this cannot capture the sensitivity of the observables on $\alpha$, as it likely involves higher order terms associated with valley-fluctuation processes. 

\subsection{\texorpdfstring{Spin model for $\text{AA t-SnSe}_2$ in the strong-coupling regime}{Spin model for AA t-SnSe2 in the strong-coupling regime}}
We now consider the model for $\text{AA t-SnSe}_2$ studied in the rest of this work, given by Eq.~\ref{eq:supp:Hamiltonian}. We take the full interaction Hamiltonian $\hat{H}_U$ as the Hamiltonian associated with the `low-energy' manifold of states.Taking $V=0$,
\begin{equation}
    \begin{split}
        \hat{H}_U &= \frac{U}{2}\sum_{\vec{R}}\left(\sum_{\eta}\hat{n}_{\vec{R},\eta}^2 + 2\alpha\sum_{\eta \neq \eta'}\hat{n}_{\vec{R},\eta}\hat{n}_{\vec{R},\eta'}\right)\\
        \hat{H}_t &= \sum_{\vec{R},\Delta \vec{R}}\sum_{\eta,s} t^{\eta}_{\Delta \vec{R}} \hat{d}^\dagger_{\vec{R},\eta,s}\hat{d}_{\vec{R}+\Delta \vec{R},\eta,s} z,.\\
    \end{split}
    \label{supp:eqn:large U starting hamiltonian}
\end{equation}
When the local repulsions dominate ($U/t \rightarrow \infty$), the low energy Hilbert space consists of states which have exactly three electrons per site, which we denote as $\{ |\{\eta^i_\vec{R},s^i_\vec{R}\}_{i=1}^{3}\rangle \}$ with \begin{equation*}
     |\{\eta^i_\vec{R},s^i_\vec{R}\}_{i=1}^{3}\rangle = \prod_{\vec{R}}\prod_{i=1,2,3} \hat{d}^\dagger_{\vec{R},\eta_\vec{R}^i,s_\vec{R}^i}|0\rangle \,.
\end{equation*}
The local anisotropy $\alpha$ places further restrictions on the low energy Hilbert space: Consider $\alpha=1$, i.e. the case where the interaction is isotropic. In this case, all  $\begin{pmatrix} 6 \\ 3\end{pmatrix}=20$ ways of placing three electrons at each site are degenerate and furnish the low-energy subspace. For \textit{physical} anisotropic interactions $\alpha <1$, configurations with one electron per valley per site are preferred, leading to $\begin{pmatrix} 2\\ 1\end{pmatrix}^3 = 8$ low-energy states per site. Finally, for the case of the \textit{unphysical} anisotropic interactions with $\alpha >1$, the low-energy subspace is made up locally of the  $\begin{pmatrix} 3\\ 1\end{pmatrix}\begin{pmatrix} 2\\ 1\end{pmatrix} \begin{pmatrix} 2\\ 1\end{pmatrix}=12$ valley-imbalanced local states. Following a general Raleigh-Schrodinger perturbation theory (see Ref.~\cite{auerbach}), the effective Hamiltonian at second order will involve the hopping operator $H_t$ taking a state from the \textit{low-energy} subspace $P$ into the \textit{high-energy} subspace $Q$ (with associated projectors $\mathcal{P},\mathcal{Q}$). These excitations correspond to transitions between processes involving two sites with valley-occupations $\{n_{\eta}\}$ and $\{n_{\eta}'\}$. They can be described schematically, without loss of generality, as
\begin{equation}
    (n_1,n_2,n_3)\&(n'_1,n_2',n_3')\longrightarrow_{H_t} (n_1-1,n_2,n_3)\&(n'_1+1,n_2',n_3')
\end{equation}
with $\sum_{\eta}n_{\eta}=\sum_{\eta}n'_\eta=3$, as is appropriate for states in the low-energy subspace. The excitation energy of such processes is
\begin{equation}
    E_{\text{exc}} = U \left[1+(\alpha-1)(n_1-n_1')\right] \,.
\end{equation}
For the cases of $\alpha \leq 1$, while there might be multiple processes ($\alpha=1$) or only a single process (for $\alpha<1$, only $n_1 = n_1'$ is allowed), the excitation energy turns out to always be $E_{\text{exc}}=U$. For the $\alpha>1$ case, there are different processes with different excitation energies, which leads to quite a different effective spin model. While the system with unphysical anisotropy is quite interesting to study in its own right, we will ignore it beyond this point. For the $\alpha \leq 1$ case, the effective Hamiltonian can be written as:
\begin{equation}
\hat{H}_{\text{eff}} = \mathcal{P} \hat{H}_U\mathcal{P} - \mathcal{P} \hat{H}_t \sum_h \frac{|h\rangle \langle h |}{E_h}\hat{H}_t\mathcal{P}\,.
    \label{eq:supp:effective Hamiltonian} 
\end{equation}
This can be simplified by considering that $\hat{H}_U$ acts in the same way for all states in the low energy subspace, so it can be replaced by a number and absorbed as an overall constant. Further, since all exited states reachable via $\hat{H}_t$  have excitation energy $E_h=U$, the denominator can be taken out of the sum and
\begin{equation}
\hat{H}_{\text{eff}} =- \frac{1}{U}\mathcal{P}\hat{H}_t \mathcal{Q} \hat{H}_t\mathcal{P}\,.
    \label{eq:supp:effective Hamiltonian 2}
\end{equation}
Lastly, $\hat{H}_t$ either kills a state lying in $\mathcal{H}_l$ or raises it to the high-energy subspace $Q$, the $\mathcal{Q}$ projector acts trivially:
\begin{equation}
\hat{H}_{\text{eff}} = - \frac{1}{U}\mathcal{P} \hat{H}_t \hat{H}_t\mathcal{P} \,.
    \label{eq:supp:effective Hamiltonian 3}
\end{equation}
Expanding the kinetic Hamiltonian,
\begin{equation*}
    \hat{H}_{\text{eff}} = -\frac{1}{U}\mathcal{P}\sum_{(\vec{R},\Delta \vec{R}),(\vec{R}',\Delta \vec{R}')}\sum_{\eta,\eta'}\sum_{s,s'} t^{\eta}_{\Delta \vec{R}}t^{\eta'}_{\Delta \vec{R}'}\left(\hat{d}^\dagger_{\vec{R},\eta,s} \hat{d}_{\vec{R} + \Delta \vec{R},\eta,s}\hat{d}^\dagger_{\vec{R}' + \Delta \vec{R}',\eta',s'} \hat{d}_{\vec{R}',\eta',s'}\right)\mathcal{P}\,.
\end{equation*}
For this process to be non-zero, an electron must move from $\vec{R}\rightarrow \vec{R}+\Delta \vec{R}$ and then back $\vec{R}\leftarrow \vec{R}+\Delta \vec{R}$, meaning $\vec{R}=\vec{R}'$ and $\Delta \vec{R} = \Delta \vec{R}'$. Further, since $t^{\eta}_{\Delta \vec{R}}$'s direction is locked to $\eta$, the valleys are also fixed: $\eta = \eta'$.  Hence
\begin{equation}
    \hat{H}_{\text{eff}} = -\frac{1}{U}\mathcal{P}\sum_{\eta}\sum_{\vec{R},\Delta \vec{R}}\sum_{s,s'} (t^{\eta}_{\Delta \vec{R}})^2\left(\hat{d}^\dagger_{\vec{R},\eta,s} \hat{d}_{\vec{R}+\Delta \vec{R},\eta,s}\hat{d}^\dagger_{\vec{R}+\Delta \vec{R},\eta,s'} \hat{d}_{\vec{R},\eta,s'}\right)\mathcal{P}
    \label{eq:supp:effective Hamiltonian 4}
\end{equation}
and after re-ordering:
\begin{equation}
    \hat{H}_{\text{eff}} = -\frac{2(t^2+t_\perp^2)N_{\text{el}}}{U}+\frac{1}{U}\mathcal{P}\sum_{\eta}\sum_{\vec{R},\Delta \vec{R}}\sum_{s,s'} (t^{\eta}_{\Delta \vec{R}})^2\left(\hat{d}^\dagger_{\vec{R},\eta,s} \hat{d}_{\vec{R},\eta,s'}\hat{d}^\dagger_{\vec{R}+\Delta \vec{R},\eta,s'} \hat{d}_{\vec{R} + \Delta \vec{R},\eta,s}\right)\mathcal{P}\,.
    \label{eq:supp:effective Hamiltonian 6}
\end{equation}
Omitting the constant term and using a Fierz Identity for the SU(2) generators:
\begin{equation}
 \hat{H}_{\text{eff}} = \mathcal{P}\sum_{\eta}\sum_{\vec{R},\Delta \vec{R}}\sum_{s,s'} \frac{2(t^{\eta}_{\Delta \vec{R}})^2}{U} \left[ \hat{\mathbf{S}}^{(\eta)}_{\vec{R}}\cdot \hat{\mathbf{S}}^{(\eta)}_{\vec{R} + \Delta \vec{R}} + \frac{1}{4} n_{\vec{R},\eta}n_{\vec{R} + \Delta \vec{R},\eta}\right]\mathcal{P} \,.
    \label{eq:supp:effective Hamiltonian 5}
\end{equation}
Now, there is a major simplification for the case of $\alpha<1$: In this case, the low energy subspace always contains one electron per valley per site and so the density-density term is a constant and can be ignored. Overall:

\begin{equation}
    \hat{H}_{\text{eff}} =
\begin{cases}
\mathcal{P}\sum_{\eta}\sum_{\vec{R},\Delta \vec{R}}\sum_{s,s'} \frac{2(t^{\eta}_{\Delta \vec{R}})^2}{U} \left[ \hat{\mathbf{S}}^{(\eta)}_{\vec{R}}\cdot \hat{\mathbf{S}}^{(\eta)}_{\vec{R} + \Delta \vec{R}} \right] \mathcal{P}, & \alpha<1,\\[12pt]
\mathcal{P}\sum_{\eta}\sum_{\vec{R},\Delta \vec{R}}\sum_{s,s'}  \frac{2(t^{(\eta)}_{\Delta \vec{R}})^2}{U} \left[ \hat{\mathbf{S}}^{(\eta)}_{\vec{R}}\cdot \hat{\mathbf{S}}^{(\eta)}_{\vec{R} + \Delta \vec{R}} + \frac{1}{4} n_{\vec{R},\eta}n_{\vec{R} + \Delta \vec{R},\eta}\right] \mathcal{P}, & \alpha = 1 \,.
\end{cases}
\label{eq:supp::effective Hamiltonian 6}
\end{equation}
Strictly speaking, for the case of $\alpha\rightarrow1^-$, the gap between the valley-imbalanced and valley-singlet states goes to zero and $\hat{H}_{\text{eff}}^{(\alpha<1)}$ would acquire additional corrections. But the expression in Eq.~\ref{eq:supp::effective Hamiltonian 6} is enough for our purposes. It should also be noted that each active bond is counted twice: by $(\vec{R},\vec{R}+\Delta\vec{R})$ and $(\vec{R}+\Delta\vec{R},\vec{R}+\Delta\vec{R} - \Delta\vec{R})$ and so the super-exchange term associated with each bond is $2\times 2t^{\eta}_{\Delta\vec{R}}/U$, matching the familiar $J\equiv 4t^2/U$ scale.
\subsection{Local quantities deep in the strong-coupling regime}
In Fig.~\ref{fig:supp:additional data:local fluctuations} (b) of Sec.~\ref{section:supp:additional data}, we plot the local moment $\langle \hat{Z}^{(\eta)}_{\vec{R}}\hat{Z}^{(\eta)}_{\vec{R}}\rangle$ deep in the strong-coupling regime, where the itinerant quasiparticles present at weaker interactions are gapped, and the state is unambiguously insulating, with N\'eel AFM order. Despite this, the ground state physics still draw a sharp distinction between the $\alpha<1$ and $\alpha=1$ case, the latter showing a noticeably more depleted local moment. Even for $\alpha<1$, the local moment is not $1/2$, due to a small weight of doubly-occupied sites. Such processes are not unique to the M-point model as they are also present in the familiar Hubbard model. So, given the relevance of the valley-fluctuations to the system, we may ask the following question: Can the depleted local moment for $\alpha=1$ be understood through the same mechanisms ---of doublon/holon excitations--- as for $\alpha<1$, or is there another mechanism related to valley-fluctuations which leads to the depletion?
\subsubsection{Perturbation theory around the Atomic limit}
One approach to calculating the local moment, or equivalently, the local intra-valley double occupancy
\begin{equation*}
    \begin{split}
        D&=\langle \hat{n}_{\eta,\uparrow} \hat{n}_{\eta,\downarrow}\rangle \,, \\
    \end{split}
\end{equation*}
is to consider corrections away from the atomic limit, generated by hopping. To do that, we utilize a Fermionic path integral formulation:
\begin{equation}
    \begin{split}
        S^{\text{At}}[\bar{d},d]&=\int_0^\beta d\tau \{ \sum_{\vec{R},\eta,s}\bar{d}_{\vec{R},\eta,s}\partial_\tau d_{\vec{R},\eta,s}+\sum_\vec{R} \left[U\sum_\eta n_{\vec{R},\eta}^2+V \sum_{\eta'>\eta}n_{\vec{R},\eta}n_{\vec{R},\eta'}-\mu n_\vec{R}\right]\}\\
        \delta S[\bar{d},d]&=\int_0^\beta d\tau \sum_{\vec{R},\Delta \vec{R},\eta,s}t^\eta_{\Delta \vec{R}}\bar{d}_{\vec{R},\eta,s}d_{\vec{R}+\Delta \vec{R},\eta,s} \\
    \end{split}
    \label{eq:supp:perturbation theory: action}
\end{equation}
We take the ‘bare' action to be $S_0=S^{\text{At}}$ and treat the hopping as a perturbation. The goal is to compute the correlations up to order $(t/U)^2$:
\begin{equation}
    \langle \hat{n}_{\eta,\uparrow} \hat{n}_{\eta,\downarrow} \rangle =\langle \hat{n}_{\eta,\uparrow} \hat{n}_{\eta,\downarrow} \rangle^{\text{At}}+\delta \langle \hat{n}_{\eta,\uparrow} \hat{n}_{\eta,\downarrow} \rangle
\end{equation}
To calculate this, we follow Ref.~\cite{Pairault_2000} and use the Grand Canonical potential $\Omega = -\frac{1}{\beta}\ln Z$ and the identity 
\begin{equation}
    \sum_\eta\langle \hat{n}_{\eta,\uparrow} \hat{n}_{\eta,\downarrow}\rangle = \frac{1}{N_\text{sites}}\frac{\partial \Omega}{\partial U}
\end{equation}
This can be seen by inspecting the action of Eq.~\ref{eq:supp:perturbation theory: action}: Crucially, this relation only holds when $V,\mu$ are only set to their ‘physical’ values $V=\alpha U,\mu=U(\alpha+1/2)$ \textit{after} the derivative $\frac{\partial}{\partial U} $ is taken.
With that in mind, we have:
\begin{equation}
    \begin{split}
        Z& =\int \mathcal{D}[\bar{d},d]e^{-S}\\
        &=Z_0 \langle e^{-\delta S}\rangle_0\\
        &=Z_0 \left(1-\langle\delta S\rangle_0+\frac{1}{2}\langle\delta S^2\rangle_0 +\dots\right) \,.
    \end{split}
\end{equation}
As we will shortly show, $\langle\delta S\rangle_0=0$ and hence:
\begin{equation}
\begin{split}
    \Omega &=-\frac{1}{\beta}\ln Z_0 -\frac{1}{2\beta}\langle\delta S^2\rangle_0+\dots\\
    \Omega &= \Omega_0 + \Omega_2 +\dots \hspace{6em}\,.
\end{split} 
\end{equation}
To show that $\langle\delta S\rangle_0=0$ and to simplify  $\langle\delta S^2\rangle_0=0$, we can take advantage of the locality of the $S_0$ action.:For a local operator,
\begin{equation}
    \langle \hat{O}_\vec{R} \rangle_{0} = \langle  \hat{O}_\vec{R} \rangle_{0,\vec{R}} \,,
\end{equation}
where $\langle \rangle_{0,\vec{R}}$ refers to averages with respect to the action at the $\vec{R}$ site only. Further, for $\vec{R}\neq \vec{R}' \neq \dots$, there exists a Wick's rule:
\begin{equation}
    \langle \hat{O}_\vec{R} \hat{O}_{\vec{R}'} \dots\rangle_{0} = \langle  \hat{O}_\vec{R} \rangle_{0,\vec{R}} \langle  \hat{O}_{\vec{R}'} \rangle_{0,\vec{R}'}\dots
\end{equation}
This immediately implies that $\langle \delta S\rangle_0 \sim \sum_{\vec{R} \neq \vec{R}'} \int d\tau \langle \bar{d}_{\vec{R},\tau} d_{\vec{R}',\tau}\rangle_0=\sum_{\vec{R}\neq \vec{R}'} \int d\tau \langle \bar{d}_{\vec{R},\tau}\rangle_{0,\vec{R}} \langle d_{\vec{R}',\tau}\rangle_{0,\vec{R}'} =0$
Further, the only non-vanishing term for $\langle (\delta S)^2\rangle_{0}$ is the process where an electron hops $\vec{R}\rightarrow \vec{R}'$ and then back $\vec{R} \leftarrow \vec{R}'$. Hence: 
\begin{equation}
\begin{split}
    \langle (\delta S)^2\rangle_0=&\sum_{\vec{R},\Delta \vec{R}}\sum_{\eta}\sum_{s,s'}t^{\eta}_{\Delta \vec{R}}t^{\eta}_{-\Delta \vec{R}} \int d\tau_1 d\tau_2 \langle \bar{d}_{\vec{R},\eta,s,\tau_1} d_{\vec{R}+\Delta \vec{R},\eta,s,\tau_1}\bar{d}_{\vec{R}+\Delta \vec{R},\eta,s',\tau_2}d_{\vec{R},\eta,s',\tau_2}\rangle_0\\
    =&-\sum_{\vec{R},\Delta \vec{R}}\sum_{\eta}\sum_{s,s'} t^{\eta}_{\Delta \vec{R}}t^{\eta}_{-\Delta \vec{R}} \int d\tau_1 d\tau_2 \langle \bar{d}_{\vec{R},\eta,s,\tau_1} d_{\vec{R},\eta,s',\tau_2}\rangle_{0,\vec{R}} \langle \bar{d}_{\vec{R}+\Delta \vec{R},\eta,s',\tau_2} d_{\vec{R}+\Delta \vec{R},\eta,s,\tau_1}\rangle_{0,\vec{R}+\Delta \vec{R}}\\
    =&-\sum_{\vec{R},\Delta \vec{R}}\sum_{\eta}\sum_{s,s'} t^{\eta}_{\Delta \vec{R}}t^{\eta}_{-\Delta \vec{R}} \int d\tau_1 d\tau_2 \mathcal{G}^{(\vec{R},\eta)}(\tau_1 - \tau_2)_{0,\vec{R}} \mathcal{G}^{(\vec{R}+\Delta \vec{R},\eta)}(\tau_2 - \tau_1)_{0,\vec{R}+\Delta \vec{R}}\delta_{s,s'}\\
    =&-2\beta\sum_{\vec{R},\Delta \vec{R}}\sum_{\eta} t^{\eta}_{\Delta \vec{R}}t^{\eta}_{-\Delta \vec{R}} \int d\tau \mathcal{G}^{(\eta)}(\tau) \mathcal{G}^{(\eta)}(- \tau)\\
    =&-2\beta N_{\text{sites}}\left(zt^2+z_\perp t_\perp^2\right)\sum_{\eta} \int d\tau \mathcal{G}^{(\eta)}(\tau) \mathcal{G}^{(\eta)}(- \tau) \,,\\
\end{split}
\end{equation}
where $\mathcal{G}$ is the \textit{single site} Green's function (with respect to $S_0$) and $z,z_\perp=2$ are the coordination numbers associated with the $t,t_\perp$ hoppings. Note that the equation knows about the valley-selective hopping structure of the model as $\eta,\eta'$ of the Green's functions are locked to each other: $\eta=\eta'$, as is the case in the effective spin model.

\begin{figure}
    \centering
    \includegraphics[width=0.8\linewidth]{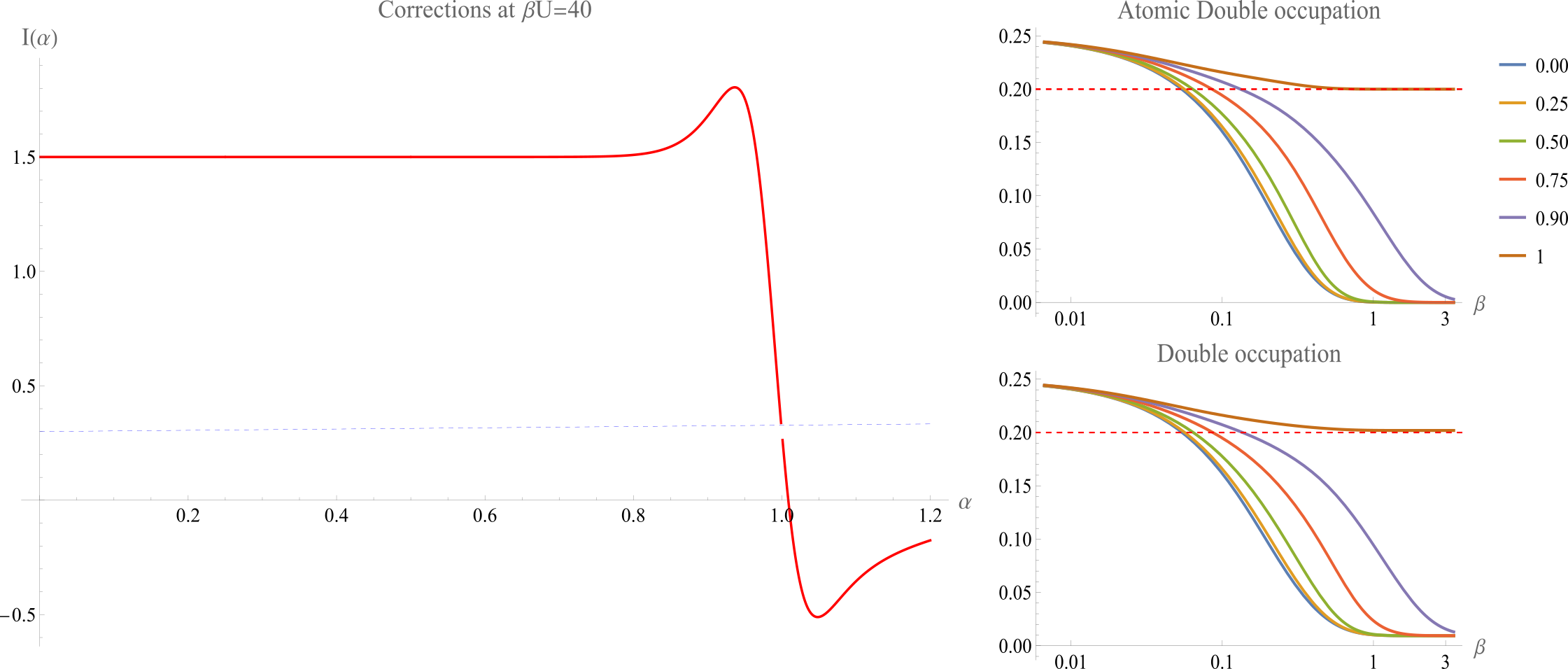}
    \caption{\textbf{Left}: The $I$ integral for $\beta U=40$, as a function of $\alpha$. For $\alpha <1$, $I$ takes the same value as in the single-orbital Hubbard model (times $3$, for the three orbitals), but Its behavior changes drastically around $\alpha=1$. \textbf{Right}: The Atomic Limit Double occupation (top) and the double occupation including corrections (bottom), for $U=15t,t_\perp=0.25t$ and different $\alpha$. }
    \label{fig:supp:perturbation theory: exact double occupancy}
\end{figure}

\noindent
The second-order corrections to the correlator are 
\begin{equation}
\begin{split}
     \delta \sum_{\eta}\langle n_{\eta \uparrow} n_{\eta\downarrow}\rangle &= \frac{1}{N_{\text{sites}}}\frac{\partial \Omega_2}{\partial U}\\
     &= \frac{1}{N_{\text{sites}}}\frac{\partial}{\partial U}\left[-\frac{1}{2\beta}\langle (\delta S)^2\rangle_0\right] \,.\\
\end{split}
\end{equation}
Hence, the double-occupancy is, to order $t^2$:
\begin{equation}
    \begin{split}
        \langle n_{\eta \uparrow} n_{\eta\downarrow}\rangle &= \langle n_{\eta \uparrow} n_{\eta\downarrow}\rangle^{\text{At}}+\frac{z(t^2+t_\perp^2)}{N_{\eta}}\int_{0}^{\beta} d\tau \sum_{\eta}\left[\frac{\partial}{\partial U}\mathcal{G}^{(\eta)}(\tau)\mathcal{G}^{(\eta)}(-\tau)\right]_{V=\alpha U,\mu=U(\alpha+1/2)}\\
        \langle n_{\eta \uparrow} n_{\eta\downarrow}\rangle &= \langle n_{\eta \uparrow} n_{\eta\downarrow}\rangle^{\text{At}}+\frac{z(t^2+t_\perp^2)}{N_{\eta}}I(\alpha,\beta,U)\\
        &= \langle n_{\eta \uparrow} n_{\eta\downarrow}\rangle^{\text{At}}+\frac{z(t^2+t_\perp^2)}{N_{\eta}U^2}f(\beta U,\alpha) \,.\\
    \end{split}
\end{equation}
The $I$ integral has dimensions of $E^{-2}$ and can be written as $U^{-2} f(\beta U,\alpha)$. This expression, with the analytical form of $I$, has been checked against Eq.(33) of Ref.~\cite{Pairault_2000} for the $N_\eta=1$ case, and they indeed agree. The results are summarized in Fig.~\ref{fig:supp:perturbation theory: exact double occupancy}, which should be contrasted with Fig.~\ref{fig:main: local charge correlations}. The $\beta U \rightarrow \infty$ values of the double occupancy for $\alpha<1$ agree: $\langle n_{\eta \uparrow} n_{\eta\downarrow}\rangle \rightarrow 0.01$, and the energy scales at which the double occupation changes match. However, the perturbation theory here does not capture the second depletion of $\langle n_{\eta \uparrow} n_{\eta\downarrow}\rangle$  for $\alpha=1$, around $T \lesssim J$. Hence, we conclude that the mechanism (which we argue is valley-fluctuations) that leads to the depletion of  the local moment in the isotropic interaction limit, cannot be captured by second order perturbation theory.
\subsubsection{Valley-fluctuation ring-exchange processes}
\begin{figure}
    \centering
    \includegraphics[width=0.4\linewidth]{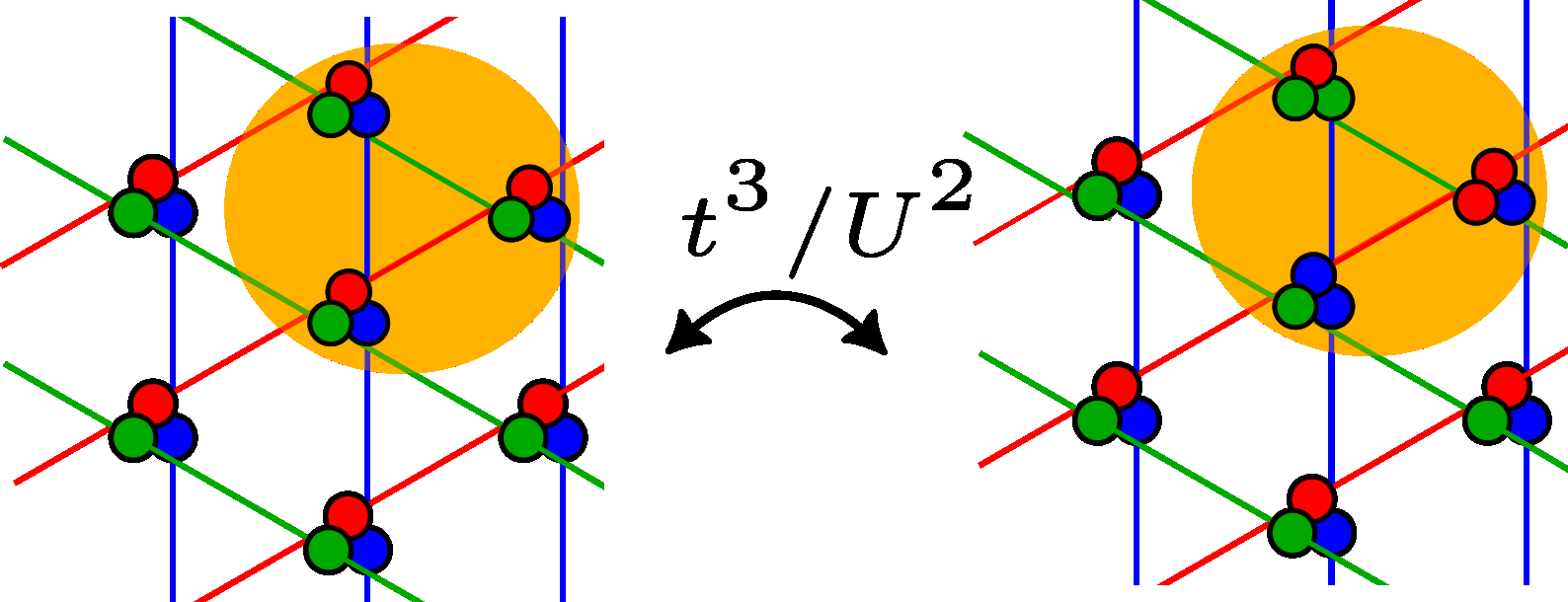}
    \caption{An example of a third-order process in perturbation theory involving hoppings of all three valleys, connecting states with $(111)$ and $(210)$ local configurations. Such a process should be though of as creating local defects.}
    \label{fig:supp:strong coupling: mechanism}
\end{figure}

Deep in the strong-coupling regime, the data shows a very ‘clean' $(t/U)^2$ decay in the ground state for the valley-coupling correlator $C^{\text{conn.}}$ of Eq.~\ref{eq:supp:additonal:C_connected} suggesting a simple mechanism is at play. If we work with the assumption that this behavior is due to valley-fluctuations -- as we have seen they are ubiquitous in the system for $\alpha \approx 1$ -- then we can consider a higher-order term arising from the perturbation theory. The effective spin model derived in this section via perturbation theory has the following issue: The number of valley-$\eta$ electrons on each site ($\{\hat{n}_{\vec{R},\eta}\}$) is conserved. Within this model, the ground state is a state with $(111)$ electrons per site, or $n_{\vec{R},\eta}=1$. For the \textit{locally degenerate} $(210)$ configurations to affect the ground state physics, one needs to go to third order in perturbation theory. While we do not explicitly derive the full third order perturbation theory of the model, we can consider one of the terms it generates, which is illustrated in Fig.~\ref{fig:supp:strong coupling: mechanism}. Such ring-exchange terms connect triangles of $(111),(111),(111)$ states (labeled $|0\rangle$) with triangles of  $(210),(102),(012)$ states (labeled $|1\rangle$). The latter can be thought of as defects, with an energy $\epsilon$ above the $(111)$ states. The energy cost is associated both with breaking the bonds of the triangle connecting to the background state, and also the local anisotropy cost:
\begin{equation*}
\epsilon = c_1 t^2/U + c_2 (1-\alpha)U \,.
\end{equation*}
Crucially, in the strong-coupling regime $U/t \gg 1$, this cost is huge and hybridization is essentially zero unless $\alpha=1$ (or $(1-\alpha)U \lesssim t^2/U$). In this case, the  ground state will be hybridized due to the valley fluctuation processes. We focus here on the (nearly) isotropic interaction case. A simple toy model Hamiltonian is thus:
\begin{equation}
H = \begin{pmatrix}
 0 & \tilde{t} \\ 
 \tilde{t} & \epsilon
\end{pmatrix} 
\end{equation}
with $\tilde{t}$ the matrix element between $|0\rangle$ and $|1\rangle$. From simple perturbation-theory arguments, this should be $\tilde{t}\sim t^3/U^2$. Solving for the ground state,
\begin{equation}
|\text{GS}\rangle = \frac{1}{\sqrt{1+(\lambda_{-}/\tilde{t})^2}}\left( |0\rangle + (\lambda_{-} / \tilde{t}) |1\rangle \right)
\end{equation}
with $\lambda_- = \frac{\epsilon}{2}[1-\sqrt{1+(2\tilde{t}/\epsilon)^2}]$.  The \textit{change} in the expectation value of an operator due to the hybridization is 
 \begin{equation}
 \delta \langle \hat{O}\rangle = \langle \text{GS}| \hat{O}|\text{GS} \rangle - \langle 0| \hat{O}|0 \rangle = \left(\frac{\lambda_-}{\tilde{t}}\right)^2\left(\langle 1| \hat{O}|1 \rangle - \langle 0| \hat{O}|0 \rangle\right)+\dots
 \end{equation}
 For the chain-correlator $\hat{O}=\frac{1}{3}\sum_{\eta>\eta'}\sum_{s,s'}\hat{n}_{\eta,s}\hat{n}_{\eta',s'}$, we can easily show that $\langle 0| \hat{O}|0 \rangle=1$ and $\langle 1| \hat{O}|1 \rangle=2/3$ and so:
 \begin{equation}
 \delta C^{\text{conn.}} = -|c| \left(\frac{t}{U}\right)^2 \,,
 \end{equation}
 with  a $\mathcal{O}(1)$ \textit{positive} prefactor $|c|$. This reproduces the behavior seen in the DQMC data (Fig.~\ref{fig:supp:additional data: large-U valley-coupling}). While this calculation does not utilize the full third order perturbation theory, it is quite suggestive that this is the mechanism at play, leading to the distinct behavior of the isotropic-interaction system in the strong-coupling limit. Hence, valley-fluctuations seem to play an important role not only in the intermediate-coupling regime or the high-$T$ strong-coupling regime, but also in the strong-coupling ground state physics. 
  \section{Parton Mean field}
\label{section:supp:parton mean field}
In this work so far, we have identified the valley-fluctuations near the interaction isotropic ($\alpha=1$) point as having a central role in the physics of $\text{AA t-SnSe}_2$; they allow for the metallic state to be stabilized up to unusually large interaction strengths $\tilde{U}/W$. All this is based on the DQMC data, which does paint a mostly-complete and self-consistent picture of the effect of valley fluctuations. However, It is always instructive to recover these effects via analytic means from simpler, approximate models, which is the purpose of this Appendix. Here, we construct a Parton decomposition of the electrons into charge-carrying rotors and charge-neutral spinons, inspired by the work of Florens and Georges~\cite{Florens_2002,Florens_2004}, hereby referred to as FG. This decomposition places the charge- and valley-fluctuations front and center and --- by solving the resulting model via Mean-Field --- qualitatively recovers many of the physics we see via DQMC. As a historical note, the original method developed in Refs.~\cite{Florens_2002,Florens_2004} strictly holds for $U(N)$-Hubbard models. In this work, we straightforwardly modify their approach to fit our $U(2)^{\otimes 3}$ model. Such a modification has (to the best of our knowledge) not been considered before in the literature pertaining to multi-orbital Hubbard models, many of the works opting instead to use constrained auxiliary spin operators~\cite{DeMedici_2005,Hassan_2010,Yu_2012,Crispino_2023}, doing away with the quantum rotor. One plausible explanation, which once again highlights the uniqueness of moir\'e materials, is that a multi-orbital Hubbard model originating from atomic orbitals will generically have significant pair-hopping and Hund's coupling terms, of magnitude comparable to the Hubbard $U$. In such a setting, the rotor description loses its computational and explanatory advantages. For $\text{AA t-SnSe}_2$ however, the moir\'e scale naturally suppresses pair-hopping and Hund's which then means the rotor representation described below becomes the correct description of the physics. \par
Being more concrete, we start by showing in detail the exact Hilbert-space mapping between electrons and rotors/spinons and how well the mapping holds on the atomic level once certain constraints are relaxed. We then take on the full lattice problem, solving it via the simplest possible mean-field decoupling. The resulting Hamiltonian already makes the physical mechanism extremely clear: charge-fluctuations are tied to valley-fluctuations via the hopping processes. As the interaction anisotropy $\alpha$ is tuned to $1$, the isotropic point, the valley-fluctuation stiffness becomes soft, leading to large charge-fluctuations and stabilization of the metallic phase.\par

\subsection{\texorpdfstring{Quantum Rotor description of an atomic system with $U(2)^{\otimes N_\eta}$ symmetry}{Quantum Rotor description of an atomic system with U(2) N eta symmetry}}
The starting point of this discussion is to consider the atomic problem and  construct a new basis for the local Hilbert space which is suggestive of the valley-fluctuation physics that we think are driving the metal to insulator transition. Consider the atomic problem for $N_\eta$ valleys of spinful electrons $\hat{H}^{\text{At}}= U/2\sum_{\eta,\eta'}\hat{n}_{\eta}\mathcal{V}_{\eta,\eta'}\hat{n}_{\eta'}$, with $\hat{n}_{\eta}=\sum_{s}\hat{n}_{\eta,s}$. The interaction matrix $\mathcal{V}$ takes the form
\begin{equation}
    \mathcal{V} = \begin{pmatrix}
    1 & \alpha & \alpha &\dots \\
    \alpha & 1 & \alpha & \dots \\
    \alpha & \alpha & 1 & \dots \\
    \dots & \dots & \dots & \dots
    \end{pmatrix}
\end{equation}
For $\alpha \neq 1$ this matrix is full-rank and has an eigenvalue $1+(N_\eta-1)\alpha$, as well as an $\left(N_\eta-1\right)$-fold degenerate eigenvalue $1-\alpha$. The first eigenvalue is associated with the valley-space vector $\mathbf{n} = (1,1,\dots)$ and corresponds to the total charge mode. The Hamiltonian can be written in terms of independent modes as:
\begin{equation}
    \hat{H}^{\text{At}} = \frac{U}{2}\left(1+(N_\eta-1)\alpha\right)\frac{1}{N_\eta}\hat{N}_{\text{tot}}^2 + \frac{U}{2}\left(1-\alpha\right)\sum_{m=1}^{N_\eta-1}\frac{1}{m\left(m+1\right)}\hat{Q}_m^2 \,,
    \label{eq:supp:PartonMeanField: Atomic Hamiltonian 1}
\end{equation}
with \begin{equation}
\begin{split}
    \hat{Q}_m &=  \sum_{\kappa=1}^{m}\hat{n}_{\kappa} - m\hat{n}_{m+1}\\
    \hat{N}_{\text{tot}} &= \sum_{\eta=1}\hat{n}_\eta\,.\\
\end{split}
\label{eq:supp:PartonMeanField: Atomic Hamiltonian transformation}
\end{equation}
Note that for $\alpha=1$, the rank of the matrix collapses and Eq.~\ref{eq:supp:PartonMeanField: Atomic Hamiltonian 1} reduces to 
\begin{equation*}
    \hat{H}^{\text{At}}=\frac{U}{2}\hat{N}_{\text{tot}}^2\,
\end{equation*}
the spectrum now depending only on the $\hat{N}_{\text{tot}}$ operator, despite the local Hilbert space being $2^{2N_{\eta}}$-dimensional. For $\alpha \neq 1$, the Hamiltonian of Eq.~\ref{eq:supp:PartonMeanField: Atomic Hamiltonian 1} breaks up into independent fluctuation modes and makes clear how to introduce rotors to the system. If one includes a chemical potential term on top of this, with the convention such that $\epsilon_0=0$ corresponds to half-filling: 
\begin{equation}
    \hat{H}^{\text{At}} = \epsilon_0 \hat{N}_{\text{tot}}+\frac{U}{2}\left(1+(N_\eta-1)\alpha\right)\frac{1}{N_\eta}\left(\hat{N}_{\text{tot}}-N_\eta\right)^2 + \frac{U}{2}\left(1-\alpha\right)\sum_{m=1}^{N_\eta-1}\frac{1}{m\left(m+1\right)}\hat{Q}_m^2\,,
    \label{eq:supp:PartonMeanField: Atomic Hamiltonian 1p5}
\end{equation}
and the spectrum can be reproduced using $N_\eta$ quantum rotors and the following Hamiltonian:
\begin{equation}
    H^{\text{At}} = \epsilon_0 \sum_{\eta,s}\hat{f}^\dagger_{\eta,s}\hat{f}_{\eta,s}+\frac{U}{2}\frac{\left(1+(N_\eta-1)\alpha\right)}{N_\eta}\hat{L}^2 + \frac{U}{2}\left(1-\alpha\right)\sum_{m=1}^{N_\eta-1}\frac{1}{m\left(m+1\right)}\hat{L}_m^2 \,.
    \label{eq:supp:PartonMeanField: Atomic Hamiltonian 2}
\end{equation}
Here, $\hat{f}_{\eta,s}$ are the \textit{charge-neutral} fermions (spinons) and $\hat{L}_{(m)} \equiv -i \partial /\partial \hat{\vartheta}_{(m)}$ is the angular momentum conjugate to the $\hat{\vartheta}_{(m)}$ rotor. Since the angular momentum spectrum is $\mathbb{Z}$ while the spectrum of $\hat{Q}_m$ is finite (the spectrum of each $\hat{n}_\eta$ is $\{0,1,2\}$), constraints must be imposed on the operators to insure the number of spinons in the system matches the angular momenta values:
\begin{equation}
    \begin{split}
    \hat{L} &=\sum_{\kappa}\sum_s(\hat{f}^\dagger_{\kappa,s}\hat{f}_{\kappa,s}-1/2)\\
        \hat{L}_m &= \sum_{\kappa=1}^{m}\sum_s\hat{f}^\dagger_{\kappa,s}\hat{f}_{\kappa,s}-m\sum_s\hat{f}^\dagger_{m+1,s}\hat{f}_{m+1,s}\,.\\
    \end{split}
    \label{eq:supp:PartonMeanField: Atomic Hamiltonian 3:constraint}
\end{equation}
To raise the angular momentum of a mode by $+1$, one can utilize $[\hat{L}_{(m)},e^{i\hat{\vartheta}_{(m)}}] = e^{i\hat{\vartheta}_{(m)}}$
which implies that $e^{i\hat{\vartheta}_{(m)}}$ act as \textit{raising operators}. Expressing the number operators $\hat{n}_\eta$ in the eigenmode basis as $\hat{n}_{\eta}=\mathcal{U}_{\eta m}\hat{Q}_m$,\footnote{with $\hat{Q}_0 \equiv \hat{N}_\text{tot}$} one gets the Hilbert space mapping
\begin{equation}
    \hat{d}^\dagger_{\eta,s} = \hat{f}^\dagger_{\eta,s}e^{i\mathcal{U}_{\eta m}\hat{\vartheta}_m}\,.
\end{equation}\par
\noindent
We now specialize to the model for $\text{AA t-SnSe}_2$ with $N_\eta=3$ valleys. For convenience, here we use slightly different notation compared to above. The Atomic Hamiltonian takes the form
\begin{equation}
    \begin{split}
        \hat{H}^{\text{At}} &=\epsilon_0 \sum_{\eta,s}\hat{d}^\dagger_{\eta,s}\hat{d}_{\eta,s} + \frac{U}{2}\sum_{\eta}\left(\hat{n}_\eta -1\right)^2 + \alpha U\sum_{\eta>\eta'}\left(\hat{n}_{\eta}-1\right)\left(\hat{n}_{\eta'}-1\right)\\
          &= \epsilon_0 \sum_{\eta,s}\hat{d}^\dagger_{\eta,s}\hat{d}_{\eta,s} + \frac{U(1+2\alpha)}{6}(\hat{n}_0+\hat{n}_1+\hat{n}_2-3)^2+\frac{U(1-\alpha)}{12}(\hat{n}_0+\hat{n}_1-2\hat{n}_2)^2+\frac{U(1-\alpha)}{4}(\hat{n}_0-\hat{n}_1)^2\\
          &= \epsilon_0 \hat{Q}_{\text{ch}} + \frac{U_{\text{ch}}}{2}(\hat{Q}_{\text{ch}}-3)^2+\frac{U_+}{2}\hat{Q}_{+}^2+\frac{U_-}{2}\hat{Q}_{-}^2 \,.\\
    \end{split}
    \label{eq:supp:parton:ham}
\end{equation}
Here, $\hat{Q}_{\text{ch}} = \sum_{\eta}\hat{n}_\eta$, $\hat{Q}_{+} = \hat{n}_0+\hat{n}_1 - 2\hat{n}_2$, $\hat{Q}_{-} = \hat{n}_0-\hat{n}_1$,$U_{\text{ch}} = \frac{1+2\alpha}{3}U$, $U_+ = \frac{1-\alpha}{6}U$ and $U_- = \frac{1-\alpha}{2}U$. Notice that these charge modes are not normalized, in contrast to Eq.~\ref{eq:supp:PartonMeanField: Atomic Hamiltonian 1p5}.
This Hamiltonian has a spectrum identical to the following Hamiltonian of rotors and spinons:
\begin{equation}
\begin{split}
    \hat{H}^{\text{At}} &= \epsilon_0 \sum_{\eta,s}\hat{f}^\dagger_{\eta,s}\hat{f}_{\eta,s} + \frac{U_{\text{ch}}}{2}\hat{L}_{\text{ch}}^2+\frac{U_+}{2}\hat{L}_+^2+\frac{U_-}{2}\hat{L}_-^2 \,,\\
    \label{eq:supp:PartonMeanField: Atomic Hamiltonian 4}
\end{split}
\end{equation}
subject to the constraints
\begin{equation}
    \begin{split}
        \hat{L}&=\sum_{\eta,s}\left(\hat{f}^\dagger_{\eta,s}\hat{f}_{\eta,s}-1/2\right)\\
        \hat{L}_+&=\sum_s\left(\hat{f}^\dagger_{0,s}\hat{f}_{0,s} + \hat{f}^\dagger_{1,s}\hat{f}_{1,s}-2\hat{f}^\dagger_{2,s}\hat{f}_{2,s}\right)\\
        \hat{L}_-&=\sum_s\left(\hat{f}^\dagger_{0,s}\hat{f}_{0,s} - \hat{f}^\dagger_{1,s}\hat{f}_{1,s}\right) \,.\\
    \end{split}
    \label{eq:supp:PartonMeanField: Atomic Hamiltonian 4:constraints}
\end{equation}
Being explicit, the $\hat{L}_{x}$ (with $x \in \{\text{ch},+,-\}$) are the angular momenta associated with the $O(2)$ rotors $\hat{\vartheta}_x$, with $\hat{L}_x \equiv -i\partial/\partial \hat{\vartheta}_x$, representing the charge-fluctuations (the \textit{ch} channel) and the charge-neutral valley-fluctuations (the $\pm$ channels). They satisfy the following commutation relations:
\begin{equation}
    [\hat{L}_{x},\hat{\vartheta}_y] = -i \delta_{xy},\qquad [\hat{L}_{x},\hat{L}_y] = 0, \qquad [\hat{\vartheta}_{x},\hat{\vartheta}_y] = 0, \qquad [\hat{L}_{x},e^{i\hat{\vartheta}_y}] = \delta_{xy}e^{i\hat{\vartheta}_y} \,.
    \label{eq:supp:PartonMeanField:commutation relations 1}
\end{equation}
The last commutation relation of Eq.~\ref{eq:supp:PartonMeanField:commutation relations 1} implies that $\hat{L}_x$ should be understood as raising-operators associated with charge channel $x$. An equivalent representation of the physics exists in the `valley' basis:
\begin{equation}
    \begin{split}
        \hat{L}_{x}& =A_{x\eta}\hat{\ell}_{\eta}\\
        \hat{\vartheta}_{x}&=[A^T]^{-1}_{x \eta} \hat{\varphi}_\eta\\
        &\\
        \text{ with } A &= \begin{pmatrix}
            +1 & +1 & +1 \\ +1 & + 1 & -2 \\ + 1 & -1 & 0 \\
        \end{pmatrix} \,,
    \end{split}
    \label{eq:supp:PartonMeanField:change of basis}
\end{equation}
where, again, $x \in \{\text{ch},+,-\}$ and $\eta \in \{0,1,2\}$. The $\eta$-indexed rotors satisfy commutation relations analogous to those of Eq.~\ref{eq:supp:PartonMeanField:commutation relations 1}:
\begin{equation}
    [\hat{\ell}_{\eta},\hat{\varphi}_{\eta'}] = -i \delta_{\eta \eta'},\qquad [\hat{\ell}_{\eta},\hat{\ell}_{\eta'}] = 0, \qquad [\hat{\varphi}_{\eta},\hat{\varphi}_{\eta'}] = 0, \qquad [\hat{\ell}_{\eta},e^{i\hat{\varphi}_{\eta'}}] = \delta_{\eta \eta'}e^{i\hat{\varphi}_{\eta'}} \,.
    \label{eq:supp:PartonMeanField:commutation relations 2}
\end{equation}
Here, $\hat{\ell}_{\eta}$ has the clear interpretation of adding charge with valley $\eta$. Further, the `valley' basis allows for a straightforward representation of the creation operators
\begin{equation}
    \hat{d}_{\eta,s}^\dagger \equiv \hat{f}^\dagger_{\eta,s}e^{i\hat{\varphi}_{\eta}}\,.
     \label{eq:supp:PartonMeanField:operator level map 1}
\end{equation}
\par
\begin{figure}
    \centering
    \includegraphics[width=0.8\linewidth]{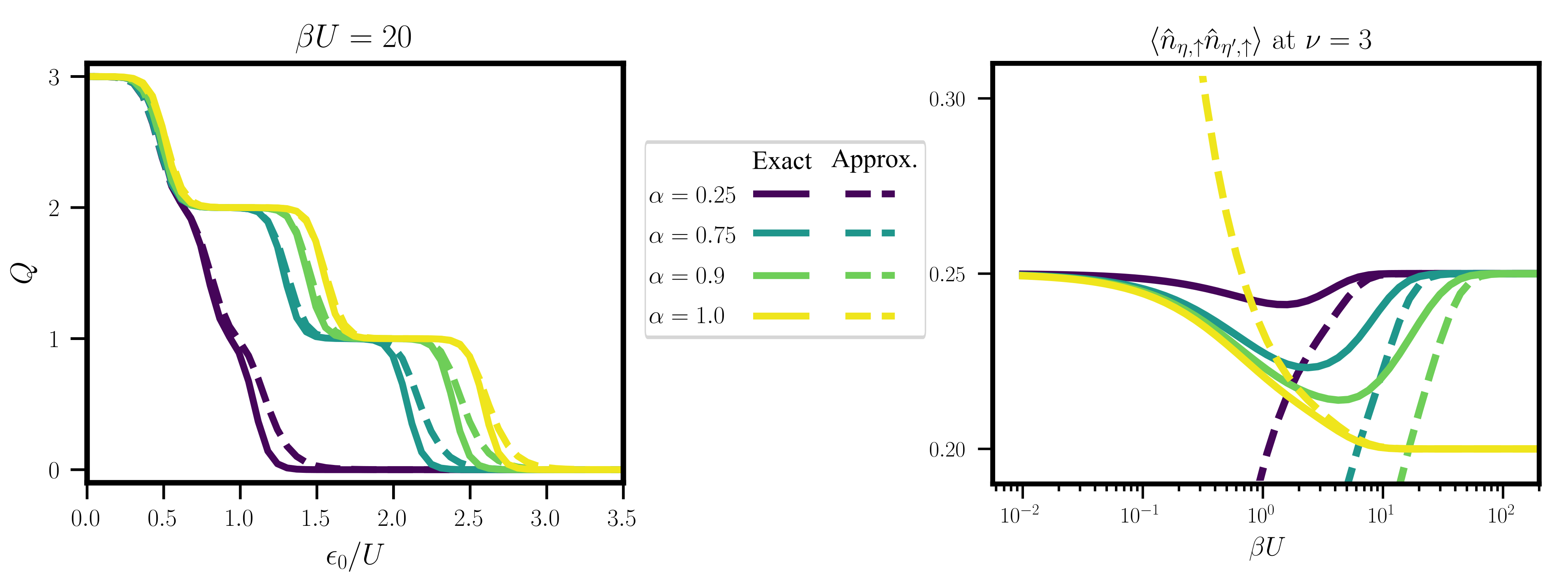}
    \caption{Atomic-limit observables calculated both in the exact prescription (full lines) and in the approximate rotor description (dashed lines). \textbf{Left:} The total charge occupation $Q$ of a single site as a function of $\alpha$ and on-site energy $\epsilon_0$. \textbf{Right:} The inter-flavor correlations at half-filling as a function of inverse temperature $\beta U$.}
    \label{fig:supp:perturbation theory:atomic rotor results}
\end{figure}

\noindent
Up to this point, the two Hilbert spaces are identical and the mapping is exact. We now proceed by imposing the constraints of Eq.~\ref{eq:supp:PartonMeanField: Atomic Hamiltonian 4:constraints} \textit{only on average}. This means that the spectra of the rotor angular momenta include all values in  $\mathbb{Z}$, including the unphysical values. As we proceed to show, it is important that the weight of a state in these unphysical values is suppressed. To implement the constraints on average, we introduce Lagrange multipliers $h_{\text{ch}},h_+,h_-$ and consider the following decoupled Hamiltonians:
\begin{equation}
    \begin{split}
        \hat{H}^{\text{At}}_f &= \left(\epsilon_0-(h_{\text{ch}}+h_++h_-)\right)\sum_s \hat{f}^\dagger_{0,s}\hat{f}_{0,s}+\left(\epsilon_0-(h_{\text{ch}}+h_+ -h_-)\right)\sum_s \hat{f}^\dagger_{1,s}\hat{f}_{1,s}+\left(\epsilon_0-(h_{\text{ch}}-2h_+)\right)\sum_s \hat{f}^\dagger_{2,s}\hat{f}_{2,s}\\
        &\\
        \hat{H}^{\text{At}}_{\theta} &= \frac{U_{\text{ch}}}{2}\hat{L}_{\text{ch}}^2 + \frac{U_+}{2}\hat{L}^2_+ +  \frac{U_-}{2}\hat{L}^2_- + \left(h_{\text{ch}}\hat{L}_{\text{ch}} + h_+\hat{L}_+ +  h_-\hat{L}_-\right)\\
    \end{split}
    \label{eq:supp:PartonMeanField: Atomic Hamiltonian 5}
\end{equation}
with the Lagrange multipliers determined via the average constraint equations
\begin{equation}
    \begin{split}
        \langle \hat{L}_{\text{ch}}\rangle_{\theta}&=2n_{\text{F}}[\epsilon_0-\left(h_{\text{ch}}+h_++h_-\right)]+2n_{\text{F}}[\epsilon_0-\left(h_{\text{ch}}+h_+ -h_-\right)] +2n_{\text{F}}[\epsilon_0-\left(h_{\text{ch}}-2h_+\right)]-3\\
       \langle \hat{L}_+\rangle_{\theta}&=2n_{\text{F}}[\epsilon_0-\left(h_{\text{ch}}+h_++h_-\right)]+2n_{\text{F}}[\epsilon_0-\left(h_{\text{ch}}+h_+-h_-\right)]-4n_{\text{F}}[\epsilon_0-\left(h_{\text{ch}}-2h_+\right)]\\
        \langle \hat{L}_-\rangle_{\theta}&=2n_{\text{F}}[\epsilon_0-\left(h_{\text{ch}}+h_++h_-\right)]-2n_{\text{F}}[\epsilon_0-(h_{\text{ch}}+h_+-h_-)] \hspace{4cm}\,.\\
    \end{split}
    \label{eq:supp:PartonMeanField: Atomic Hamiltonian 5: constraint}
\end{equation}
Here, $n_F(\epsilon)=\left(1+e^{-\beta\epsilon}\right)^{-1}$ is the Fermi-Dirac distribution. The $C_{3z},C_{2x}$ symmetries present in the model (See SI Sec.~\ref{section:supp:model}) take the form of a valley-permutation $\hat{d}^\dagger_{\eta,s}\rightarrow d^\dagger_{\eta',s}$ in the atomic limit. Since we do not expect (or observe) any \textit{spontaneous valley-polarization} forming in the system for realistic parameters corresponding to $\text{AA t-SnSe}_2$ samples, we can restrict our treatment to $h_\pm=0$. At the atomic limit, this means there is a single self-consistent equation to solve: from Eqs.~\ref{eq:supp:PartonMeanField: Atomic Hamiltonian 5},\ref{eq:supp:PartonMeanField: Atomic Hamiltonian 5: constraint}, setting $h_\pm=0$:
\begin{equation}
    \langle \hat{L}_{\text{ch}}\rangle_\theta = 6n_{\text{F}}[\epsilon_0 - h_{\text{ch}}]-3 \,.
    \label{eq:supp:PartonMeanField: Atomic Hamiltonian final}
\end{equation}
In the valley-symmetric case of $h_\pm=0$,  it follows that $\langle\hat{L}_{\pm}\rangle_{\theta}=0$. Solving the atomic-limit problem in the rotor description while treating the constraints only on average, we can ask how well it matches the exact solution. To answer this, we compare the two approaches for some simple observables: (1)~the average occupation as a function of chemical potential $Q=6n_{\text{F}}[\epsilon_0 - h_{\text{ch}}(\epsilon_0;\alpha,T/U)]$ and (2)~the inter-valley double occupancy at half-filling $\langle \hat{n}_{\eta,\uparrow} \hat{n}_{\eta',\uparrow}\rangle$, computed here via the identity
\begin{equation}
\begin{split}
    \langle \hat{n}_{\eta,\uparrow} \hat{n}_{\eta',\uparrow}\rangle &= \frac{1}{4}\left(\frac{\langle \hat{N}^2\rangle}{9} - \frac{\langle (\hat{n}_0+\hat{n}_1-2\hat{n}_2))^2\rangle}{36}- \frac{\langle (\hat{n}_0-\hat{n}_1)^2\rangle}{12}\right)\\
    &=\frac{1}{4}\left(\frac{\langle \hat{L}_{\text{ch}}^2\rangle_\theta}{9}- \frac{\langle \hat{L}_+^2\rangle_\theta}{36} -\frac{\langle \hat{L}_-^2\rangle_\theta}{12}\right)\,.\\
    \end{split}
\end{equation}
The results are presented in Fig.~\ref{fig:supp:perturbation theory:atomic rotor results}. As in Ref.~\cite{Florens_2004}, for $T=0$ the two methods agree exactly, but as $T$ increases, the approximate method produces worse and worse results, because the `unphysical' parts of the rotor spectra contribute more. As a note: For $\alpha \rightarrow1^-$, the `cost' of large, unphysical values of $\hat{L}_\pm$ becomes vanishingly small and hence care must be taken to ensure an adequately large cutoff in the spectrum is used. At the $U(6)$ interaction-symmetric point of $\alpha=1$, the valley-fluctuation modes cost zero energy and hence one must revert back to the approach of FG. Strictly speaking, this specific Parton Mean field construction holds for $\alpha<1$. Any results presented here with $\alpha=1$ use the fully-$U(6)$ symmetric approach of FG.
\subsection{Mean-Field treatment of Parton Hamiltonian}
We can go beyond the atomic limit and write down the full lattice Hamiltonian using the parton description outlined above. While in parton language the Hamiltonian may appear more complicated that the original one,  we can solve it using a simple Mean-Field approximation decoupling the rotors from the spinons. On an operator level, we generalize the commutator relations of Eqs.~\ref{eq:supp:PartonMeanField:commutation relations 1} and~\ref{eq:supp:PartonMeanField:commutation relations 2} to include a lattice-site $\vec{R}$ index and Eq.~\ref{eq:supp:PartonMeanField:operator level map 1} to
\begin{equation}
    \hat{d}^\dagger_{\vec{R},\eta,s} \equiv \hat{f}^\dagger_{\vec{R},\eta,s}e^{i\hat{\varphi}_{\vec{R},\eta}}
     \label{eq:supp:PartonMeanField:operator level map 2} \,.
\end{equation}
The parton Hamiltonian takes the form
\begin{equation}
\begin{split}
    \hat{H} &= \hat{H}^{\text{At}} + \sum_{\vec{R},\Delta\vec{R}}\sum_{\eta,s}t^{\eta}_{\Delta \vec{R}}\hat{d}^{\dagger}_{\vec{R},\eta,s}\hat{d}_{\vec{R}+\Delta\vec{R},\eta,s}\\
    &=\hat{H}^{\text{At}}_{f} + \hat{H}^{\text{At}}_{\theta}+\sum_{\vec{R},\Delta\vec{R}}\sum_{\eta,s}t^{\eta}_{\Delta \vec{R}}\hat{f}^{\dagger}_{\vec{R},\eta,s}\hat{f}_{\vec{R}+\Delta\vec{R},\eta,s} e^{i\left(\hat{\varphi}_{\vec{R},\eta}-\hat{\varphi}_{\vec{R}+\Delta\vec{R},\eta}\right)}\\
    &\approx\hat{H}^{\text{At}}_{f} + \hat{H}^{\text{At}}_{\theta}+\sum_{\vec{R},\Delta\vec{R}}\sum_{\eta,s}t^{\eta}_{\Delta \vec{R}}\left(\langle \hat{f}^{\dagger}_{\vec{R},\eta,s}\hat{f}_{\vec{R}+\Delta\vec{R},\eta,s}\rangle e^{i\left(\hat{\varphi}_{\vec{R},\eta}-\hat{\varphi}_{\vec{R}+\Delta\vec{R},\eta}\right)}+\hat{f}^{\dagger}_{\vec{R},\eta,s}\hat{f}_{\vec{R}+\Delta\vec{R},\eta,s}\langle e^{i\left(\hat{\varphi}_{\vec{R},\eta}-\hat{\varphi}_{\vec{R}+\Delta\vec{R},\eta}\right)}\rangle\right)\\
    &=\hat{H}^{\text{At}}_{f} + \hat{H}^{\text{At}}_{\theta}+\sum_{\vec{R},\Delta\vec{R}}\sum_{\eta,s}t^{\eta}_{\Delta \vec{R}}\left(\langle \hat{f}^{\dagger}_{\vec{R},\eta,s}\hat{f}_{\vec{R}+\Delta\vec{R},\eta,s}\rangle \cos{\left(\hat{\varphi}_{\vec{R},\eta}-\hat{\varphi}_{\vec{R}+\Delta\vec{R},\eta}\right)}+\hat{f}^{\dagger}_{\vec{R},\eta,s}\hat{f}_{\vec{R}+\Delta\vec{R},\eta,s}\langle \cos{\left(\hat{\varphi}_{\vec{R},\eta}-\hat{\varphi}_{\vec{R}+\Delta\vec{R},\eta}\right)}\rangle\right) \,.\\
\end{split}
\label{eq:supp:PartonMeanField: Hamiltonian 1}
\end{equation}
In Eq.~\ref{eq:supp:PartonMeanField: Hamiltonian 1} between the second and third line we performed a mean-field decomposition and on the fourth line we assumed $\langle \hat{f}^\dagger_{\vec{R},\eta,s}\hat{f}_{\vec{R}+\Delta\vec{R},\eta,s}\rangle$ and $\langle e^{i\left(\hat{\varphi}_{\vec{R},\eta}-\hat{\varphi}_{\vec{R}+\Delta\vec{R},\eta}\right)}\rangle$ are real, i.e. no charge-current order develops. At this stage, the rotor Hamiltonian $\hat{H}_\theta$ is a Hamiltonian of three coupled $XY$ Hamiltonians on the lattice. We further simplify this by treating the rotor Hamiltonian itself at the Mean-Field level, corresponding to $\langle \cos{\hat{\varphi}_{\vec{R},\eta}}\rangle =\langle \cos{\hat{\varphi}_{\eta}}\rangle $. With this, the Hamiltonian takes the form
\begin{equation}
    \begin{split}
        \hat{H}_{f}&=\sum_{\vec{R}}\hat{H}^{\text{at}}_{f;\vec{R}} + \sum_{\vec{R},\Delta\vec{R}}\sum_{\eta,s}t^{\eta}_{\Delta\vec{R}}\langle \cos{\hat{\varphi}_\eta}\rangle_\theta^2 \hat{f}^{\dagger}_{\vec{R},\eta,s}\hat{f}_{\vec{R}+\Delta\vec{R},\eta,s}\\
        \hat{H}_{\theta}&=\sum_{\vec{R}}\hat{H}^{\text{at}}_{\theta;\vec{R}} + \sum_{\vec{R}}\sum_{\eta}K_{\eta}\cos{\hat{\varphi}_\eta}\\
        \text{ with } K_{\eta}&=2\langle \cos{\hat{\varphi}_\eta}\rangle_{\theta}\sum_{\vec{R},\Delta \vec{R}} \frac{1}{L_x L_y}\sum_s t^{\eta}_{\Delta \vec{R}} \langle \hat{f}^\dagger_{\vec{R},\eta,s}\hat{f}_{\vec{R}+\Delta \vec{R},\eta,s}\rangle_f \,.\\
    \end{split}
    \label{eq:supp:PartonMeanField: Hamiltonian 2}
\end{equation}
We can simplify these coupled equations by assuming the system does not spontaneously break valley-symmetry. This implies that, as in the atomic case, $h_{\pm}=0$ and further, $\langle \cos{\hat{\varphi}_\eta}\rangle_{\theta}= Q_{\eta}\equiv Q$. Also, the density of states $D^{(\eta)}(\epsilon) \equiv \int d\hspace*{-0.08em}\bar{}\hspace*{0.1em}^2\vec{k} \delta\left(\epsilon - \epsilon^{(\eta)}(\vec{k})\right)$ is actually valley-independent by $C_{3z}$ symmetry: $D^{(\eta)}(\epsilon) =D(\epsilon)$. With these considerations, we obtain the following equations for the spinon and rotor Hamiltonians:
\begin{equation}
    \begin{split}
        \hat{H}_f&=\sum_{\vec{k},\eta,s} [\left(\epsilon_0-h_{\text{ch}}\right) + Q^2\epsilon^{(\eta)}(\vec{k})]\hat{f}^\dagger_{\vec{k},\eta,s}\hat{f}_{\vec{k},\eta,s}\\
        \hat{H}_{\theta}&=h_{\text{ch}}\hat{L}_{\text{ch}} + \sum_{x \in \{\text{ch},+,-\}}\frac{U_x}{2}\hat{L}^2_x + K\sum_{\eta\in \{0,1,2\}}\cos{\hat{\varphi}_{\eta}}
    \end{split}
    \label{eq:supp:PartonMeanField: Hamiltonian 3}
\end{equation}
with self-consistency equations, at a target filling $\nu \in [0,6]$:
\begin{equation}
    \begin{split}
        \langle \hat{L}_{\text{ch}}\rangle_{\theta} &= \left(\nu-3\right)\\
        \nu &= 6\int d\epsilon D(\epsilon)n_{\text{F}}[\left(\epsilon_0-h_{\text{ch}}\right) + Q^2\epsilon]\\
        K &= 4Q\int d\epsilon D(\epsilon)\epsilon n_{\text{F}}[\left(\epsilon_0-h_{\text{ch}}\right) + Q^2\epsilon] \,.\\
    \end{split}
    \label{eq:supp:PartonMeanField: Hamiltonian 4}
\end{equation}
At the mean-field level, Eq.~\ref{eq:supp:PartonMeanField: Hamiltonian 3} describes charge- and valley-fluctuations, each with its own stiffness, coupled to each other via hopping. $Q^2$ should be interpreted as the quasiparticle weight and hence $Q=0$ signals the onset of a Mott-insulating phase. When $K\sim Q=0$, the rotor Hamiltonian $\hat{H}_{\theta}$ reverts to its atomic form and hence this approach does not treat the system deep in the Mott phase particularly well. Through these equations, the mechanism for the stabilization of metallic phase as $\alpha \rightarrow 1^-$ becomes clear: Despite the charge fluctuations consistently having a `large' stiffness $\sim (1+2\alpha)U$, hopping couples the charge-fluctuations with the valley fluctuations which themselves become soft as $\alpha \rightarrow 1^-$ since their stiffness is $\sim (1-\alpha)U$. Hence, via this coupling, the valley fluctuations make it harder for the charge fluctuations to quench, leading to the enhanced region of metallicity observed in the $\text{AA t-SnSe}_2$ system!\par
\begin{figure}
    \centering
    \includegraphics[scale=0.75]{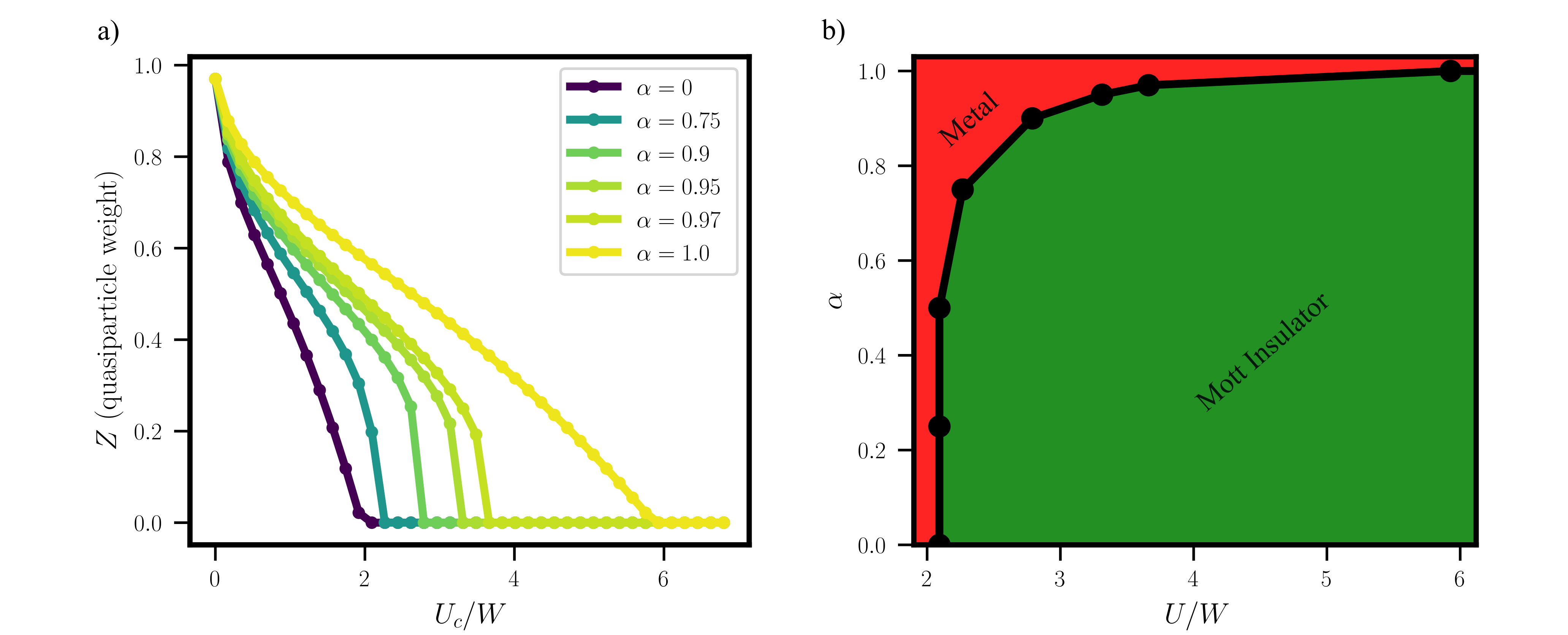}
    \caption{\textbf{Parton Mean Field calculations} at half-filling for $T=0$ and $t_\perp/t=0.25$.~(a) shows the spinon quasi-particle weight $Z=Q^2$ as a function of the Hubbard-$U$ over the bandwidth $W$ for different values of the interaction anisotropy;~(b) shows the critical $U_c$ at which system goes from metal to Mott, and should be compared to the DQMC crossover diagram Fig.~\ref{fig:main:Phase diagram}(c). For these simulations, the local Hilbert space of each valley-basis rotor $\hat{\ell}_\eta$ was truncated to a maximum angular momentum of $8$: $\{|\ell_\eta \rangle\}_{-8}^{+8}$}
    \label{fig:supp:PartonMeanField:MF rotor results 2}
\end{figure}

\noindent
Solving these equations at different fillings $\nu$, we get good qualitative agreement with the exact Monte Carlo results presented here and also in the companion paper~\cite{Calugaru_2026}. In this appendix, we concentrate at the $\nu=3$ case, with the results presented in Fig.~\ref{fig:supp:PartonMeanField:MF rotor results 2}. We see the dramatic enhancement of the metallic region as $\alpha$ approaches the isotropic value, upon which point the critical interaction needed to induce a Mott insulator approaches $U_c/W\approx6$, the $U(6)$ value also found in FG. It should be noted that this approach cannot account for any magnetic order in the spinon channel, whereas in the $\text{AA-tSnSe}_2$ system we know that the formation of magnetic long-range-order is a significant ingredient of the system. This is presumably also why the parton approach \textit{overestimates} the critical $U_c/W$.
 \section{Order-by-Disorder phenomena in the model}
\label{section:supp:ObD}
Towards the end of the main text, we briefly commented on the nature of the strong-coupling  $T=0$ state. In this work, we have so far ignored processes such as the $\tilde{M}_z$ breaking hopping $t'$ along $\Delta \vec{R} = \pm C^{\eta}_{3z}\left(\mathbf{a}_{M_1}\right)$ and $\Delta \vec{R} = \pm C^{\eta}_{3z}\left(\mathbf{a}_{M_1} + \mathbf{a}_{M_2}\right)$, as well as the next-nearest intra-chain hopping $t''$ along $\Delta \vec{R} = \pm C^{\eta}_{3z}\left(2\mathbf{a}_{M_2}\right)$ and the Hund's coupling $J_H$ (see also the discussion in SI Sec.~\ref{section:supp:model}). These terms are ---albeit being quite small--- will be present in a more realistic model of $\text{AA t-SnSe}_2$. While we do not expect them to qualitatively alter the intermediate-coupling picture, they will in general have a more significant effect in the strongly-coupled regime. As we have seen, at strong coupling our DQMC reveals robust long-range AFM order on each rectangular sublattice within a valley, leading to $3 \times 2$ independent order parameters. This however gets frustrated upon the inclusion of $J',J'',J_H$. In this section, we concentrate on the special case of $J''=J_H=0$, which can be analytically explored at lowest order in the quantum fluctuations\footnote{This could be further relaxed by allowing a small but non-zero $J''$}. The Hamiltonian describing the system is then
\begin{equation}
    H_{\text{eff}} = \sum_{\eta}\sum_{\mathbf{R},\Delta \mathbf{R}} J_{\Delta \mathbf{R}} \mathbf{S}^{(\eta)}_\mathbf{R} \cdot \mathbf{S}_{\mathbf{R}+\Delta\mathbf{R}}^{(\eta)}  \hspace{0.2cm} \text{w/}\hspace{0.2cm} J_{\Delta \mathbf{R}} \in \{J,J_\perp,J'\} \,,
    \label{eq:supp:odb:Ham}
\end{equation}
and the allowed processes are shown in Fig.~\ref{fig:supp:ObD: diagram and  Classical Phase Diagram}(a). Note that for $J_H=0$, the valleys remain spin-decoupled and hence the three-valley spin model can be reduced to a single valley one, taking $\eta=0$ from now on. From the strong-coupling description of Sec.~\ref{section:supp:perturbation theory}, the spin-couplings are related to the fermion hoppings through $J\sim t^2/(U-V)$ (and similarly for $J_\perp,J'$).\par
First, some motivation: It is common wisdom that in most many-body systems, fluctuations (be it thermal, classical,etc) tend to suppress order: A $T=0$ classical ground state of the system is unique, besides other `trivial' ground states related to it via the action of some symmetry subgroup $H \leq G$, with $G$ the symmetry group of the Hamiltonian. In such cases, fluctuations can only suppress order, by adding other states to the mix. In certain systems however, frustrated spin systems being the best example, the classical ground state manifold can have `larger' symmetry compared to the quantum Hamiltonian, i.e. there are classical ground states that are non-trivially degenerate. In such cases,  fluctuations (disorder) may lift this degeneracy, reinstating a ground state manifold that does not have more symmetries than the Hamiltonian. This phenomenon has been termed \textit{Order-by-Disorder} by Villain et al in Ref.~\cite{Villain_80} for the classical Ising model and extended to vector spins by Henley and Sheng in Ref.~\cite{Henley_89}.\par
With this perspective in mind, we can approach solving the frustrated Hamiltonian of Eq.~\ref{eq:supp:odb:Ham}. While triangular lattice systems involving order-by-disorder have been studied before in Refs.~\cite{Jolicoeur_1989,Jolicoeur_1990,Chubukov_1992,Henley_89}, the classical state we are expanding around is different, since in the $\text{AA t-SnSe}_2$ model, $t'\ll t$ due to the $\tilde{M}_z$ symmetry. We begin by studying the classical model, and identifying the degenerate manifold of ground states which posses an additional $U(1)$ symmetry not present in the Hamiltonian of Eq.~\ref{eq:supp:odb:Ham}. We then perturb around these states using linear spin-wave theory and analyze what configuration is energetically favored. We find that the fluctuations favor a stripe order, with $\phi=0,\pi$ in Fig.~\ref{fig:supp:ObD: diagram and  Classical Phase Diagram}(a).
\subsection{Classical Ground state}
\label{subsection:Classical Ground State}
\begin{figure}
    \centering
    \includegraphics[width=0.75\linewidth]{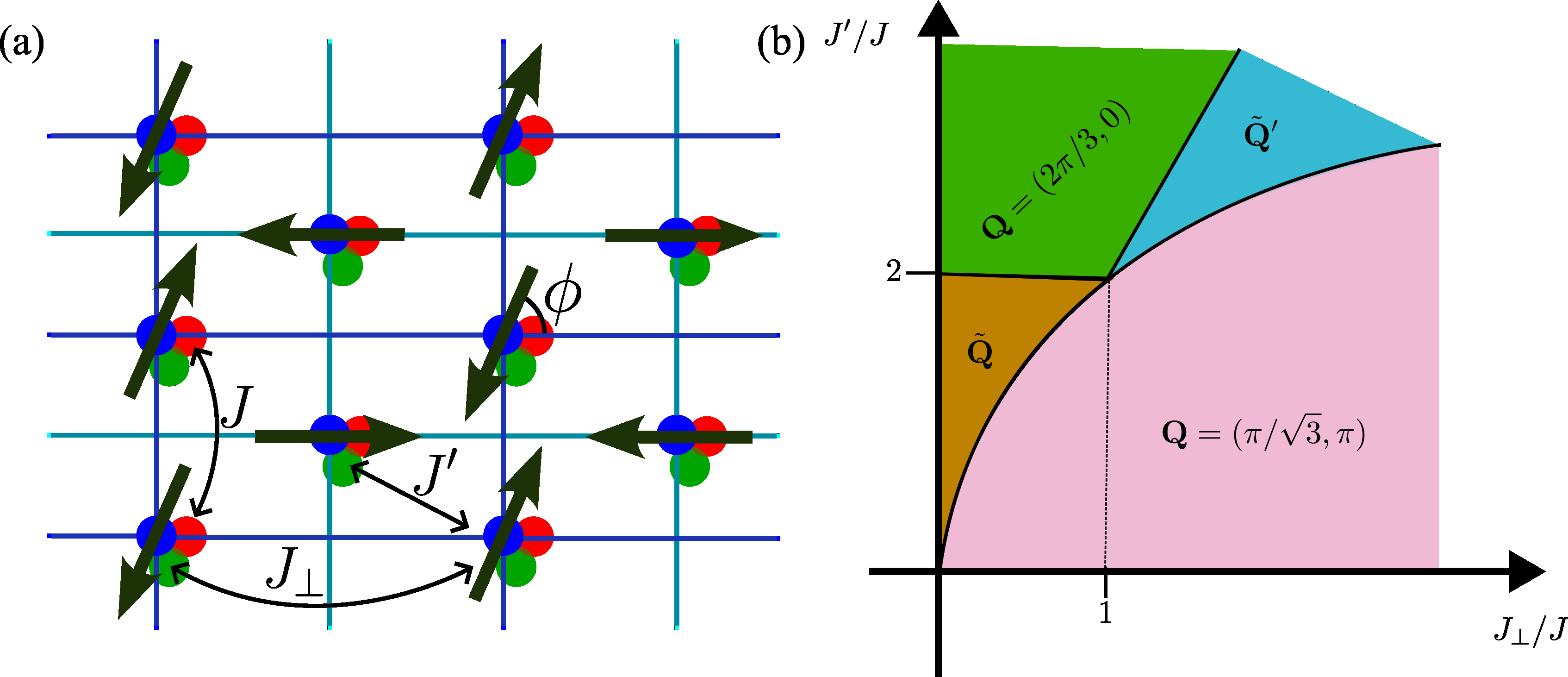}
    \caption{\textbf{(a)} The spin model considered in this Section, for a single valley $\eta=0$. The spin configuration shown here is that of the N\'eel phase in the main text, with an arbitrary orientation $\phi$ between the two AFM orders. \textbf{(b))} The classical phase diagram, for a single-spiral ansatz. The $\mathbf{Q}=(\pi/\sqrt{3},\pi)$ order corresponds to the N\'eel AFM phase. The $\tilde{\mathbf{Q}},\tilde{\mathbf{Q}}'$ phases are incommensurate spirals. For $\text{AA-tSnSe}_2$, the relevant parameter region lies within the $\mathbf{Q}=(\pi/\sqrt{3},\pi)$ classical region, as seen in Table ~\ref{supp:model:tab:parameters_master}.}
    \label{fig:supp:ObD: diagram and  Classical Phase Diagram}
\end{figure}

What is the classical ground state of the spin model? Assuming a single-spiral solution~\cite{Lyons_1960}, we can take the ansatz $\mathbf{S}(\mathbf{r})=S\left(\hat{e}_1\cos{(\mathbf{Q}\cdot \mathbf{r})}+\hat{e}_2\sin{(\mathbf{Q}\cdot \mathbf{r})}\right)$ for which the energy becomes
\begin{equation}
    E_{\mathbf{Q}}/N = 2S^2J_1\left(\cos{(Q_y)+2\left(\frac{J'}{J}\right)\cos{(Q_y/2)}\cos{(\sqrt{3}Q_x/2)}+\left(\frac{J_\perp}{J}\right) \cos{(\sqrt{3}Q_x)}}\right)\,,
\end{equation}
where $N$ is the number of sites. Minimizing the energy leads to the classical phase diagram of Fig.~\ref{fig:supp:ObD: diagram and  Classical Phase Diagram}(b). The classical ground state relevant to our model is the $\mathbf{Q}=(\pi/\sqrt{3},\pi)$ phase, which is the correct classical ground state as long as $(J_\perp/J)>(J'/2J)^2$. As a brief aside, the $\tilde{\mathbf{Q}}$ incommensurate spiral phase has a pitch $2\arccos{(-J'/2J)}$. This pitch starts from $\pi$ at $J'/J=0$, goes to the $K$ point at $J'/J=1$ and ends up at the $M$ point at $J'/J=2$. The $J'/J=1$ value, corresponding to the $120^\circ$ AFM order, is the classical ground state studied in Ref.~\cite{Jolicoeur_1989}.

In the `pink' region of Fig.~\ref{fig:supp:ObD: diagram and  Classical Phase Diagram}(b), the energy of the $\mathbf{Q}=(\pi/\sqrt{3},\pi)$ spiral is indeed the minimum, but there are other degenerate states that are missed by this single-spiral approach. These states are characterized by an arbitrary relative orientation $\phi$ between AFM orders on two rectangular sublattices, and are exactly what is predicted by the DQMC data and the strong-coupling model. Such configurations, shown in Fig.~\ref{fig:supp:ObD: diagram and  Classical Phase Diagram}(a) have classical energy
\begin{equation*}
\begin{split}
    E(\phi) &= S^2 \sum_{\vec{R}}\sum_{\Delta \vec{R}}J_{\Delta \vec{R}}\cos({\phi_{\Delta \vec{R}}}) \\
    &= S^2\frac{N}{2}\{4J\cos{(\pi) + 4J_\perp \cos{(\pi)}+ 2J'\sum_{\pm}\left(\cos{(\pi\pm \phi)+\cos{(\pm\phi)}}\right)}\}\\
    &=-2JS^2N \left(1+J_\perp/J\right)
\end{split}
\end{equation*}
which is independent of $\phi$, and equal to the single spiral state $E_{\mathbf{Q}}$!
\subsection{Spin Wave Hamiltonian}
Any classical spin configuration $\{\phi_{\vec{R}}\}$ can be viewed as ferromagnetic in an appropriate frame. To go to this frame,we unitarily transform the spin operators
\begin{equation}
    \begin{split}
        U_{\vec{R}} = &e^{i\frac{\phi_\vec{R}}{2} \sigma^z} \\
        \mathbf{S}_\vec{R}\rightarrow&U_\vec{R}\mathbf{S}_\vec{R} U^\dagger_\vec{R}=\begin{pmatrix}
            \cos{(\phi_\vec{R})}\hat{X}_\vec{R} -\sin{(\phi_\vec{R})}\hat{Y}_\vec{R}\\ \sin{(\phi_\vec{R})}\hat{X}_\vec{R} +\cos{(\phi_\vec{R})}\hat{Y}_\vec{R} \\ \hat{Z}_\vec{R}
        \end{pmatrix} \,,\\
    \end{split}
\end{equation}
where, without loss of generality, the spin configuration is taken to be coplanar and on the XY plane.
The Hamiltonian is now defined in a frame with Ferromagnetic classical order along $\hat{x}$, at the expense of the appearance of anisotropic and DM-type terms:
\begin{equation}
\begin{split}
    H = \sum_{\vec{R}}\sum_{\Delta \vec{R}}J_{\Delta \vec{R}}[\hat{Z}_{\vec{R}}\hat{Z}_{\vec{R}+\Delta \vec{R}} &+\cos{\left(\phi_{\vec{R}} - \phi_{\vec{R}+\Delta\vec{R}}\right)}(\hat{X}_{\vec{R}}\hat{X}_{\vec{R}+\Delta \vec{R}} + \hat{Y}_{\vec{R}}\hat{Y}_{\vec{R}+\Delta \vec{R}} ) \\
    &+ \sin{\left(\phi_{\vec{R}} - \phi_{\vec{R}+\Delta\vec{R}}\right)}(\hat{X}_{\vec{R}}\hat{Y}_{\vec{R}+\Delta \vec{R}} - \hat{Y}_{\vec{R}}\hat{X}_{\vec{R}+\Delta \vec{R}})] \,.
\end{split}
    \label{eq: rotated heisenberg}
\end{equation}
The family of classical ground states under consideration and shown in Fig.~\ref{fig:supp:ObD: diagram and  Classical Phase Diagram}(a) can be captured via
\begin{equation}
    \phi_{\vec{R}} = \phi_{m\mathbf{a}_{M_1} + n\mathbf{a}_{M_2}} = n\pi +\left(m\text{ mod }2\right)\phi \,.
\end{equation}
Thus, the Spin Hamiltonian in the FM frame of Eq.~\ref{eq: rotated heisenberg} becomes:
\begin{equation}
\begin{split}
    H = \sum_{\vec{R}=m\mathbf{a}_{M_1} + n\mathbf{a}_{M_2}}&\sum_{\Delta \vec{R} ||\mathbf{a}_{M_2},2\mathbf{a}_{M_1}+\mathbf{a}_{M_2}}J_{\Delta \vec{R}}[\hat{Z}_{\vec{R}}\hat{Z}_{\vec{R}+\Delta \vec{R}} -(\hat{X}_{\vec{R}}\hat{X}_{\vec{R}+\Delta \vec{R}} + \hat{Y}_{\vec{R}}\hat{Y}_{\vec{R}+\Delta \vec{R}} )\\
    &+\sum_{\Delta \vec{R} ||\mathbf{a}_{M_1}}J_{\Delta \vec{R}}[\hat{Z}_{\vec{R}}\hat{Z}_{\vec{R}+\Delta \vec{R}} +\cos{\phi}(\hat{X}_{\vec{R}}\hat{X}_{\vec{R}+\Delta \vec{R}} + \hat{Y}_{\vec{R}}\hat{Y}_{\vec{R}+\Delta \vec{R}} ) + (-1)^m(\hat{X}_{\vec{R}}\hat{Y}_{\vec{R}+\Delta \vec{R}} - \hat{Y}_{\vec{R}}\hat{X}_{\vec{R}+\Delta \vec{R}} )\\
    &+\sum_{\Delta \vec{R} ||\mathbf{a}_{M_1}}J_{\Delta \vec{R}}[\hat{Z}_{\vec{R}}\hat{Z}_{\vec{R}+\Delta \vec{R}} -\cos{\phi}(\hat{X}_{\vec{R}}\hat{X}_{\vec{R}+\Delta \vec{R}} + \hat{Y}_{\vec{R}}\hat{Y}_{\vec{R}+\Delta \vec{R}} ) - (-1)^m(\hat{X}_{\vec{R}}\hat{Y}_{\vec{R}+\Delta \vec{R}} - \hat{Y}_{\vec{R}}\hat{X}_{\vec{R}+\Delta \vec{R}} )\,.\\
\end{split}
    \label{eq: rotated heisenberg explicit}
\end{equation}
It should be clear that as is, the Hamiltonian of Eq.~\ref{eq: rotated heisenberg explicit} is \textit{ not} translationally symmetric, and one would need to double the unit cell to make it so. That is because $\phi_{\vec{R}}-\phi_{\vec{R}+\Delta\vec{R}}$ is \textit{not} translationally symmetric. However, to treat the effect of quantum fluctuations at the lowest order we will not need to do that, as it will soon become clear.\par
\begin{figure}
    \centering
    \includegraphics[width=0.75\linewidth]{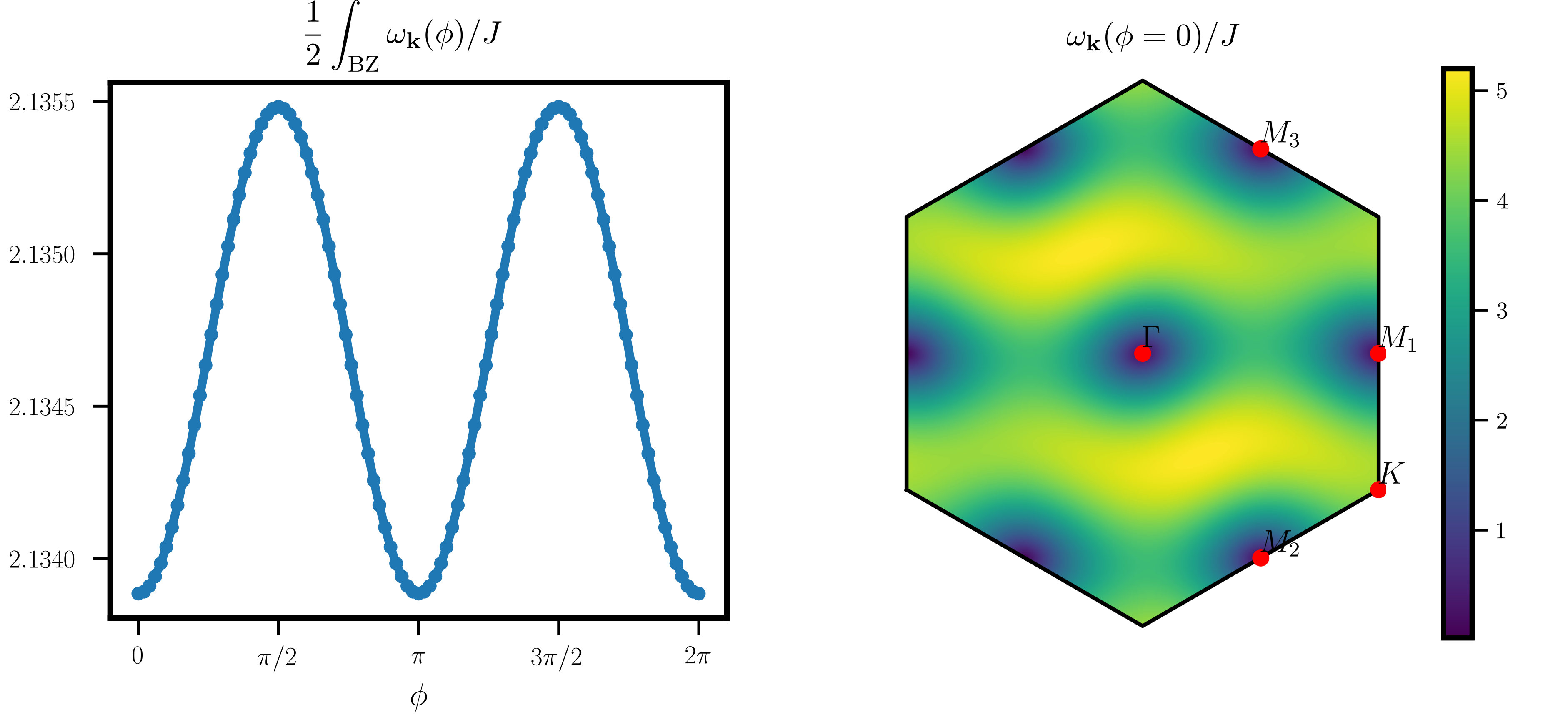}
    \caption{\textbf{Left:}The Quantum zero point energy as a function of the classical ground state parametrized by $\phi$, for $J_\perp/J=0.1,J'/J=0.05$. \textbf{Right:} The spin fluctuation spectrum for  $J_\perp/J=0.1,J'/J=0.05$ around the collinear classical state $\phi^*=0$ }
    \label{fig:supp:ObD energy and spectrum}
\end{figure}

\noindent We now introduce the Holstein-Primakoff bosons to describe fluctuations about the classical spin state:
\begin{equation*}
\begin{split}
    \hat{X}_\vec{R} =& S-\hat{n}_\vec{R}\\
    \hat{S}^+_\vec{R} =
    \hat{Y}_\vec{R} +i \hat{Z}_\vec{R}=& \sqrt{2S}\left(1-\hat{n}_\vec{R}/2S\right)^{1/2}\hat{a}_\vec{R}\\
     \hat{S}^-_\vec{R} =\hat{Y}_\vec{R} -i \hat{Z}_\vec{R}=& \sqrt{2S}\hat{a}^\dagger_\vec{R}\left(1-\hat{n}_\vec{R}/2S\right)^{1/2} \,,\\
\end{split}
\end{equation*}
where $\hat{a}^\dagger$ is the boson creation operator and $\hat{n}$ is the boson number operator. The truncation of the square-root power series can only be justified if the spin-fluctuations around the classical state are small, \textit{i.e} if $\langle \hat{n}_{\vec{R}}\rangle\ll2S=1$. Expanding the Hamiltonian in powers of the spin, one can show that $\{\phi_{\vec{R}}\}$ being a classical ground state means that the $\sim\mathcal{O}(S)$ part of $\hat{H}$ vanishes, and the generic expression is
\begin{equation*}
    \hat{H} = \hat{H}_0 + \hat{H}_2 + \hat{H}_3 + \hat{H}_4 + \mathcal{O}(S^{-1/2})
\end{equation*}
with $\hat{H}_n$ denoting an n-th order (in $\hat{a},\hat{a}^\dagger$) Hamiltonian. Note that $\hat{H}_3$, which involves terms such as $\hat{a}^\dagger \hat{a}^\dagger\hat{a}$ vanishes for collinear states but in this generic setup it does not. As we are aiming for the simplest possible treatment of this problem, at the level of linear spin-wave theory (LSWT) we ignore $\hat{H}_3 + \hat{H}_4$. This essentially treats the spin waves as non-interacting bosons. With this simplification, the effective Hamiltonian takes the form
\begin{equation}
    H=\underbrace{S^2\sum_{\vec{R}}\sum_{\vec{R}+\Delta\vec{R}}J_{\Delta\vec{R}}\cos{(\phi_{\Delta\vec{R}})}}_{E_{\text{cl}}}
     + S\sum_{\vec{R}}\sum_{\vec{R}+\Delta\vec{R}}\hat{a}^\dagger_\vec{R} A_{\Delta\vec{R}} \hat{a}_{\vec{R}+\Delta\vec{R}} +\frac{1}{2}[\hat{a}^\dagger_{\vec{R}} B_{\Delta\vec{R}} \hat{a}^\dagger_{\vec{R}+\Delta\vec{R}} +\text{h.c.}] + \mathcal{O}(S^{1/2}) \,,
    \label{eq: Spin-wave equation}
\end{equation}
with
\begin{equation}
    \begin{split}
    A_{\Delta\vec{R}} = &J_{\Delta\vec{R}}\left(\cos{\left(\phi_{\Delta\vec{R}}\right)} + 1\right) -\delta_{\Delta\vec{R} ,\mathbf{0}}\times2 \sum_{\Delta\vec{R}'}J_{\Delta\vec{R}'}\cos{\left(\phi_{\Delta\vec{R}'} \right)}\\  \\
    B_{\Delta\vec{R}} = &J_{\Delta\vec{R}}\left(\cos{\left(\phi_{\Delta\vec{R}}\right)} - 1\right)\,.
    \end{split}
    \label{eq: BdG matrices definition}
\end{equation}
Interestingly, since the $\sin{\left(\phi_{\vec{R}} - \phi_{\vec{R}+\Delta\vec{R}}\right)}$ terms of Eq.~\ref{eq: rotated heisenberg explicit} only appear at the cubic order ignored in LSWT, the quadratic Hamiltonian of Eq.~\ref{eq: Spin-wave equation} is invariant under translations and it can thus be reformulated in momentum space, with
\begin{equation}
    \hat{H}=E_{\text{cl}}
     + S\sum_{k} \hat{c}^\dagger_{\mathbf{k}}A_{\mathbf{k}}(\phi)\hat{c}_{\mathbf{k}} +\frac{1}{2}\left( \hat{c}^\dagger_{\mathbf{k}} B_{\mathbf{k}}(\phi)\hat{c}^\dagger_{\mathbf{-k}}+\text{h.c.}\right)\,.
    \label{eq: Spin-wave equation k space}
\end{equation}
Here,
\begin{equation}
    \begin{split}
        A_{\mathbf{k}}(\phi) &= +4J\left[ (1+J_\perp/J)+\left(J'/J\right)\left(\cos{(\phi)}\sin{(\frac{\sqrt{3}k_x}{2})}\sin{(\frac{k_y}{2})}+\cos{(\frac{\sqrt{3}k_x}{2})}\cos{(\frac{k_y}{2})}\right)\right]\\
        B_{\mathbf{k}}(\phi) &= -4J\left[ \cos{(k_y)+\left(J_\perp/J \right)\cos{(\sqrt{3}k_x)}}+\left(J'/J\right)\left(-\cos{(\phi)}\sin{(\frac{\sqrt{3}k_x}{2})}\sin{(\frac{k_y}{2})}+\cos{(\frac{\sqrt{3}k_x}{2})}\cos{(\frac{k_y}{2})}\right)\right] \,.\\
    \end{split}
    \label{eq: BdG matrices definition k space}
\end{equation}
A Bogoliubov transform then leads to a harmonic oscillator expression for the fluctuations, with the zero-point energy
\begin{equation}
    E(\phi) = E_{\text{cl}} + \frac{1}{2}\int_{\text{BZ}} \omega_{\mathbf{k}}(\phi) =  E_{\text{cl}} + \frac{1}{2}\int_{\text{BZ}} \left(A_{\mathbf{k}}(\phi)-B_{\mathbf{k}}(\phi)\right)^{1/2} \,.
\end{equation}
The quantum zero-point energy as a function of $\phi$ is shown in Fig.~\ref{fig:supp:ObD energy and spectrum}. As is the case other \textit{order-by-disorder} calculations (see for example Refs.~\cite{Henley_89,Jolicoeur_1990}) quantum fluctuations tend to favor a collinear order, with $\phi=0,\pi$! This is a stripe order, in the sense that if $\phi=0$ the spins are parallel along $\mathbf{a}_{M_1}$ and they are anti-parallel along $\mathbf{a}_{M_2}$ and $\mathbf{a}_{M_1} + \mathbf{a}_{M_2}$. The spin-fluctuation spectrum has zero modes at $\mathbf{q}=\Gamma$ and $\mathbf{q}=M_1 = \left(\pi/\sqrt{3},\pi\right)$, as well as the other two M points. The first two zero modes should be associated with the Goldstone modes for the $SU(2)\longrightarrow U(1)$ Spontaneous Symmetry Breaking pattern, while last two are accidental zero modes which should get lifted upon the inclusion of the $\hat{H}_3 + \hat{H}_4$ perturbations~\cite{Chubukov_1994}.

 \end{document}